\documentclass[11pt,a4paper]{article}
\pdfoutput=1
\usepackage{jheppub}
\usepackage{graphicx}
\usepackage{epsfig}
\usepackage{amsmath, amssymb}
\usepackage{dcolumn}
\usepackage{bm}
\usepackage{amsfonts}
\usepackage[toc,page]{appendix} 
\usepackage{color}

\usepackage{subfigure}
\usepackage{amssymb}
\usepackage{epstopdf}

\input epsf
\allowdisplaybreaks

\newcommand{\be}{\begin{equation}}
\newcommand{\ee}{\end{equation}}

\newcommand{\mn}{{\mu\nu}}

\begin{document}

\title{Anomalous non-conservation of fermion/chiral number in Abelian gauge theories at finite temperature}

\author{Daniel G. Figueroa$^1$\,,}
\affiliation{$^1$CERN Theory Department, CH-1211 Geneve 23, Switzerland}
\emailAdd{daniel.figueroa@cern.ch}

\author{Mikhail Shaposhnikov$^2$}
\affiliation{$^2$Institute of Physics, Laboratory of Particle Physics and Cosmology, \'Ecole Polytechnique F\'ed\'erale de Lausanne, CH-1015 Lausanne, Switzerland}
\emailAdd{mikhail.shaposhnikov@epfl.ch}

\date{\today}

\abstract{We discuss the non-conservation of fermion number (or chirality breaking, depending on the fermionic charge assignment) in Abelian gauge theories at finite temperature. We study different mechanisms of fermionic charge disappearance in the high temperature plasma, using both analytical estimates and real-time classical lattice numerical simulations. We investigate the random walk of the Chern-Simons number $Q \propto \int d^4x F_{\mu\nu}{\tilde F}^{\mu\nu}$, and show that it has a diffusive behaviour in the presence of an external magnetic field $B$. This indicates that the mechanism for fermionic number non-conservation for $B \neq 0$, is due to fluctuations of the gauge fields, similarly as in the case of non-Abelian gauge theories. We determine numerically, with lattice simulations, the rate $\Gamma$ of chirality non-conservation associated with this diffusion. We find the rate to be a factor $\sim 60$ larger compared to previous theoretical estimates, what calls for a revision of the implications of Abelian fermion number and chirality non-conservation  for baryogenesis, magnetogenesis and chiral symmetry evolution.}


\maketitle


\section{Introduction}
\label{sec:Intro}
The triangular anomaly~\cite{Adler:1969gk,Bell:1969ts} in fermionic current leads to many important consequences in particle physics. For example, in Abelian gauge theories, such as quantum electrodynamics (QED), it describes the decay $\pi^0\to2\gamma$. In non-Abelian theories like quantum chromodynamics (QCD), it plays a decisive role in the resolution of the $U_A(1)$ problem~\cite{Veneziano:1979ec,Witten:1979vv}, whereas in the electroweak  theory it leads to baryon and lepton number non-conservation~\cite{tHooft:1976rip,tHooft:1976snw}.  

In hot and dense matter in the early universe the fluctuations of gauge and scalar fields -- sphalerons \cite{Klinkhamer:1984di}  -- lead to rapid fermion number non-conservation in the Standard Model (SM) \cite{Kuzmin:1985mm}, and to chirality non-conservation in QCD~\cite{McLerran:1990de}. The existence of these transitions is associated with the non-trivial {\em vacuum} structure of non-Abelian gauge theories \cite{Callan:1976je,Jackiw:1976pf}.  The vacuum field configurations connected by large gauge transformations have the same energy but different Chern-Simons (CS) number, allowing the disappearance of fermion/chiral number.

The situation in Abelian gauge theories is visibly different.  In the electroweak theory the anomaly in the fermionic and/or chiral current contains in fact a U(1) contribution. In the symmetric phase of the SM model it is associated with the hypercharge field, and in the Higgs phase with the electomagnetic field of  QED. However, in an Abelian gauge theory, there are no large gauge transformations, nor vacuum configurations with different Chern-Simons numbers. As a result, there is no {\em irreversible} fermion (or chiral) number non-conservation associated to an Abelian gauge sector, as in the case of non-Abelian theories. This does not prevent the fermion/chiral number to be transferred into gauge configurations carrying Chern-Simons number, and to reappear back again due to the changes in the gauge field background. These processes may have an important impact on the problems of baryogenesis~\cite{Giovannini:1997eg,Kamada:2016eeb,Kamada:2016cnb}, magnetic field generation in the early Universe~\cite{Joyce:1997uy}, and chiral asymmetry evolution at temperatures in the MeV range \cite{Boyarsky:2011uy}; they may also be visible in heavy ion collisions~\cite{Fukushima:2008xe}.  For all applications it is important to have a reliable value of the rate of the anomalous U(1) processes, and it is the goal of our paper to provide such an estimate accounting for {\em small scale fluctuations} of the fields, what has never been  done before\footnote{All previous estimates of this rate were done with the use of magnetohydrodynamics, which accounts for distance scales exceeding the mean free path of the charged particles in the medium.}.

Of course, below  the electroweak cross-over temperature $T_c$, there is a perturbative contribution to the rate of chirality change, proportional to the mass squared of the charged fermion $m$, $\Gamma_{pert} \sim \alpha^2 m^2/T$, where $\alpha$ is the fine-structure constant and $T$ is the temperature. A similar contribution exists also above $T_c$, proportional to the square of the fermion Yukawa coupling $f$, $\Gamma_{pert} \sim f^2 T$. In general, both perturbative and non-perturbative rates must be taken into account.  Even for the moderate strength of the magnetic field  the anomalous rate may exceed the perturbative one for the lightest fermion - electron, leading to a number of interesting effects above~\cite{Joyce:1997uy} and below $T_c$~\cite{Boyarsky:2011uy}.

In spite of the fact that a lot of work on the dynamics of Abelian gauge theories with chiral fermions has been already done, a number of questions still remain unanswered. We briefly review these questions in what follows. To set up the scene and fix notation, let us consider scalar electrodynamics with a massless vector-like fermion field $\Psi$, described by the Lagrangian\footnote{An equivalent way is to take chiral electrodynamics and consider the fermionic charge.}
\be
\mathcal{L} = -{1\over4} F_{\mu\nu}F^{\mu\nu} - {\bar\Psi}\gamma^\mu D_\mu\Psi - (D_\mu\phi)^*(D^\mu\phi) - V(\phi)~,
\label{cl}
\ee
where $D_\mu=\partial_\mu - i e A_\mu$,  $V(\phi) = m^2|\phi|^2 + \lambda|\phi|^4$, $A_\mu$ is the Abelian gauge field, $F_{\mu\nu} = \partial_\mu A_\nu - \partial_\nu A_\mu$ its field strength, and we use metric signature $(-,+,+,+)$. The chiral fermionic current, defined as 
\be
J_\mu^5= {\bar\Psi}\gamma^\mu \gamma_5\Psi\,,
\ee
is conserved at the classical level, i.e.~$\partial_\mu J^\mu_5 = 0$. When quantum effects are taken into account, it has however an anomaly \cite{Adler:1969gk,Bell:1969ts},
\be
\partial_\mu J^\mu_5 = {e^2\over 8\pi^2} F_{\mu\nu}{\tilde F}^{\mu\nu}~,
\label{anom}
\ee
where ${\tilde F}^{\mu\nu}=\frac{1}{2}\epsilon^{\mu\nu\rho\sigma}F_{\rho\sigma}$ is the dual of the field strength, and $\epsilon^{\mu\nu\rho\sigma}$ is the Levi-Civita antisymmetric tensor in 4 dimensions, with $\epsilon^{0123} = -\epsilon_{0123} = +1$.

The chiral analogue of Lagrangian Eq.~(\ref{cl}) can describe the hypercharge sector of the SM at temperatures above the electroweak cross-over; if the mass of the scalar field $m$ is sufficiently large, it can be integrated out so that we are left with a theory including only fermions and a U(1) gauge field, hence representing ordinary quantum electrodynamics. A number of results describing the anomalous dynamics of fermions and Abelian gauge fields are available for this type of theory. \\

\noindent {\bf I) Instability domain}\vspace*{0.3cm}

\noindent ~~{\tt I.1) Symmetric phase}. Let us take first the symmetric phase of the model $m^2>0$, when the temperature is considered to be zero. We consider an initial state containing no gauge field, but a non-zero fermionic charge density. We characterize the latter by a chemical potential $\mu$, adding to the statistical sum the term $\mu J_0^5/2$, with initial value $\mu = \mu_o \neq 0$. The left- and right-handed fermionic chemical potentials  are $\mu_L=-\mu_R=\mu/2$. A state with non-zero chemical potential is unstable against creation of gauge fields with non-zero Chern-Simons number~\cite{Rubakov:1985nk}. This can be seen as follows. The (free) energy for the gauge fields with fermions integrated out, contains the Chern-Simons (CS) term \cite{Redlich:1984md, Niemi:1985yp}
\be
H_{CS} = \mu N_{CS}, ~~N_{CS}=\int d^3x n_{CS},~~n_{CS} = \frac{\alpha}{2\pi} \vec A \cdot \vec B~, ~~~~~\alpha={e^2\over 4\pi}, 
\label{ncs}
\ee
where $\vec B \equiv \vec\nabla \times \vec A$ is the magnetic field. The relation between $F_{\mu\nu}{\tilde F}^{\mu\nu}$ and the CS number is given by
\be
{e^2\over 16\pi^2} F_{\mu\nu}{\tilde F}^{\mu\nu} = \partial_\mu K^\mu,~~K^0=n_{CS}~.
\ee

In Fourier space the term (\ref{ncs}) is linear in the momentum $\vec{k}$ of the gauge field and is not positive definite, contrary to the energy density $\propto {\vec B}^2$, which is quadratic in $\vec{k}$. For sufficiently infrared (IR) modes 
\be
\label{eq:kInst}
k < k_{\rm inst} \equiv {\alpha\over \pi}\mu\,,
\ee 
the CS term $H_{CS}$ dominates the free energy, leading to an instability. Eventually, the fermionic number disappears as it is converted into long-ranged configurations of the gauge field, with a density of Chern-Simons number equal to the initial chiral density $n_F$,
\be
n_{CS} = \frac{\mu_o^3}{12\pi^2} = n_F\,,
\label{fnd}
\ee
and a scale of the order of the system size $L $, with  magnetic field $B \sim {4\pi\over e}\sqrt{n_F\over L}$ and gauge field $A\sim  \frac{4\pi}{e}\sqrt{n_F L}$  \cite{Rubakov:1985nk}. While the initial energy density of the system was $\rho_F \sim \mu_o^4$, the energy density of the gauge field carrying the same fermionic number is now smaller, as it is suppressed by the size of the system, $\rho_A \sim \mu^3/L e^2$.

\vspace*{0.3cm}

\noindent ~~{\tt II.2) Higgs phase}. In the Higgs phase, when $m^2 < 0$, an analogous instability  can only be developed if~\cite{Rubakov:1985nk} 
\be
k_{inst} > 2 M_A ~~ \Leftrightarrow ~~ \mu_o > {2\pi\over\alpha}M_A\,,
\label{hin}
\ee
with $M_A= e v$ the mass of the vector field in the broken phase, where the scalar Higgs-like field takes a vacuum expectation value $v^2 = -\frac{m^2}{\lambda}$. The unstable IR modes obey the relation 
\be\label{eq:kInstMa}
{k_{\rm inst}\over2}\left(1-\sqrt{1-{4M_A^2\over k^{2}_{\rm inst}}}\right)<k < {k_{\rm inst}\over2}\left(1+\sqrt{1-{4M_A^2\over k^{2}_{\rm inst}}}\right) \,.
\ee
For small $M_A \ll k_{\rm inst}$, we have ${M_A^2/k_{\rm inst}}<k<k_{\rm inst}-{M_A^2/k_{\rm inst}}$.\\

\noindent {\bf II) Rate of instability at finite temperature and density}\vspace*{0.3cm}

\noindent ~~{\tt II.1) Symmetric phase}. Let us take again the same initial state in the symmetric phase, but now at finite temperature. A linear analysis similar to that carried out at zero temperature, reveals the instability of the system for long-range gauge field modes with momenta $k<k_{inst}$. Since Abelian magnetic fields in a high temperature plasma are not screened (this can be proven in all orders of perturbation theory~\cite{fradkin,Kalashnikov:1979kq}, with non-perturbative lattice simulations confirming the same behavior~\cite{Kajantie:1996qd}), the instability takes place for any IR mode $k < k_{\rm inst}$. 

The rate of the instability growth depends on the relation between $k_{\rm inst}$ and other dynamical scales in the medium. One can distinguish two qualitatively different regimes, depending on whether the typical instability length scale $l_{\rm inst} \sim 1/k_{\rm inst}$ is larger or smaller than the mean free path $\lambda\sim 1/(\alpha^2 T)$ of the fermions in the plasma:\\

{\it a)} $l_{\rm inst} > \lambda$. This situation occurs when the chemical potential is sufficiently small, $\mu < e^2 T$.  For the analysis of instabilities one can use the effective description of long range modes in a plasma, namely {\it magneto-hydrodynamics} (MHD). The rate of the instability growth is suppressed in comparison with the zero temperature case by the electric conductivity $\sigma$ of the plasma~\cite{Joyce:1997uy,Giovannini:1997eg}, like
\be
A_k \sim e^{\omega_k t}\,,~~~\omega_k =  \frac{k}{\sigma}(k_{\rm inst}-k)~,
\label{small}
\ee
where $k_{\rm inst}$ is still given by Eq.~(\ref{eq:kInst}).  For QED with $N_l$  fermionic flavours  \cite{Baym:1997gq},
\be\label{cond}
\sigma \simeq \frac{3\zeta(3)}{\log 2}\frac{T}{\alpha \log(1/\alpha N_l)}~,
\ee
where $\zeta$ is the Riemann $\zeta$-function, so $3\zeta(3)/\log 2 \simeq 5.2$. The result (\ref{small}) can be derived easily from Eqns~(\ref{homog}), which we introduce later on when discussing the magnetohydro-dynamical regime of the system. In this regime, parametrically $\omega_k \propto \mu^2$.\\

{\it b)} $l_{\rm inst} < \lambda$. This situation corresponds to chemical potentials $\mu> e^2 T$. The instability growth rate can be found using the thermal photon propagator, and is equal to~\cite{Akamatsu:2013pjd}
\be
A_k \sim e^{\omega_k t}\,,~~~\omega_k = \frac{4k^2}{\pi m_D^2}(k_{inst}-k)\,,
\label{interm}
\ee
where $m_D$ is the {Debye} mass. For the theory with Lagrangian (\ref{cl}) one finds, in the one loop approximation, $m_D^2 =e^2 T^2/3+ e^2 \mu^2/\pi^2$. Thus, parametrically, $\omega_k \propto \mu^3$  when $e^2 \lesssim \mu/T \lesssim \pi^2/3$, whilst $\omega_k \propto \mu$ for  $\mu/T \gtrsim\pi^2/3$.\\

The transition between regimes $a)$ and $b)$ is smooth, and the expressions for $\omega_k$ are  parametrically the same at the matching point $\mu \sim e^2 T$. In all cases the density of real fermions eventually disappears and is replaced by a Chern-Simons condensate of the gauge field.\\

\noindent ~~{\tt ii.2) Higgs phase}. To the best of our knowledge, a detailed analysis of instabilities, in what concerns the Higgs phase, has not been carried out. We expect that the condition for the instability to develop given in Eq.~(\ref{hin}) still remains in force, but the rates Eqs.~(\ref{small},\ref{interm}) are to be modified.\\

As we have already mentioned, these behaviors are different from what happens in non-Abelian theories in similar situations. For instance, non-Abelian gauge theories in the symmetric phase are confining at zero temperatures. At non-zero temperatures, still in the symmetric phase, the non-Abelian magnetic fields acquire a ``magnetic'' mass $m_{\rm mag}\propto g^2 T$~\cite{Linde:1980ts}, where $g$ is the non-Abelian gauge coupling\footnote{An exact gauge invariant definition of the magnetic mass at very high temperatures can be given in terms of the lightest glueball mass in pure 3-dimensional gauge theory, see, e.g.~\cite{Kajantie:1997tt}.}. We expect this to lead to a threshold similar to Eq.~(\ref{hin}) with replacement of $M_A$ by the strong coupling scale $\Lambda$ similar to $\Lambda_{QCD}$ at zero $T$, or by $m_{\rm mag}$ if $T >>\Lambda$. On top of that, in the non-Abelian case the final state of the system, after the development of the instability, does not contain long-range non-Abelian fields, as they are screened by $m_{\rm mag}$. The system contains instead vacuum configurations with non-zero topological charge. 

Moreover, at non-zero temperature, in addition to classical instabilities, there are thermal fluctuations -- sphalerons -- which also drive the system to a state with zero net fermionic charge. The rate of these fluctuations $\Gamma_{sph}$ does not depend on the chemical potential of fermions (at least, for small enough $\mu$), and  the chemical potential behaves as $\mu \propto \exp(-\kappa \Gamma_{sph} t)$, with $\kappa \sim 1$ a known number, which depends on the matter sector of the corresponding gauge theory~\cite{Khlebnikov:1988sr}. 

A lot of work has been done for the study of fermion number non-conservation and sphaleron transitions in non-Abelian theories in the past.  We mention just a few. The prefactor for the sphaleron rate in the Higgs phase of the SM was found in \cite{Arnold:1987mh,Carson:1990jm}. In \cite{Arnold:1996dy} it was demonstrated that the sphaleron rate in the symmetric phase scales with the SU(2) gauge coupling like $\alpha_W^5$, contrary to $\alpha_W^4$ expected from naive scaling \cite{Khlebnikov:1988sr}. In  \cite{Bodeker:1998hm} B\"odeker argued that the rate has an extra $\log(1/\alpha_W)$ enhancement and suggested an effective field theory description accounting for this effect, which has been worked out further in \cite{Arnold:1998cy,Moore:1998zk}. The numerical simulations of sphaleron transitions in 1+1 dimensions were initiated in \cite{Grigoriev:1988bd} and carried out in \cite{Grigoriev:1989ub}.  In 3+1 dimensions  the first lattice simulations were done in  \cite{Ambjorn:1990pu}, with many improvements accounting for hard thermal loops and  B\"odeker effective theory, appearing later in \cite{Bodeker:1999gx, Moore:1999fs,Moore:1998swa}. The ultimate results for the sphaleron rate were reported  in \cite{DOnofrio:2014rug}. The combined dynamics of the fermionic chemical potential was addressed in lattice simulations in \cite{Ambjorn:1990pu} and refined considerably in \cite{Moore:1996qs}.

Much less efforts were invested to the study of the Abelian case. Besides the works we have already mentioned,  Ref. \cite{Long:2013tha} underlined the potential importance of fluctuations of the electromagnetic field for the problem of the chiral charge erasure. This question has been also studied recently in \cite{Long:2016uez}; our discussion of the same phenomena in the present paper  is  considerably different.

To summarise, the aim of the present work is to elucidate the difference between the Abelian and non-Abelian theories in what concerns the behaviour of the fermionic number at high temperatures. This happens to be not as trivial at it may seem. As we will discuss in Section~\ref{sec:analytics}, the ground state of an Abelian theory at {\em non-zero} temperatures, may have more in common with non-Abelian theories than normally considered, potentially leading to other possible mechanisms for anomalous fermion number non-conservation, in supplement to the instabilities discussed before. This certainly happens in the presence of a background magnetic field, leading to a ground state degeneracy with respect to the Chern-Simons number. To account for fluctuations of Abelian gauge theories at non-zero temperatures, we propose to study numerically a classical  Abelian Higgs model with a Chern-Simons term, replacing the theory described by Eq.~(\ref{cl}) with fermions, by a theory where fermions with non-zero density are integrated out. This theory and its lattice implementation are discussed in Section~\ref{sec:model}. In Section~\ref{sec:diff} we study the diffusion of the Chern-Simons number in the presence of external magnetic fields of various strengths. The rate of CS diffusion is known to give the rate of fermion number dilution in non-Abelian gauge theories~\cite{Khlebnikov:1988sr}, and a similar relation is expected to hold in the Abelian case as well. We find the rate of CS diffusion as proportional to the square of the magnetic field (an expected result~\cite{Joyce:1997uy,Giovannini:1997eg}  but quantified with novel inputs for the pre-factor). In Section~\ref{sec:conclus} we summarize our results, and discuss their implications and limitations.

\section{Dynamics of the chiral charge and magnetic fields: analytical estimates}
\label{sec:analytics}

\subsection{Abelian configurations with CS number}
\label{sec:conf}
The Abelian Chern-Simons number density (\ref{ncs}) vanishes for zero energy configurations. This means that the bosonic ground state of the system at zero temperature is unique (in the absence of an external magnetic field). This is opposed to the non-Abelian case, where pure gauge configurations with non-trivial topology carry an integer CS number. The situation changes if we have a non-zero magnetic field $\vec B$. A non-zero uniform magnetic field can in fact play the role of the ground state of an Abelian theory, as this configuration satisfies the equations of motion, and is stable due to magnetic flux conservation. 

For definiteness, let us consider an external magnetic field in the direction of the third axis $z$, $\vec B = (0,0,B_3)$. Now, if $B_3\neq 0$, the ground state of the system acquires a degeneracy with respect to the CS number: a non-zero constant gauge field $A_3$ does not cost any energy but leads to a  configuration with non-zero $n_{\rm CS} \propto \vec A\cdot \vec B$. Contrary to the case of a non-Abelian theory, $N_{CS} = \int d^3x n_{\rm CS}$ is not quantized and can have any value. This type of configuration can serve as an infinite reservoir to absorb the fermionic charge, exactly as for the case of non-Abelian theories. One should expect therefore, the dynamics of the fermion charge and of the Chern-Simons number, to be qualitatively similar for Abelian and non-Abelian theories at high temperatures, in the presence of a magnetic field in the symmetric phase. 

The case when an external magnetic field is absent is more subtle. Consider a gauge field with a typical amplitude $A$ and variation scale $l$. The energy density of this field is $\rho_A \sim  A^2/l^2$, whereas it can carry a CS number density 
\be
n_{CS} \sim \frac{\alpha}{2\pi} \frac{A^2}{l}.
\label{CSdens}
\ee
The dependence of the energy density on the CS number is therefore infrared-sensitive as
\be
\rho_A \sim \frac{1}{2}B^2\sim \frac{\pi n_{CS}}{\alpha l}.
\label{CSenergy}
\ee
In other words, a configuration with a fixed CS number density $n_{CS} \neq 0$ may have arbitrarily small energy density for a sufficiently long-wave vector field. Since at non-zero temperatures {\em the energy density} of the system is different from zero, the thermal  ground state may contain configurations with arbitrarily large CS numbers in the limit of infinite volume $l \to \infty$. This looks pretty much similar to the non-Abelian case, where the same is true, though not only the energy density, but also {\em the total energy} of a configuration carrying a Chern-Simons number may be vanishing. One may wonder therefore whether in the absence of background magnetic field there might be yet another mechanism for fermion number transfer to Abelian fields, not related to the instabilities discussed in the Introduction. In particular, do the fluctuations of the Abelian field, similarly as the sphalerons of non-Abelian fields, play any role? We discuss this question next.

\subsection{Instabilities or fluctuations?}
\label{subsec:inst}

The rate $\omega_k$ at which the fermion number is 'eaten up' by the gauge field due to the instabilities described in Section~\ref{sec:Intro}, scales as $\propto \mu^n$, with $n =1,~ 2$ or $3$, depending on the situation. As it has been already mentioned in the Introduction, it is well known that in non-Abelian theories there is yet another mechanism for fermion number dilution, related to thermal fluctuations -- sphalerons --, with a  $\mu$-independent rate (at least for small $\mu$). 

This can be seen at a qualitative level, from the following considerations~\cite{Arnold:1996dy,Arnold:1998cy}. Let us forget about fermions and consider a gauge field fluctuation with size $l$ and amplitude $A$, with CS number $N_{CS} \sim n_{CS} l^3$ and energy $E \sim \rho_A l^3$, see Eqs.~(\ref{CSdens}), (\ref{CSenergy}). In non-Abelian theories, fluctuations that go from one topological vacuum to another should have $\Delta N_{CS} \sim 1$, leading to a relation $ A \sim \frac{\pi}{l e}$. Their energy is, therefore, of the order $E\sim \frac{\pi}{\alpha l}$. The probability to have such a fluctuation at temperature $T$ is given by the Boltzmann exponent $\exp( -\frac{E}{T})$, so the typical size of the required fluctuation should be at least of the order of $l\sim 1/(\alpha T)$, with a number density $n_{\rm sph} \sim (\alpha T)^3$. To get the rate of the diffusion of the topological  number, what is left to do is to  divide $n_{sph}$ by the typical time $t_{\rm sph} \sim 1/(\alpha^2 T)$ of the $\Delta N_{CS} = 1$ changing transition (one power of $\alpha$ comes from the dimensional analysis of effective 3-dimensional theory~\cite{Khlebnikov:1988sr}, while the extra power accounts for slowing down of the process by the medium~\cite{Arnold:1987mh,Arnold:1996dy}). As a result, the rate (per unit volume) of fluctuations changing the CS number by one unit, is parametrically given by 
\be
\label{gammasph}
\Gamma_{\rm sph} \propto \alpha^5 T^4~,
\ee
up to a logarithmic factor that comes from a more refined treatment~\cite{Bodeker:1998hm}. In a plasma carrying a non-zero fermionic charge, the rates leading to its increment or decrement differ by $\mu/T$, as follows from Eq.~(\ref{ncs}). This leads to the conclusion that $\mu(t) \propto \exp(- \Gamma_{\rm sph} t/T^3)$, with a $\mu$-independent exponential.

In the previous consideration the non-Abelian character of the gauge theory was used only at one point, namely when we required that the change of the CS number is equal to an integer number one, as dictated by the non-trivial vacuum structure. For Abelian theories the integer changes of the CS number are however not specific, and the rate of CS number diffusion for fluctuations with size $l$, amplitude $A$, and time $\tau$, is given by 
\be
\Gamma \propto (\alpha A^2 l^2) \left[\frac{\exp (-A^2 l/T)}{l^3 \tau}\right]~,
\ee
where the first factor is a typical change of the CS number, and the second is the probability of a fluctuation per unit time and volume. The characteristic non-linear dynamics length and time scales in the Abelian Higgs with $\lambda \sim e^2, m^2 \sim e^4 T^2$, are parametrically the same as in the non-Abelian gauge theory, $l \propto 1/(\alpha T$,  $\tau \propto 1/(\alpha^2 T)$. In other words, it seems that Eq.~(\ref{gammasph}) may be equally applied to the Abelian theory as well.  Qualitatively, the difference appears in the subsequent development of the fluctuation. In the non-Abelian case, after crossing the ``sphaleron barrier'' it may evolve to a new vacuum state with the different CS number but zero energy, thermalising its energy. In the Abelian case, such a vacuum state does not exist. The discussion in Section \ref{sec:conf} indicates however, that the effective degeneracy with respect to the CS number, appears in the limit of long wavelengths of the gauge field. It is an open question whether this effect may lead to dissipation of the fermionic number in the Abelian theory, with a rate that does not depend on $\mu$. If the non-Abelian estimate were applicable to the Abelian case as well, the Abelian ``sphaleron'' rate $\Gamma_{sph} \propto \alpha^5 T^4$ would supersede the rate associated with the instability (\ref{small}), for a chemical potential $\mu < \alpha T$. We present some arguments in the next subsection \ref{sec:diffCS} that the Abelian rate may scale like $\Gamma_{sph} \propto \alpha^6 T^4$, i.e.~with one extra power of $\alpha$ compared to the non-Abelian case. If true, this will exceed the rate associated to the instability for small $\mu < e \alpha T$. Whether this is correct or not, it can be studied, in principle, in real time lattice simulations. Unfortunately, due to limited computer resources, we could not get an answer to this question. We thus leave open the investigation of this matter for the future. 

\subsection{Diffusion of CS number and chirality non-conservation}
\label{sec:diffCS}

For physics applications, the main quantity of interest is the time evolution of the fermion number and of the magnetic fields. This is a complicated problem involving many different time and length scales operating at different stages of the equilibration process. However, in the limit of small chemical potential $\mu$, the rate of fermion number non-conservation can be found within the pure bosonic theory, by considering the {\em diffusion} of the CS number and the use of the fluctuation-dissipation theorem~\cite{Khlebnikov:1988sr,Moore:1996qs}.

To make a proper comparison between Abelian and non-Abelian cases, consider bosonic theories, without fermions, based on either a SU(2) gauge theory with Higgs doublet, or on a U(1) theory. The main quantity which determines the dynamics of the topological transitions is the CS diffusion rate $\Gamma$~\cite{Khlebnikov:1988sr}, which characterizes the expectation value 
\be
\langle  Q(t)^2 \rangle~,
\label{qsquare}
\ee
where
\be
Q(t) \equiv \frac{e^2}{16\pi^2}  \int_0^t dt \int d^3x  F^a_{\mu\nu}{\tilde F}_a^{\mu\nu} = N_{CS}(t)-N_{CS}(0)~,
\ee
is a gauge-invariant quantity related to the Chern-Simons number, with the index $a$ omitted in the U(1) case, and running like $a=1,2,3$ in the SU(2) case. Homogeneity in time and space leads to 
\be
\langle  Q(t)^2 \rangle = \Gamma V t~,
\label{qQ2beh}
\ee
where for $t\to\infty$,
\be
\Gamma= \frac{\alpha^2}{16\pi^2} \int_0^\infty dt \int d^3x \left\langle F^a_{\mu\nu}({\bf x},t){\tilde F}_a^{\mu\nu}({\bf x},t)    
 F^a_{\alpha\beta}(0){\tilde F}_a^{\alpha\beta}(0) \right\rangle ~.
\label{qcorr}
\ee

As we briefly reviewed in Section~\ref{subsec:inst}, it is well known  that in the symmetric phase of a non-Abelian theory, $\Gamma$ depends on the coupling constants as\,
\be
 \Gamma \propto \alpha^5 \log(1/e) T^4\,.
\label{gnonab}
\ee
In the Higgs phase it is suppressed by the Boltzmann exponent $\exp\left(-M_{\rm sph}/T\right)$ \cite{Kuzmin:1985mm}, where $M_{\rm sph}\propto M_A/e^2$ is the sphaleron mass, and $M_A$ is the temperature dependent mass of the vector boson. 

In Abelian theories, and in the absence of a magnetic field, we expect $\langle Q(t)^2 \rangle$ to become constant at large times $t$, because contrary to the non-Abelian case, to have a non-zero CS number now costs energy. We expect however a diffusive behavior if an external magnetic field is present, since the CS number can grow without limit with the energy fixed. 

We show now that the perturbative contributions to the diffusion rate at $B\neq 0$ vanish. For this end  let us consider the expansion of the correlator Eq.~(\ref{qcorr}) with respect to an external magnetic field in the $\hat z$-direction, writing $B_z = \bar B +\delta B$. The lowest order term in $\bar B$ is  zero due to the Abelian character of the theory, while the first order term vanishes as well, due to symmetry considerations (isotropy). We are thus left with the second order contribution in $\bar B$, 
\be
\Gamma=\left( \frac{\alpha \bar B}{\pi}\right)^2 \left\langle  \int_0^\infty dt \int d^3x E_3(t,x) E_3(0) \right\rangle ~.
\label{corrmagn}
\ee
In perturbation theory this correlator is zero. This is most easily seen in the gauge $A_o=0$, where the electric field can be written as $E_3(x,t) = \partial_o A_3$. The correlator in Eq.~(\ref{corrmagn}) becomes
\be
 \left\langle  \int d^3x (A_3(\infty,x)-A_3(0,x)) \partial_0 A_3(0) \right\rangle  = 0~,
\label{corrmagnpt}
\ee
where the first term is zero due to the cluster property, while the second can be written as  $\frac{1}{2}\partial_0  \langle  A_3(0)^2 \rangle$, and hence it also vanishes. 

To get a non-perturbative contribution, we start from an estimate of the diffusion rate which can be obtained from previous results~\cite{Joyce:1997uy, Giovannini:1997eg}, based on the equations of magnetohydrodynamics.  In the presence of a homogeneous chemical potential for the chiral charge, the effective action for the electromagnetic fields, leads to a modification of Maxwell equations for wavelengths larger than the fermions mean free path $\lambda\sim 1/(\alpha^2 T)$ in the plasma. Defining the electric and magnetic fields as $E^i = E_i = \dot{A}_i - \partial_i\phi$, and $B^i = B_i = \epsilon_{ijk}\partial_jA_k$, the modified equations read~\cite{Joyce:1997uy, Giovannini:1997eg} [we use metric signature (--,+,+,+)]
\be
\frac{\partial{{\vec{B}}}}{\partial t} = \vec{\nabla}\times {\vec{E}}
,~~~~~
\frac{\partial{\vec{E}}}{\partial t} + {\vec{\nabla}}\times{ \vec{B}}= {e\vec{j}} - \frac{e^2}{4\pi^2} \mu {\vec{B}}, ~~~~~
e {\vec{J}}= - \sigma {\vec{E}},
\label{homog}
\ee
where $\sigma$ is the electric conductivity of the plasma, and we have assumed that the density of electric charge is zero, and the plasma has zero velocity. This system of equations is complemented by an anomaly equation like Eq.~(\ref{anom}), which rewritten in terms of the chiral chemical potential reads
\be
\frac{d\mu}{dt} = \frac{3e^2}{T^2\pi^2} {1\over V}\int d^3x\,\vec{E}\vec{B}\,.
\label{mu5eq}
\ee
Neglecting the time derivative of the electric field in Eq.~(\ref{homog}), one gets an equation for $\mu$ as
\be
\frac{d\mu}{dt} =  -\frac{3e^2}{\pi^2 \sigma T^2} {1\over V}\int d^3x \left(\frac{e^2}{4\pi^2} \mu  {\vec{B}} +  {\vec{\nabla}}\times{ \vec{B}}\right)\vec{B}\,.
\label{mu3eff}
\ee
This shows that, in the presence of an external homogeneous magnetic field, the contribution of long-range fluctuations of gauge fields to the rate of chirality non-conservation, is 
\be
\Gamma_5 = \frac{3e^4}{4\pi^4\sigma T^2}B^2 \propto \alpha^3 B^2~.
\label{ratemu5}
\ee
The second term in (\ref{mu3eff}) can be considered as a source leading to non-zero chemical potential in the presence of helical magnetic fields, with interesting dynamics  of exchange between the chiral charge and magnetic fields \cite{Joyce:1997uy,Giovannini:1997eg,Boyarsky:2011uy,Kamada:2016eeb,Kamada:2016cnb}. 

The fluctuation-dissipation theorem allows to relate the diffusion rate $\Gamma$ (per unit time and volume) of the CS number with the dilution rate $\Gamma_5$ (per unit time) of the chemical potential. Standard considerations following~\cite{Khlebnikov:1988sr,Moore:1996qs}, give for the theory (\ref{cl}) the following relation
\be
\Gamma_5 = 12 \frac{\Gamma}{T^3}~.
\label{5tog}
\ee
Using  (\ref{5tog}) and Eq.~(\ref{ratemu5}), we finally arrive at
\be
\Gamma =\frac{\alpha^2T}{ \pi^2\sigma}B^2 \propto \alpha^3 B^2~.
\label{ralp3}
\ee
The rate scales as $\alpha^3$ and is proportional to $B^2$.

We present below yet another estimate inspired by \cite{Arnold:1996dy}, making use of the fluctuations discussed in Section~\ref{subsec:inst} with typical size parametrically smaller than the mean free path, $l_{\rm sph}\sim  1/(\alpha T)\ll \lambda$. For this end we take as in Section \ref{subsec:inst} $A \sim \frac{\pi}{e\,l_{\rm sph}}$ and $\partial_o \sim 1/t_{\rm sph} \sim \alpha^2 T$, leading to the rate 
\be
\Gamma_f \sim \left( \frac{\alpha \bar B}{\pi}\right)^2 \frac{l_{\rm sph}}{\alpha t_{\rm sph}} \sim \left( \frac{\alpha \bar B}{\pi}\right)^2~.
\label{corrmagnfluct}
\ee
This result has the same parametric dependence on the magnetic field as Eq.~(\ref{ralp3}). However, it contains the square of the fine structure constant contrary to $\alpha^3$  in (\ref{ralp3}).  In principle, there is no contradiction, as (\ref{ralp3}) only accounts for sufficiently long range fluctuations (for which the Eqs.~(\ref{homog}) are valid) and when ${\partial \vec E \over \partial t} = 0$. Still, this difference is perturbing and calls for reconsideration of the typical time scale in Abelian gauge theories, which may be different from the non-Abelian case because of the different vacuum structure in the absence of a magnetic field. Assuming that Eqs.~(\ref{homog}) with $\mu=0$ and $e\vec j = -\sigma \vec E$ were valid (at least parametrically) for short ranged fields as well, we get that the typical electric field $E$ associated with magnetic field of strength $B \propto A/l \propto e^3 T^2$ and size $l_{\rm sph}$, is of the order $E \sim B/(\sigma l) \propto A/(\sigma l^2)\propto \omega A$ with $\omega \propto \alpha^3 T$, i.e. $\alpha$ times slower than the non-Abelian one. If correct, the rate from fluctuations is proportional to an extra power in $\alpha$,
\be
\Gamma_f  \sim \alpha \left( \frac{\alpha \bar B}{\pi}\right)^2~,
\label{corrmagnfluct1}
\ee
and has the same parametric dependence as (\ref{ralp3}). If we replace the external magnetic field by a  typical magnetic field $B \sim A/l \propto e^3 T^2$ of the  fluctuation which carries $N_{\rm CS} \sim 1$,  we will get an ``Abelian sphaleron rate'' as
\be
\Gamma_{\rm sph} \sim \alpha^6 T^4~,
\label{absph}
\ee
which is suppressed in comparison with the non-Abelian one by an extra power of $\alpha$. In the estimates above we assumed that the electric conductivity scale like $\sigma \propto 1/\alpha$ even at small distances.

The real time lattice simulations of the CS diffusion in the presence of magnetic field are considered in Section~\ref{sec:diff}. We will see there that the parametric dependence on $B$ and $\alpha$ coincides with (\ref{ratemu5}, \ref{corrmagnfluct1}), but with a numerical pre-factor exceeding substantially that in~(\ref{ralp3}), indicating that the small scale fluctuations are more important. 

\section{A model for real-time classical simulations}
\label{sec:model}

It is notoriously difficult to make lattice simulations with dynamical massless fermions. One can use an effective field theory approach, where fermions are integrated out, but their presence is accounted for by a homogeneous chemical potential $\mu$ for their chiral charge. As mentioned before, in this approximation the energy of the system acquires the term Eq.~(\ref{ncs}), and the theory becomes purely bosonic, so that it becomes suitable for classical real time simulations. We can then put the resulting bosonic theory in a finite volume $V$, with periodic boundary conditions for gauge-independent quantities in all spatial directions. 

An economic way to derive, in an explicit gauge-invariant way, the equations of motion for the different fields involved in the bosonic approach, is to use an (auxiliary) {\it axion} field $a(x)$ with action
\begin{eqnarray}
\label{eq:ActionContinuum}
S &=& -\int d^4x\left(\mathcal{L}_{\varphi} + {1\over 4}F_\mn F^\mn - {1\over 2c_s^2}(\partial_0 a)^2 + {1\over 2}(\partial_i a)(\partial_i a) - {e^2\over (4\pi)^2}{a\over M}F_\mn \tilde{F}^\mn\right)\,,
\end{eqnarray}
where $c_s^2$ and $M$ are respectively dimensionless and dimensionfull parameters, to be fixed later on. Due to the fact that $F_{\mu\nu}\tilde{F}^{\mu\nu}$ is a 4-divergence, this action is invariant under the shift symmetry $a \to a + const$. The equations of motion following from action~(\ref{eq:ActionContinuum}) are
\begin{eqnarray}
\label{eq:EOM1}
D_\mu D^\mu \varphi &=& V,_{\varphi^*}\,,\\
\label{eq:EOM2}
\partial_\nu F^\mn &=& e j^\mu + {e^2\over 4\pi^2M}\tilde{F}^{\mu \nu}{\partial_\nu a} \,,\\
\label{eq:EOM3}
c_s^{-2}\partial_0\partial_0 a &=& \partial_i\partial_i a + {e^2\over (4\pi)^2 M} {F}_\mn\tilde{F}^\mn\,,
\end{eqnarray}
where the (unit-charge) current is defined as $j^\mu = 2{\rm Im}\lbrace\varphi^* D^\mu\varphi\rbrace$. As expected, these equations only contain derivatives of the field $a(x)$. Now, the anomaly equation Eq.~(\ref{anom}) for the chiral current $J_\mu^5$ can be compared with Eq.~(\ref{eq:EOM3}), leading to an identification 
\begin{eqnarray}
\partial_0 a \to {c_s^2\over2} {J_0^5\over M}\,,~~~~~\partial_i a \to {J_5^i\over 2M}\,.
\end{eqnarray}
We consider only the homogeneous\footnote{In this paper we only consider space-independent chemical potentials. As we discuss in detail~\cite{Figueroa:2017qmv}, accounting for a space-dependent chemical potential (or axion field) in a consistent way in lattice simulations, requires a much more elaborated framework, so we postpone an analysis of such case for future work.} case $\vec\nabla a = 0$, so that $\vec J_5 = \vec 0$. As a result, the equation of motion for the axion field Eq.~(\ref{eq:EOM3}), turns into
\begin{eqnarray}
\partial_0 J_5^0 = \frac{e^2}{8\pi^2} {1\over V}\int d^3x\,F_{\mu\nu}\tilde F^{\mu\nu}\,.
\end{eqnarray}
The equation of motion of the gauge field, in the presence of a homogeneous axion (chemical potential), changes to
\begin{eqnarray}
\partial_\nu F^{\mu\nu} &=& ej^{\mu} + {c_s^2 e^2\over 8\pi^2M^2}J_5^0\tilde{F}^{\mu 0}
\,.
\label{eq:EOM3c}
\end{eqnarray}
The relation between the chemical potential and the chiral charge at $T \neq 0$, for small $\mu$, reads
\be
J_5^0=\frac{1}{6} \mu T^2~.
\ee
In order to have Eq.~(\ref{eq:EOM3c}) mimicking Eq.~(\ref{homog}), we simply need to make the identification $c_s^2T^2 = 12M^2$. A simple solution is then to take $c_s^2 = 1$ and $M^2 = T^2/12$. With this choice we conclude that the equations of motion for a homogeneous axion $\vec\nabla a = 0$, Eqs.~(\ref{eq:EOM2})-(\ref{eq:EOM3}), together with the identification $\partial_0 a = M\mu$, mimic exactly the equations~(\ref{homog})-(\ref{mu5eq}) in the presence of a homogeneous chemical potential.

The evolution equations in terms of electric and magnetic fields  $E^i = E_i = \dot{A}_i - \partial_i\phi$ and $B^i = B_i = \epsilon_{ijk}\partial_jA_k$, taking the Coulomb gauge $A_o = 0$, and expressing them in a vector form, read
\begin{eqnarray}
\label{eq:EOMChPot0}
\ddot\varphi - \vec D \vec D\varphi &=& -V_{,|\varphi|^2}\varphi\,,\\
\label{eq:EOMChPot1}
\dot{\vec E} + \vec\nabla \times \vec B &=& + e\vec{j} - {e^2\over 4\pi^2}{\mu}\vec B\,, \\
\label{eq:EOMChPot2}
\vec\nabla \vec E &=& e j^0  ~~{\rm (Gauss\,\,Constraint)}\,,\\
\label{eq:EOMChPot}
\dot \mu &=& {3e^2\over \pi^2T^2}{1\over V}{\int_V d^3x ~\vec E \cdot \vec B}\,,
\end{eqnarray}
where we have used $F_{\mu\nu}\tilde F^{\mu\nu} = 4\vec E \vec B$ and $\tilde{F}^{i0} = - B_i$.

Besides, the system is characterized by an integral of motion -- the total energy -- given by
\be
E_{tot}= \int d^3x\left[\frac{1}{2} \vec E^2 +\frac{1}{2} \vec B^2+  |D_o\varphi|^2+|D_j\varphi|^2 +V(|\varphi|^2)\right]+ \frac{1}{24}\mu^2 V \, .
\label{eq:energy}
\ee
This quantity is important for monitoring the numerics as any time discretization induces always some degree of violation of energy conservation. The system of equations ensures also the magnetic flux conservation, for example 
\be
\frac{\partial}{\partial t} \int dx dy B_3=0\, .
\label{eq:flux}
\ee
Thus, the initial conditions for the time evolution can be characterised by giving the energy of the system and of the magnetic flux through one of the planes, which we can always choose to be that of $xy$.  

The equations of motion Eqs.~(\ref{eq:EOMChPot0})-(\ref{eq:EOMChPot}) and the Hamiltonian Eq.~(\ref{eq:energy}), serve as the theoretical basis for our numerical simulations. The way how these equations can be put on the lattice while keeping the gauge invariance, the topological character of $F\tilde F$, and a non-zero magnetic flux, is discussed in detail in~\cite{Figueroa:2017qmv}.  In Sect.~\ref{subsec:LatticeFormulation} we summarize the lattice formulation of the theory based on~\cite{Figueroa:2017qmv}, whereas in Sect.~\ref{subsec:parameters} we discuss the choice of parameters to capture appropriately the correct physical regimes in the numerical simulations. In Sect.~\ref{subsec:InitialCond} we discuss how we set up the initial conditions of the system in the lattice. 

In Section~\ref{sec:diff} we present our numerical results, studying the diffusion of the Chern-Simons number in the absence of fermions, i.e. with $\mu=0$\footnote{We  plan to study the dynamics of the chiral charge $\mu$ with an initial value $\mu_0 \neq 0$ in a follow-up publication.}. With standard Monte-Carlo techniques we create a set of configurations with probability $\exp(-E/T)$, which are then evolved in time with the use of the lattice equations of motion. We also consider an external magnetic field to be present, and determine the correlator Eq.~(\ref{qQ2beh}), from which one can extract the diffusion rate.

\subsection{Lattice formulation}
\label{subsec:LatticeFormulation}

We work in a periodic cubic lattice of length $L = Ndx$, with $dx$ the lattice spacing and $N$ the number of points per dimension. We do not consider summation over repeated indexes. A lattice point $n = (n_o,\vec n) = (n_o,n_1,n_2,n_3)$ displaced in the $\mu-$direction by one unit lattice spacing, $n + \hat\mu$, will be referred simply as $n+\mu$ or by $+\mu$, e.g.~$\varphi_{+\mu} \equiv \varphi(n+\hat\mu)$, $A_{\mu,+\nu} \equiv A_{\mu}(n+{1\over2}\hat\mu + \hat\nu)$, etc. We define lattice ordinary and covariant derivatives, forward (+) and backward (-), as
\begin{eqnarray}\label{eq:CovDer}
\begin{array}{rclr}
\Delta_\mu^\pm\phi \equiv {\pm 1\over dx}(\phi_{\pm \mu} - \phi)\,,~~~~~~~~~~
(D_\mu^\pm\varphi) \equiv {\pm 1\over dx}(U_{\pm \mu}\varphi_{\pm \mu}-\varphi)\,,
\end{array}
\end{eqnarray}
where a {\it link} is defined as $U_{\mu} \equiv U_\mu(n+{1\over 2}\hat\mu) \equiv e^{-ie\int_{x(n)}^{x(n+\hat\mu)} A_\mu(x')dx'^\mu}$ $\simeq e^{-iedx^\mu A_\mu(n+{1\over 2}\hat\mu)}$, and we have also defined $U_{-\mu} \equiv U_{\mu,-\mu}^* \equiv U_\mu^*(n-{1\over 2}\hat\mu) \simeq e^{+iedx^\mu A_\mu(n-{1\over 2}\hat\mu)}$. An Abelian $U(1)$ gauge transformation in the lattice corresponds to 
\begin{eqnarray}
\varphi(n) ~~\longrightarrow~~ e^{+i\beta(n)}\varphi(n)\,,~~~~~~ A_{\mu}(n+{1\over 2}\hat\mu) ~~\longrightarrow~~ A_{\mu}(n+{1\over 2}\hat\mu) \,+\, {1\over e}\Delta_{\mu}^+\beta(n+{1\over 2}\hat\mu),
\end{eqnarray}
with $\beta(n)$ an arbitrary function of the space-time site, so that the links and covariant derivatives transform as
\begin{eqnarray}
U_{\pm\mu,n} ~~\longrightarrow~~ e^{i\beta}\,U_{\pm\mu,n}\,e^{-i\beta_{\pm\mu}}\,,~~~~~~ D_\mu^\pm\varphi ~~\longrightarrow~~ e^{i\beta}\,D_\mu^\pm\varphi\,.\hspace*{4cm}
\end{eqnarray}
Using these transformation rules we can build a gauge invariant lattice action mimicking the continuum action Eq.~(\ref{eq:ActionContinuum}). We first build gauge invariant electric and magnetic fields
\begin{eqnarray}\label{eq:ElecMagLatt}
E_i \equiv (\Delta_o^+ A_i - \Delta_i^+ A_o)\,,~~~~~~~~~~ B_i \equiv \sum_{j,k}\epsilon_{ijk}\Delta_j^+A_k\,.
\end{eqnarray}

For convenience, we also define the following field combinations
\begin{eqnarray}
A_i^{(2)} &\equiv& {1\over2}(A_i+A_{i,-i})\,,\\
E_i^{(2)} &\equiv& {1\over2}(E_i+E_{i,-i})\,,\\
E_i^{(4)} &\equiv& {1\over4}(E_i+E_{i,-i}+E_{i,-0}+E_{i,-i-0})\,,\\
B_i^{(4)} &\equiv& {1\over4}(B_i+B_{i,-j}+B_{i,-k}+B_{i,-j-k})\,,\\
B_i^{(8)} &\equiv& {1\over2}\left(B_i^{(4)}+B_{i,+i}^{(4)}\right)\,.
\end{eqnarray}
We introduce a homogeneous axion as a homogeneous auxilairy field so that its time derivative represents the chemical potential of the system, 
\begin{equation}
\mu(t) \equiv {\dot a\over M}\,,~~~~ M^2 = {T^2\over 12}\,.
\end{equation}
We derived in Ref.~\cite{Figueroa:2017qmv} a lattice action built out of two pieces, an {\it Abelian-Higgs} part and a {\it Chemical Potential} part,
\begin{eqnarray}\label{eq:LatticeActionTot}
S^L_{\rm tot} = S_{{\rm AH}}^{L} + S_{{\alpha(t)}}^{L}\,,
\end{eqnarray}
given by
\begin{eqnarray}\label{eq:ActionLAH}
S_{\rm AH}^{L} &=& dt dx^3\sum_{\vec n,t} \Big[ 
(D_o^{+}\varphi)^\dag(D_o^{+}\varphi) - \sum_j(D_j^{+}\varphi)^\dag(D_j^{+}\varphi) - V(\varphi\varphi^*,\phi)  \\ 
&& \hspace*{2.0cm} +~ \frac{1}{2}\sum_{j}\left(\Delta_o^+A_i-\Delta_i^+A_o\right)^2  - \frac{1}{4}\sum_{i,j}(\Delta_i^+A_j-\Delta_j^+A_i)^2 \Big]\,,\nonumber\\
\label{eq:LatticeHomAxionAction}
S_{\alpha(t)}^{L} &\equiv& dt dx^3\sum_{n_o}\left\lbrace {N^3\over 2}\left(\Delta_o^-a\right)^2 + {e^2\over 4\pi^2}{a\over M}\sum_{\vec n}\sum_i {1\over2}E_i^{(2)}\left(B^{(4)}_i + B_{i,+0}^{(4)}\right)\right\rbrace\,.
\end{eqnarray}
The lattice action Eq.~(\ref{eq:LatticeActionTot}) exhibits exact gauge and shift symmetries in the lattice, and reproduces the continuum action Eq.~(\ref{eq:ActionContinuum}) to order $\mathcal{O}(dx_\mu^2)$. Besides it gives rise to a set of lattice equations of motion compatible with $i)$ the Bianchi identities\footnote{In Ref.~\cite{Figueroa:2017qmv} we showed how other lattice representations of $F\tilde F$ often fail to fulfill the Bianchi identities, introducing extra terms in the lattice equations of motion which do not represent well the continuum limit.}, and $ii)$ an explicit iterative scheme to solve them. We note that the choice of the lattice operator $\sum_i E_i^{(2)}\left(B^{(4)}_i + B_{i,+0}^{(4)}\right)$ in Eq.~(\ref{eq:LatticeActionTot}) to mimic a continuum term $F_{\mu\nu}\tilde{F}^{\mu\nu}$, is crucial in order to obtain an explicit iterative scheme, as shown below in Eq.~(\ref{eq:EOMlatticeChemPot}). Besides, Eq.~(\ref{eq:LatticeHomAxionAction}) naturally leads to a lattice definition of the topological number density $\mathcal{K} \equiv {e^2\over (4\pi)^2}F_{\mu\nu}\tilde{F}^{\mu\nu}$, that admits\footnote{For this it is crucial that we use a non-compact formulation for the gauge sector.} an exact total (lattice) derivative representation as $\mathcal{K} = \sum_\mu \Delta_\mu^+K^\mu$, see Eq.~(\ref{eq:QequalDK}). 

Varying $S^L_{\rm tot} = S_{{\rm AH}}^{L} + S_{{\alpha(t)}}^{L}$, one obtains the set of lattice equations of motion mimicking a continuum system with chemical potential at finite temperature $T$. In the Coulomb gauge $A_o = 0$ (so that $U_o = 1$), these equations read~\cite{Figueroa:2017qmv}
\begin{eqnarray}
\vspace*{1.5cm}
{\rm Equation} \hspace*{7.8cm} {\rm Natural~Site}\hspace*{1.3cm}\nonumber\vspace*{0.5cm}\\
\begin{array}{rclcl}
\pi &\equiv& \Delta_o^+\varphi\,, & \rightarrow & l \equiv (n_o+{1\over2},\vec n)\\
E_i &\equiv& \Delta_o^+A_i\,, & \rightarrow & l \equiv (n_o+{1\over2},\vec n+{1\over2}\hat i)\\
\mu &\equiv& \Delta_o^-(a/M)\,,~~{\rm (Chemical~Potential)} & \rightarrow & l \equiv (n_o,\vec n)\\
\Delta_o^-\pi &=& \sum_iD_i^-D_i^+\varphi - V_{,\varphi^*} = 0\, & \rightarrow & l \equiv (n_o,\vec n)\\
\Delta_o^- E_i &=& {2e}\,{\rm Im}\lbrace\varphi^*D_i^+\varphi\rbrace -\sum_{j,k}\epsilon_{ijk}\Delta_{j}^-B_k
 - {e^2\over 4\pi^2}\mu B_i^{(8)}\,, & \rightarrow & l \equiv (n_o,\vec n + {1\over2}\hat i)\\
\sum_i \Delta_i^- E_i &=& {2e}\,{\rm Im}\lbrace\varphi^*\pi\rbrace\hspace*{0.5cm}(\rm Gauss\,\,Constraint)\,, & \rightarrow & l \equiv (n_o+{1\over2},\vec n)\\
\Delta_o^+\mu &=& {3\over \pi^2} {1\over T^2}{e^2\over N^3}\sum_{\vec n} {1\over2}\sum_i E_i^{(2)}(B_i^{(4)}+B_{i,+0}^{(4)})\,, & \rightarrow & l \equiv (n_o+{1\over2},\vec n)
\end{array}\label{eq:EOMlatticeChemPot}
\end{eqnarray} 
We have indicated in the $rhs$ the common natural space-time site to all terms belonging to a given equation of motion, around which we can expand each equation and reproduce the continuum analogue Eqs.~(\ref{eq:EOMChPot0})-(\ref{eq:EOMChPot}) to order $\mathcal{O}(dx_\mu^2)$. 

A lattice definition of the Chern-Simons number follows naturally from Eq.~(\ref{eq:LatticeHomAxionAction}) as
\begin{eqnarray}
\label{eq:EBdx4EqualsABdx3Discrete2}
{4\pi^2\over e^2}Q_{L} \equiv dt dx^3\sum_{n_o=0}^{p-1} \sum_{\vec n, i} {1 \over 2}E_i^{(2)}(B_i^{(4)}+B_{i,+0}^{(4)}) = {dx^3\over 2}\sum_{\vec n, i} \sum_i A_{i,+p\hat 0}^{(2)}B_{i,+p \hat 0}^{(4)} ~~+~ \mathcal{D}_o\,,
\end{eqnarray}
where $\mathcal{D}_o = - {1\over 2}dx^3\sum_{\vec n}\sum_{i}A_{i}^{(2)}B_{i}^{(4)}$ is an initial constant. This representation of the Chern-Simons number reproduces to order $\mathcal{O}(dx_\mu^2)$ the continuum expression ${4\pi^2\over e^2}Q \equiv \int d^4x \vec E\,\vec B = {1\over 2}\int d^3x \,\vec A\,\vec B\big|_{0}^{t}$. It represents in fact the U(1) limit of the SU(2) expression for the Chern-Simons number introduced in~\cite{Moore:1996qs,Moore:1996wn}. Furthermore, Eq.~(\ref{eq:EBdx4EqualsABdx3Discrete2}) admits a total (lattice) derivative representation, mimicking the continuum relation $\mathcal{K} \equiv {e^2\over(4\pi)^2} F_{\mu\nu}\tilde{F}^{\mu\nu} = \partial_\mu K^\mu$, with $K^\mu$ the CS current. In particular, 
\begin{eqnarray}\label{eq:QequalDK}
\mathcal{K}_L \equiv {e^2\over4\pi^2}\sum_i {1\over2}E_i^{(2)}(B_i^{(4)}+B_{i,+0}^{(4)}) = \sum_{\mu}\Delta_\mu^+ K_L^\mu 
\end{eqnarray}
represents a local identity at every lattice site, with $K_L^\mu$ defined by components like
\begin{eqnarray}\label{eq:K0def}
K^0_L &=& - K_0^L \equiv {e^2\over8\pi^2}\sum_i A_{i}^{(2)}B_{i}^{(4)}\,,\\
\label{eq:KiDef}
K^i_L &=& K_i^L \equiv -{e^2\over16\pi^2}\sum_{j,k}\epsilon_{ijk}\left(E_{j}^{(2)}A_{k,-i}^{(2)}+E_{j,-i}^{(2)}A_{k}^{(2)}\right)\,.
\end{eqnarray}
Using that, for periodic boundary conditions, $\sum_{\vec n} \sum_{i}\Delta_i^+K^i = 0$, we then arrive at a expression for the Chern-Simons number (say after $p$ time iterations), as
\begin{eqnarray}\label{eq:CSnumLat}
Q_{L} = {dt dx^3}\sum_{n_o,\vec n}\sum_\mu \Delta_\mu^+K^\mu = {dt dx^3}\sum_{n_o,\vec n} \Delta_0^+K^0 = dx^3\left(\sum_{\vec n}K^o_{+p\hat 0} - \sum_{\vec n}K^o\right),
\end{eqnarray}
depending only (as it should) on the difference between the final and initial values of $K^o$, mimicking the continuum result $Q = {e^2\over 8\pi^2}\left(\int d^3x \,\vec A\,\vec B\big|_{t}-\int d^3x \,\vec A\,\vec B\big|_{0}\right)$.

Let us also note that Eq.~(\ref{eq:QequalDK}) represents an exact solution to the lattice equation of motion for the chemical potential (say after $p$ time steps) as 
\begin{eqnarray}\label{eq:MuCSrelationContinuum}
\mu_{+p\hat 0} = \mu_o + {12\over T^2L^3}Q_{L}\,,
\end{eqnarray}
which mimics the continuum relation given by $\mu(t) = \mu(0) + {12\over T^2}\lim_{V\rightarrow \infty}{Q\over V}$. In practice, as in this paper we only study numerically the case $\mu = 0$, we simply use Eq.~(\ref{eq:EBdx4EqualsABdx3Discrete2}) to obtain the CS number in our lattice simulations. In a forthcoming publication we plan to study situations where $\mu$ is a dynamical degree of freedom. In such a case, whenever we want to obtain the Chern-Simons number, instead of calling Eq.~(\ref{eq:EBdx4EqualsABdx3Discrete2}), we can simply read the chemical potential amplitude as we solve the lattice equation of motion for $\mu$ [last equation in Eq.~(\ref{eq:EOMlatticeChemPot})].

Finally, let us mention that in order to introduce an external magnetic field in the lattice, we use {\it twisted} boundary conditions for the (relevant) component of the gauge field, following~\cite{Kajantie:1998rz}. For instance, without loss of generality, we can consider an external magnetic field in the $\hat z$ direction, $\vec{B}_{\rm ext} = (0, 0, B_{\rm ext})$. In this case we only need twisted boundary conditions for $A_1$ at a given x-site, say $n_1 = 1$. Whenever we want to calculate a (forward) magnetic field $B_3^+ = (\Delta_1^+ A_2 - \Delta_2^+ A_1)$ at the location $(n_1,n_2,n_3) = (1,N-1,n_3)$ [or equivalently a (backward) magnetic field $B_3^- = (\Delta_1^- A_2 - \Delta_2^- A_1)$ at the location $(n_1,n_2,n_3) = (1,0,n_3)$], we must really make the substitution
\begin{eqnarray}\label{eq:ExtMag}
B_3^+(n_1 = 1, n_2 = N-1, n_3) ~~\longrightarrow~~ 
 (\Delta_1^+ A_2 - \Delta_2^+ A_1) + {2\pi n_{\rm mag}\over dx^2}\,,\\
\label{eq:ExtMagII}
B_3^-(n_1 = 1, n_2 = 0, n_3) ~~\longrightarrow~~ 
 (\Delta_1^- A_2 - \Delta_2^- A_1) + {2\pi n_{\rm mag}\over dx^2}\,,
\end{eqnarray}
where $n_{\rm mag}$ is a positive integer number. This is equivalent to assume that the first component of the gauge field makes a 'discrete jump' from $n_2 = N-1$ to $n_2 = N$ as $A_1(1,N,n_3) = A_1(1,0,n_3) - {2\pi n_{\rm mag}\over dx}$, and from $n_2 = 0$ to $n_2 = N$ as $A_1(1,N,n_3) = A_1(1,N-1,n_3) + {2\pi n_{\rm mag}\over dx}$. This twisted boundary prescription represents precisely the condition required to introduce an external magnetic field in the $\hat z$ direction, with magnitude $B_{\rm ext} \equiv {2\pi n_{\rm mag}\over (dx N)^2}$~\cite{Kajantie:1998rz}. 

The introduction of an external magnetic field requires a correction in the expression of the CS number as~\cite{Figueroa:2017qmv}
\begin{eqnarray}\label{eq:EBdx4EqualsABdx3Discrete3}
Q_{L}\Big|_{\rm mag} = Q_{L} + \mathcal{M}_p - \mathcal{M}_0\,,
\end{eqnarray}
where $Q_{L}$ in the $rhs$ is the CS number given by Eq.~(\ref{eq:EBdx4EqualsABdx3Discrete2}) in the absence of an external magnetic field, and $\mathcal{M}_0, \mathcal{M}_p$ are initial and final (after $p$ time steps) magnetic field corrections, given by
\begin{eqnarray}\label{eq:MagConst}
\mathcal{M}_{q} &\equiv& - {e^2\over 8\pi}{n_{\rm mag}\over dx^2}\sum_{n_3} \left\lbrace A_{3,+q\hat 0}^{(2)}(1,0,n_3)+A_{3,+q\hat 0}^{(2)}(1,N-1,n_3) \right. \nonumber\\ && \left. \hspace*{3cm} + A_{3,+q\hat 0}^{(2)}(2,0,n_3)+A_{3,+q\hat 0}^{(2)}(2,N-1,n_3) \right\rbrace\,.
\end{eqnarray}
For further details and clarifications about of the lattice formulation presented above, both in the absence and presence of a magnetic field, we refer the reader to Ref.~\cite{Figueroa:2017qmv}.

\subsection{Choice of parameters for lattice simulations}
\label{subsec:parameters}

For the discretised Eqs.~(\ref{eq:EOMlatticeChemPot}) to capture well the physics described by the continuum set of  Eqs.~(\ref{eq:EOMChPot0})-(\ref{eq:EOMChPot}), a number of conditions must be met in the lattice. In particular, to reduce the effects of discretisation, all relevant length scales of the continuum system $\lambda_i$, must be larger than the lattice spacing $dx$. To reduce finite volume effects, the same scales must also be smaller than the lattice size $L = Ndx$, where $N$ is the number of lattice points per dimension. In other words, relevant scales $\lambda_i$ must fulfill the condition
\be
1 \ll {\lambda_i\over dx} \ll N~.
\ee
In this section we discuss the choice of parameter values so that these two simple conditions are satisfied as best as possible.

Let us recall that we consider an effective potential for the Higgs-like field as
\begin{eqnarray}
V_{\rm eff}(\phi) = m_{\rm eff}^2|\phi|^2 + \lambda_{\rm eff}|\phi|^4\,,
\end{eqnarray}
with $\lambda_{\rm eff}, m^2_{\rm eff}$ the physical self-coupling and physical mass of the scalar excitation. In the simplest case of the Higgs phase of the theory, when the mass parameter is $m_{\rm eff}^2 < 0$ (at tree level $m_{\rm eff}^2 \equiv -\lambda v^2 < 0$), we have two particle excitations: one corresponding to the Higgs boson with the mass $m_H^2= 2\lambda v^2$, and another to the vector boson with the mass $M_A^2=e^2 v^2$, both given at tree level approximation. To minimize the number of scales, it is then natural to make the choice $m_H^2= M_A^2$, leading to $2\lambda=e^2$. We will keep this relation all through our investigation, even when considering the symmetric phase. 

We are actually mostly interested in the symmetric phase, when the effective mass parameter of the Higgs field is $m_{\rm eff}^2 > 0$. At tree level, the 3d Higgs model in this case contains a scalar excitation with mass $m_{\rm eff}^2$ and a massless photon. We require therefore
\begin{eqnarray}
\label{higgs}
1\ll {1\over m_{\rm eff}dx}\ll N\,.
\end{eqnarray}
In addition, we expect to have a length scale $l \simeq {\pi^2\over e^2} {1\over T}$ associated with the fluctuations carrying a CS-number $N_{\rm cs} \sim 1$, see discussion in Section~\ref{sec:Intro}. This requires
\begin{eqnarray}
\label{sphal}
1\ll\frac{\pi^2}{e^2}{1\over dx T}\ll N\,.
\end{eqnarray}
For instance, for $e^2 \simeq 1$, $dx T = 1$ satisfies that $Tdx = 1 \ll {\pi^2\over e^2} \simeq {10\over e^2}$, and at the same time that $LT = N \gg {\pi^2\over e^2} \simeq {10\over e^2}$ (for as long as $N > 10$). The smaller we make $dx T$ the larger we need $N$ in the right hand side of the inequality~(\ref{sphal}). However, due to the need to simulate some cases up to very large physical times, very large values for $N$ (typically expressed in powers of $2^n$) require (unfeasible) long simulation times. Hence, as long as we take values of $e^2$ not much smaller or larger than $\sim 1$, a good compromise is to fix the lattice spacing to the value $dx T = 1$. If instead we choose typical values of the gauge coupling say as $\sim 10^{-n}$ with $|n| \gg 1$ (with $n$ either positive or negative), then we simply need to re-scale the lattice spacing to $dx T = 10^{n}$. From now on we set $e^2 = 1$ as a canonical value, and hence consider values for the gauge coupling somewhat smaller or larger than unity, but not orders of magnitude different. We then fix $dx T = 1$ for all our simulations. In that way we verify the left hand side inequality (\ref{sphal}) with a factor of margin $1 \lesssim 10/e^2$, while the right hand side with a factor $1 \gtrsim 10/(Ne^{2}) \simeq 0.62/e^2, 0.31/e^2, 0.16/e^2, ...$ for $N = 16, 32, 64, ...$ respectively. If we literally take $e^2 = 1$, then $N = 16$ is really border line for verifying the $rhs$ of inequality Eq.~(\ref{sphal}), as 1 is only marginally bigger than $0.62$. Therefore, for $N = 16$ we will not simulate smaller gauge coupling values than unity, and hence $e^2 \geq 1$. Similarly, if we double the points per dimension to $N = 32$, we must consider gauge coupling values $e^2 \geq 0.5$, for $N = 64$ then $e^2 \geq 0.25$, and so on. We have thus considered a sample of gauge coupling values taken by powers of $2$, i.e. $e^2 = 2^p$, $p = ..., -2, -1, 0, 1, ...$. In practice, $N = 128$ is the maximum number of points per dimension we take (otherwise the simulation time becomes too large), so the smallest coupling value we have really considered is $e^2 = 2^{-3} = 0.125$. 

Furthermore, and independently of $N$, the gauge coupling cannot be very large (given that we have already fixed $dxT = 1$), because otherwise the $lhs$ of inequality (\ref{sphal}) becomes only roughly satisfied. Since we are taking the gauge coupling values in powers of $2$, $e^2 = 2^2 = 4$ makes already too rough the verification of the $lhs$ of the inequality. Hence, in practice, the maximum value we have considered is $e^2 = 2^1 = 2$. In summary, for the lattice spacing $dx T = 1$ we chose, we can capture well the scales of the fluctuations carrying out a CS-number of the order $\sim 1$, for the parameters
\begin{eqnarray}
N = 16&:& e^2 = 2,1\label{eq:Param1}\\
N = 32&:& e^2 = 2, 1, 1/2\label{eq:Param2}\\
N = 64&:& e^2 = 2, 1, 1/2, 1/4\label{eq:Param3}\\
N = 128&:& e^2 = 2, 1, 1/2, 1/4, 1/8\label{eq:Param4}~.
\end{eqnarray}

We need of course that such choice of parameters verifies as well the inequality (\ref{higgs}). Some care is however needed when identifying in the lattice the Higgs mass parameter in the symmetric phase. The classical Abelian Higgs model is equivalent to the quantum Abelian 3d theory with an extra field: the zero component of the gauge field $A_0$ interacting with the  scalar $\phi$ as $e^2 A_o^2 |\phi^2|$. This theory is super-renormalisable, with the only parameter requiring renormalisation being the Higgs mass. As immediately seen by the use of power counting, the Higgs mass is linearly divergent at the one-loop level and log-divergent at the two loop level. The relationship between the parameters in the ${\overline{MS}}$ renormalisation scheme $m^2$ and the lattice $m_{\rm lat}^2$ has been worked out in~\cite{Laine:1997dy}, and reads in one-loop approximation accounting for a linear divergence
\begin{eqnarray}\label{cc}
m_{\rm lat}^2 &=& m^2 - (2e^2 +4\lambda +e^2)\frac{3.176 T}{4\pi dx} ,
\end{eqnarray}
where the  term $\propto 2e^2$ is coming from the gauge field contribution, and those  $\propto 4\lambda$ and $\propto e^2$ are coming from the scalar and $A_0$ tadpole graphs. The two-loop logarithmic contribution can be neglected if $m^2 \gg \left({e^2\over 4\pi}\right)^2 T^2$. Even though this condition is not really necessary, we will adopt it to make our analysis more transparent. The relation between the physical and the lattice expectation value of the Higgs squared amplitude (which represents the 'order parameter' in the phase transition), was also computed in~\cite{Laine:1997dy}, in one-loop approximation, as
\begin{eqnarray}\label{cc2}
\langle |\phi|^2 \rangle &=& \langle |\phi|^2 \rangle_{\rm lat} - \frac{3.176 T}{4\pi dx} \,.
\end{eqnarray}
The non-perturbative lattice simulations of~\cite{Dimopoulos:1997cz,Kajantie:1997vc,Kajantie:1997hn} have demonstrated that the physical correlation length in the symmetric phase is close to the ${\overline{MS}}$ mass $m^2$. So, in order to be in the symmetric phase but still sufficiently close to the phase transition, the lattice mass $m_{\rm lat}^2$ should be tuned to the linear counter-term $\propto {1\over dx}$, to get the physical mass $m^2(\mu_*)$ positive but sufficiently small.  For most of the simulations we  chosen  the physical scalar mass to be  $m^{2}(\mu) = 0.25 e^2 T^2$, close to the hard thermal loop contribution to the scalar mass,  $m^2 = e^2 T^2/4+\lambda T^2/12$.   According to the phase diagram of the 3d Higgs + U(1) model~\cite{Dimopoulos:1997cz,Kajantie:1997vc,Kajantie:1997hn}, for this value of $m^2$ we are in the symmetric phase of U(1) theory, but still sufficiently close to the phase transition line to satisfy the inequality (\ref{higgs}).  We also varied the mass towards larger and smaller values but did not see any strong dependence of the CS diffusion rate on this parameter. In summary,  our set of $\lbrace e^2, N \rbrace$ values listed in Eqs.~(\ref{eq:Param1})-(\ref{eq:Param3}), satisfy essentially both inequalities~(\ref{higgs}) and~(\ref{sphal}) in a reasonable way, given our choice $dxT = 1$. We will therefore stick to such parameter values in the following sections.

As a final remark about parameters, let us note that in all simulations we have chosen a time step $dt$ sufficiently small, so that when decreasing it further, the results would be the same within the statistical error. Whereas $dtT \simeq 0.1$ was borderline for some cases, $dtT \simeq 0.01$ proved to be a good choice in all cases. As a compromised we adopted $dtT = 0.05$ as our fiducial value, and only increased it (hence speeding up our runs) when we saw explicitly (by trial and error) that the observables were not sensitive to it. For our $N  = 128$ runs we had however to take always a large time step as $Tdt = 0.2$, as to run up to the large physical times needed to probe the diffusion regime we describe in Section~\ref{sec:diff} (e.g.~up to $T\Delta t \sim 50000$), already took around 6 days of computing time for each (serial) job in a single core processor. Thus, contrary to most modern applications of lattice field theory in early Universe contexts, like the study of topological defects, phase transitions, gravitational wave generation, or preheating and post-inflationary dynamics, see e.g.~\cite{Bevis:2006mj,Figueroa:2012kw,Daverio:2015nva,Hindmarsh:2017qff,Figueroa:2015rqa,Enqvist:2015sua,Figueroa:2016wxr,Easther:2006gt,GarciaBellido:2007dg,GarciaBellido:2007af,Dufaux:2007pt,Dufaux:2008dn,Figueroa:2011ye,Figueroa:2016ojl,Figueroa:2017vfa,Hindmarsh:2013xza,Hindmarsh:2015qta,Cutting:2018tjt}, in our present case we are limited by the integration time, rather than by memory. Even though the number of field degrees of freedom in an Abelian-Higgs set-up like ours can be easily accommodated nowadays in simulations with $N = 256-512$, such large lattice sizes would require, in our case, unreasonably long integration times for each serial job to run up to a sufficiently large physical time. We postpone for future work a parallelization and vectorization of our code that will allow us to probe the diffusion regime for long enough while using large lattices.

\subsection{Initial conditions}
\label{subsec:InitialCond}

In Section~\ref{sec:diff} we start our numerical study by considering the diffusion of the Chern-Simons number in the absence of fermions, i.e.~fixing $\mu=0$. In order to set up the initial thermal configuration, we use a standard Monte-Carlo Metropolis-Hasting algorithm to create a set of configurations according to the Gibbs distribution $\exp(-H_L/T)$, where $H_L$ is the lattice Hamiltonian
\begin{eqnarray}\label{eq:LatticeHamiltonian}
H_{L} &=& dx^3\sum_{\vec n} \mathcal{H}_L = dx^3\sum_{\vec n} \Big[ 
(D_o^{+}\varphi)^\dag(D_o^{+}\varphi) + \sum_j(D_j^{+}\varphi)^\dag(D_j^{+}\varphi) + V(\varphi\varphi^*,\phi)  \nonumber\\ 
&& \hspace*{2.0cm} +~ \frac{1}{2}\sum_{j}\left(\Delta_o^+A_i-\Delta_i^+A_o\right)^2  + \frac{1}{4}\sum_{i,j}(\Delta_i^+A_j-\Delta_j^+A_i)^2 \Big]\,,
\end{eqnarray}
which reproduces the continuum Hamiltonian Eq.~(\ref{eq:energy}) [for $\mu = 0$] to order $\mathcal{O}(dx_\mu^2)$. The updates are such that we first update, in a given site, a given field amplitude, and then we determine whether we accept the change or not (if not, we undo the update). Following, we update in the same site, another field amplitude, and determine whether we accept it or not. We repeat this procedure in the same lattice site until we have considered all field amplitudes $\lbrace \phi, \vec A \rbrace$. Afterwards, we do the analogous updates, still in the same site, of the conjugate momenta $\lbrace \pi, \vec E \rbrace$. Only after we have updated, in a give site, all field and conjugate momenta variables $\lbrace \phi, \vec A, \pi, \vec E \rbrace$, we move to the next lattice site, and repeat the procedure. We perform the updates at sites selected sequentially, i.e.~at each sweep we survey the $N^3$ lattice sites always in the same order. By tuning (by trial and error) the typical size of the field and conjugate momenta updates, we achieve a $\sim 50\%$ acceptance rate, and typically within $\mathcal{O}(10^2)-\mathcal{O}(10^3)$ sweeps, we obtain the desired configuration. As we consider the presence of an external magnetic field, say in the $\hat z$ direction $B_{\rm ext} = B \hat e_{z}$, we simply repeat the procedure just described, but imposing the conditions Eqs.~(\ref{eq:ExtMag}), (\ref{eq:ExtMagII}).

The configurations created with the Monte-Carlo method just described, do not preserve the Gauss law $G \equiv \sum_i (E_i - E_{i,-i}) - {2e^2}dx{\rm Im}\lbrace\varphi^*\pi\rbrace = 0$. As this is a constraint between the field amplitudes and conjugate momenta, the Metropolis-Hasting updates does not know about this {\it apriori}. In order to enforce the Gauss law during the Monte-Carlo process, we have performed updates with respect to a new lattice Hamiltonian density $\mathcal{H}_L \rightarrow \mathcal{H}_L + \xi G^2$, with $\xi$ a dimensionless parameter. By tuning $\xi$ to increasingly large values, one obtain configurations that respect the Gauss law to an increasingly better and better degree. We have used $\xi = 1, 10$ and 100. In practice, the larger the value of $\xi$, the larger the required Monte-Carlo thermalization time, thus making the method impractical if one desires to obey the Gauss law to a high degree of accuracy. In order to achieve a better accuracy we have also added a {\it cooling} procedure, immediately after the  the Monte-Carlo sweeps. In particular, we have solved the equations of motion derived from considering an effective action $S_{\rm cooling} = G^2$. Such a cooling evolution hardly changes the energy of the gauge field transverse modes if the configuration was generated by the Metropolis-Hasting algorithm, but it kills the longitudinal components responsible for the violation of Gauss law~\cite{Ambjorn:1990pu}. In this way we can get the Gauss law satisfied to any desired accuracy for the initial configuration (all our simulations reached machine precision). After the cooling, we finally start the real time evolution of the fields, dictated by the set of discrete lattice equations of motion Eqs.~(\ref{eq:EOMlatticeChemPot}) [with the chemical potential fixed to $\mu = 0$]. 

Let us note that we have verified explicitly that, within the statistical error, our results (once deep inside the diffusion regime) are insensitive to the choice of $\xi = 1, 10$ or $100$. In particular, our main observable, the statistical average of the Chern-Simons number squared $\left\langle Q^2 \right\rangle$, is actually insensitive to even setting $\xi = 0$ during the Monte-Carlo, as quickly after the end of the cooling period, the real time dynamics finish thermalizing the system, achieving an equivalent configuration as if $\xi$ had been non-zero during the Monte-Carlo. We have checked explicitly that neither the time onset of diffusion, nor the properties of the diffusion regime itself, are affected the choice of $\xi$, independently or whether the latter vanishes or is non-zero. The thermalisation algorithm we used is not as efficient as that proposed in \cite{Moore:1996qs}. Still, it satisfied our needs as in any event the most time consuming part of our analysis was associated with real-time evolution rather than preparation of an initial state.

\section{Chern-Simons number diffusion} 
\label{sec:diff}

In this section we consider the diffusion of the CS number in the absence of fermions, fixing $\mu = 0$ at all times. We will consider from the beginning that there is an external magnetic field present. If there was no external magnetic field the correlator $\left\langle Q^2 \right\rangle$ would become constant after some time, but the amplitude and time scale to develope such $pleateau$, is dominated by ultraviolet contributions, and hence dictated by the lattice details~\cite{Ambjorn:1990pu,Ambjorn:1995xm}. The absence of an external magnetic field is therefore of no particular physical interest. Our interest is rather to 'switch on' an external magnetic field, and study the diffusion of $\left\langle Q^2 \right\rangle$. In that case, the correlator is then expected to grow continuously as time goes by. Based on the numerical outcome from our lattice simulations, we will be able to obtain fits to the appropriate coefficients characterizing the diffusive behavior of $\left\langle Q^2 \right\rangle$. We will then compare these with analytical estimates. Through the following sections, where we report our numerical results, we will consider the quantity $Q$ defined as
 \begin{eqnarray}\label{eq:Qlattice}
Q \equiv {e^2\over 4\pi^2}\int d^4x \,\vec E \,\vec B = N_{\rm CS}(t) - N_{\rm CS}(0)\,,~~~~~~~~ N_{\rm CS} \equiv {e^2\over 8\pi^2}\int d^4x \,\vec A \,\vec B\,.
\end{eqnarray}

Let us therefore turn our attention into the case where we consider a constant external magnetic field. As we mentioned in the previous section, without loss of generality we can consider a magnetic field in the $\hat z$ direction, $\vec B = (0,0,B)$, with a flux at a given orthogonal lattice area $A \equiv (Ndx)^2$
\begin{eqnarray}\label{eq:Bmag}
\int_A \vec B d^2\vec x = \int_A B dx_1dx_2 = B (Ndx)^2 \equiv 2\pi n_{\rm mag}\,,~~~\Rightarrow~~~ B \equiv {2\pi n_{\rm mag}\over (dx N)^2}\,,
\end{eqnarray}
with $n_{\rm mag}$ an integer number. For a given magnetic seed $n_{\rm mag}$, we have obtained $\langle Q^2\rangle$ for a fixed lattice spacing $Tdx = 1$, varying the gauge coupling $e^2$ and the number of points per dimension $N$ [hence varying the lattice volume\footnote{From now on, as $Tdx = 1$ is fixed, as discussed in Sect.~\ref{subsec:parameters}, we will interchangeably refer to the the lattice volume by either $V$, or loosely simply by $N$.} $V = N^3dx^3$]. For each set of parameters $(n_{\rm mag},e^2,N)$, we have obtained $N_R$ independent realizations of $Q^2$ as a function of time (with $N_R \geq 20$). We have built the desired correlator $\left\langle Q^2 \right\rangle$ simply by statistical averaging,
\begin{eqnarray}\label{eq:Q2averaged}
\langle Q^2\rangle \equiv {1\over N_R}\sum_{i} Q_i^2\,,
\end{eqnarray}
where the index $i$ labels the different realizations. The statistical error in the measurement of $\langle Q^2\rangle$ is determined by
\begin{eqnarray}
\Delta_{Q^2} \equiv {\Sigma_{Q^2}\over\sqrt{N_R}}\,,~~~\Sigma_{Q^2} \equiv \sqrt{{1\over N_R}\sum_{i} (Q_i^2-\langle Q^2\rangle)^2}\,,
\end{eqnarray}
where $\Sigma_{Q^2}^2$ represents the statistical variance of the ensemble of $N_R$ realizations. In most plots we represent with a central solid line the mean value $\langle Q^2\rangle$, whereas a shaded area between some upper $\langle Q^2\rangle + \Delta_{Q}$ and and lower $\langle Q^2\rangle - \Delta_{Q}$ curves around a given central line, represents the error in the assessment of $\langle Q^2\rangle$. See for instance Fig.~\ref{fig:nMag4Q2_vs_e2_and_t}. We will maintain this plotting style all through the paper.

In all the figures of this section we plot $\langle Q^2\rangle$ and its statistical error $\Delta_{Q^2}$, as obtained from our simulations with an external magnetic field with magnetic seeds $n_{\rm mag} = 4, 16$ and $64$, and gauge coupling values listed in Eqs.~(\ref{eq:Param1})-(\ref{eq:Param4}), depending on $N$. Let us notice that due to the scaling of the magnetic field as $B \propto N^{-2}$, the strength of the magnetic field weakens the larger the $N$ (for a fixed magnetic seed $n_{\rm mag}$). Hence, if we take e.g.~$n_{\rm mag} = 4$, the magnetic field for the simulations with $N = 32$ will be a factor ${1\over 4}$ weaker than the magnetic field for the simulations with $N = 16$ for the same magnetic seed. However, the strength of the magnetic field will be the same for $N = 16$ and $N = 32$ simulations if we take $n_{\rm mag} = 4$ for the former, but $n_{\rm mag} = 16$ for the latter.

\subsection{Gauge coupling dependence} 
\label{sec:GaugeCouplingScaling}

\begin{figure}[t]
\begin{center}
\includegraphics[width=7.5cm]{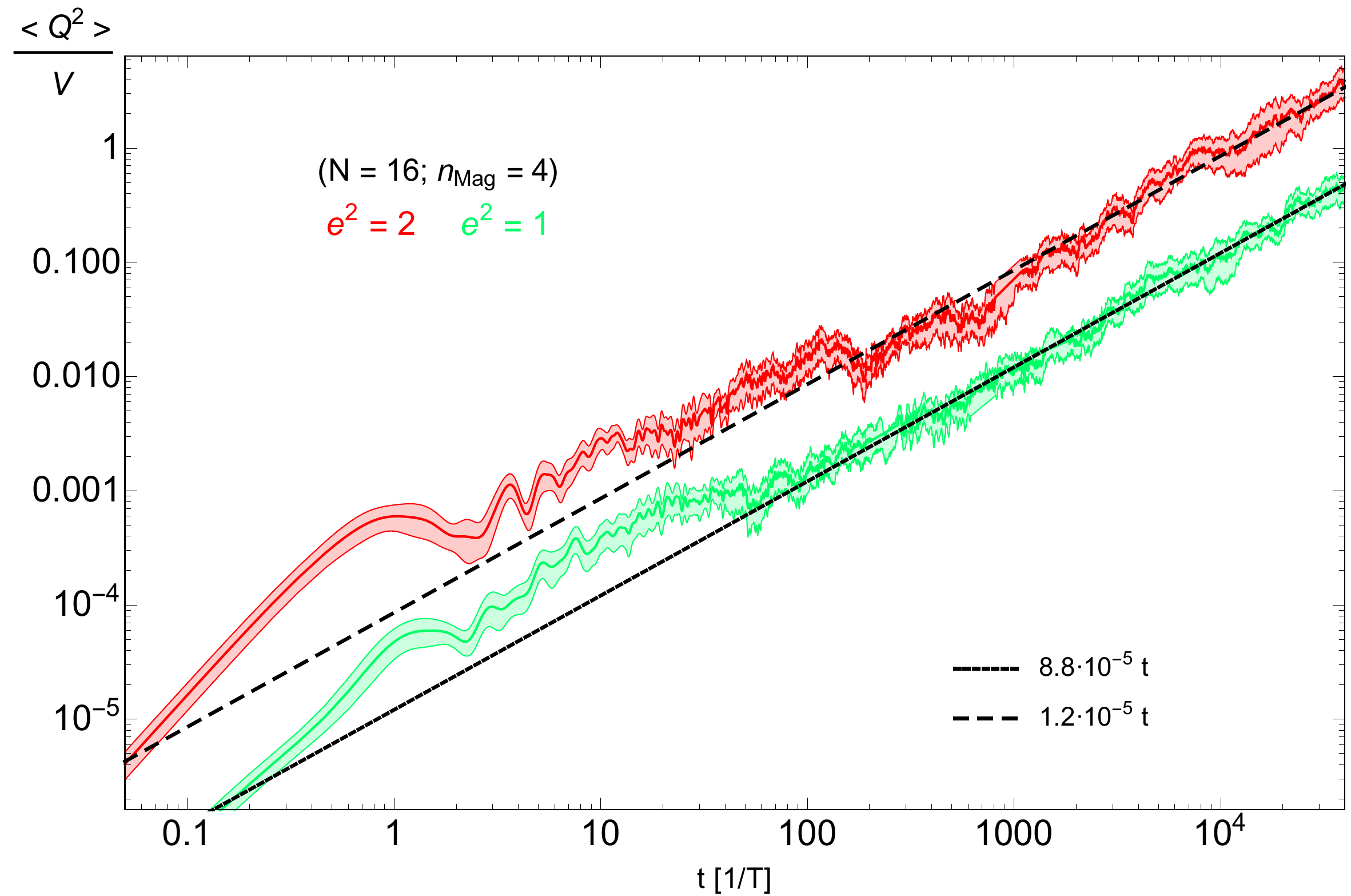}
\includegraphics[width=7.5cm]{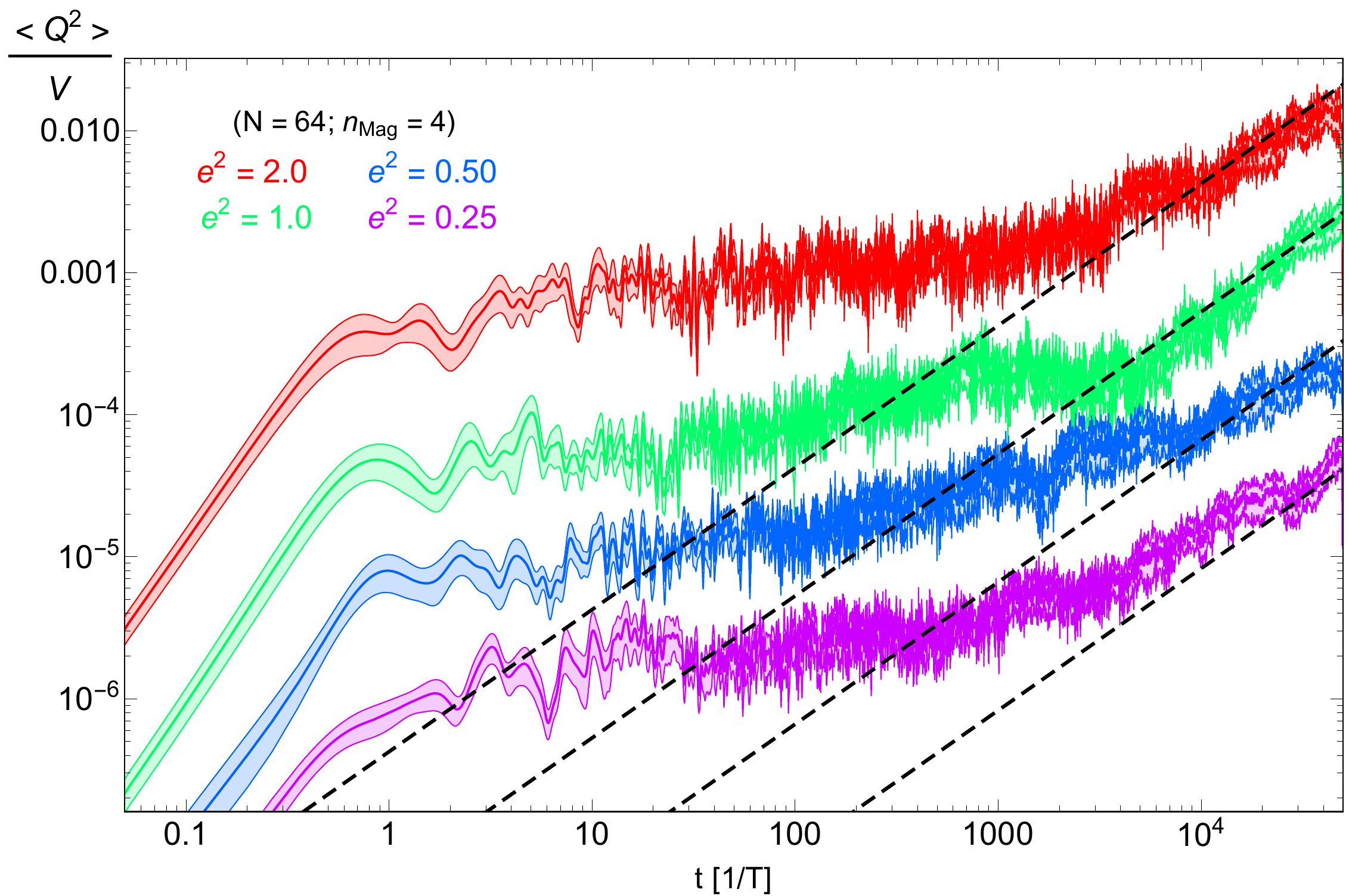}
\end{center}
\caption{The volume normalized correlator $\langle Q^2\rangle/V$, obtained from our simulations with $n_{\rm mag} = 4$. {\it Left}: $\langle Q^2\rangle/V$ for a fixed volume ($Tdx = 1, N = 16)$ and two gauge coupling values $e^2 = 1, 2$. The two straight lines correspond to fits to the data, denoting a linear growth. {\it Right}: $\langle Q^2\rangle/V$ for a fixed volume ($Tdx = 1, N = 64)$ and four gauge coupling values $e^2 = 0.25, 0.5, 1, 2$. The four straight lines are not fits in this case, but aid the eye to compare the correlators' growth against a linear behavior.}
\label{fig:nMag4Q2_vs_e2_and_t}
\end{figure}

Let us start by considering a fixed magnetic seed $n_{\rm mag}$. We will vary also $n_{\rm mag}$ later on, in Section~\ref{sec:MagneticFieldScaling}, but for the time being we shall fix it to a given value, say e.g.~$n_{\rm mag} = 4$. In Fig.~\ref{fig:nMag4Q2_vs_e2_and_t} we then plot $\langle Q^2\rangle$ as obtained from our simulations with $n_{\rm mag} = 4$. In the left panel we show ${\langle Q^2\rangle\over V}$ for a fixed volume ($Tdx = 1, N = 16)$ and two gauge coupling values $e^2 = 1, 2$. As expected the correlator amplitude grows linearly in time as $\propto t$, corresponding to the diffusive regime of CS number in the presence of a constant external magnetic field. The system actually attempts first to relax into a constant asymptotic amplitude (as expected in the case of absent magnetic field), until the diffusion becomes noticeable. For such small volume simulations with $N = 16$, at a time $t \sim \mathcal{O}(1) T^{-1}$ the system has already entered into some kind of diffusion. However it is not until a time $t \sim \mathcal{O}(100) T^{-1}$ (gauge coupling dependent), that $\langle Q^2 \rangle$ enters into its final asymptotic regime, linearly growing in time. We have fitted this growth (averaging over the wiggly behavior) as ${\langle Q^2\rangle\over V} = \Gamma t$, with $\Gamma_{1} \simeq 1.2\cdot 10^{-5}~T^4$ for $e^2 = 1$ and $\Gamma_2 \simeq 8.8\cdot 10^{-5}~T^4$ for $e^2 = 2$. As appreciated in the plot, the linear behavior (once started) is well sustained during all the  simulation time, and it seems a robust feature against the oscillatory behavior of $\langle Q^2 \rangle$. The amplitude of the slopes is clearly bigger the larger the coupling $e^2$, however the ratio $\Gamma_{2}/\Gamma_1 \simeq 7.3$ deviates from the theoretical value $2^3 = 8$, based on the expected behavior $\Gamma \propto (e^2)^3$ (for fixed $n_{\rm mag}$). We will quantify this deviation later when exploring other volumes and gauge coupling values, properly taking into account statistical and averaging errors.

In the right panel of Fig.~\ref{fig:nMag4Q2_vs_e2_and_t} we plot ${\langle Q^2\rangle\over V}$ for $n_{\rm mag} = 4$, $N = 64$ ($Tdx = 1$), and gauge coupling values $e^2 = 0.25, 0.5, 1$ and $2$. The volume in these simulations is a factor $(64/16)^3 = 2^6 = 64$ larger than in the simulations used in the left panel, so the physical magnetic field is, correspondingly, a factor ${1\over 16}$ weaker ($n_{\rm mag} = 4$ is the same for both panels). This fact can be clearly appreciated by noting that the moment when the correlator amplitude starts growing linearly, is visibly delayed in the right panel, compared to the left panel smaller volume simulations. In the larger volume simulations it is clearly appreciated as well that $\langle Q^2 \rangle$ grows in time and its amplitude is larger the bigger the coupling $e^2$. By eye, it is not clear however what is the exact scaling with $e^2$. Quantifying this dependence requires a proper parameter fit which we will present shortly.

Let us make before a small digression about one of the technical difficulties we have encountered in this work: the onset of the diffusion regime was delayed in some cases up to very large times, often up to physical times $Tt \sim 10^3-10^4$. The weaker the physical magnetic field the longer it took in general to reach the onset of a diffusion regime, where the correlator truly grows linearly in time. As we are only interested in such regime, in some cases\footnote{Particularly in those with large volume $N$, as that implies smaller magnetic field $B \propto 1/N^2$.} we could only start measuring the linear growth at very late times, for a period marginally longer than a decade in time. This really limited our capabilities to study systems with a large volume, as the simulation time needed to reach and probe the difussion regime, was unfeasible for large volumes, given our computing resources. In practice, for a case with magnetic seed $n_{\rm mag} = 4$, volume $N^3 = 64^3$, time step $Tdt = 0.05$ and final physical time $Tt_f = 10^5$, we could only measure the linear regime for roughly $\sim 1-2$ decades in time, and this took around $\sim 6$ days of computing time for each individual job. For $n_{\rm mag} = 4$, we could not obtain useful data from our $N = 128$ runs, as the diffusion stage had barely started by the end of simulations that had run for around $\sim 1$ week of computing time. Only for simulations with larger magnetic seeds, as those we present in Sect.~\ref{sec:MagneticFieldScaling}, we could reach $N = 128$ lattices, as the onset of diffusion was reached at earlier physical times (yet we had to enlarge the time step to $Tdt = 0.2$ and to reduce the final time to $Tt_f = 5\cdot 10^4$, in order to run these simulations under $\sim 1$ week of computing time).

\begin{figure}[t]
\begin{center}
\includegraphics[width=7.5cm]{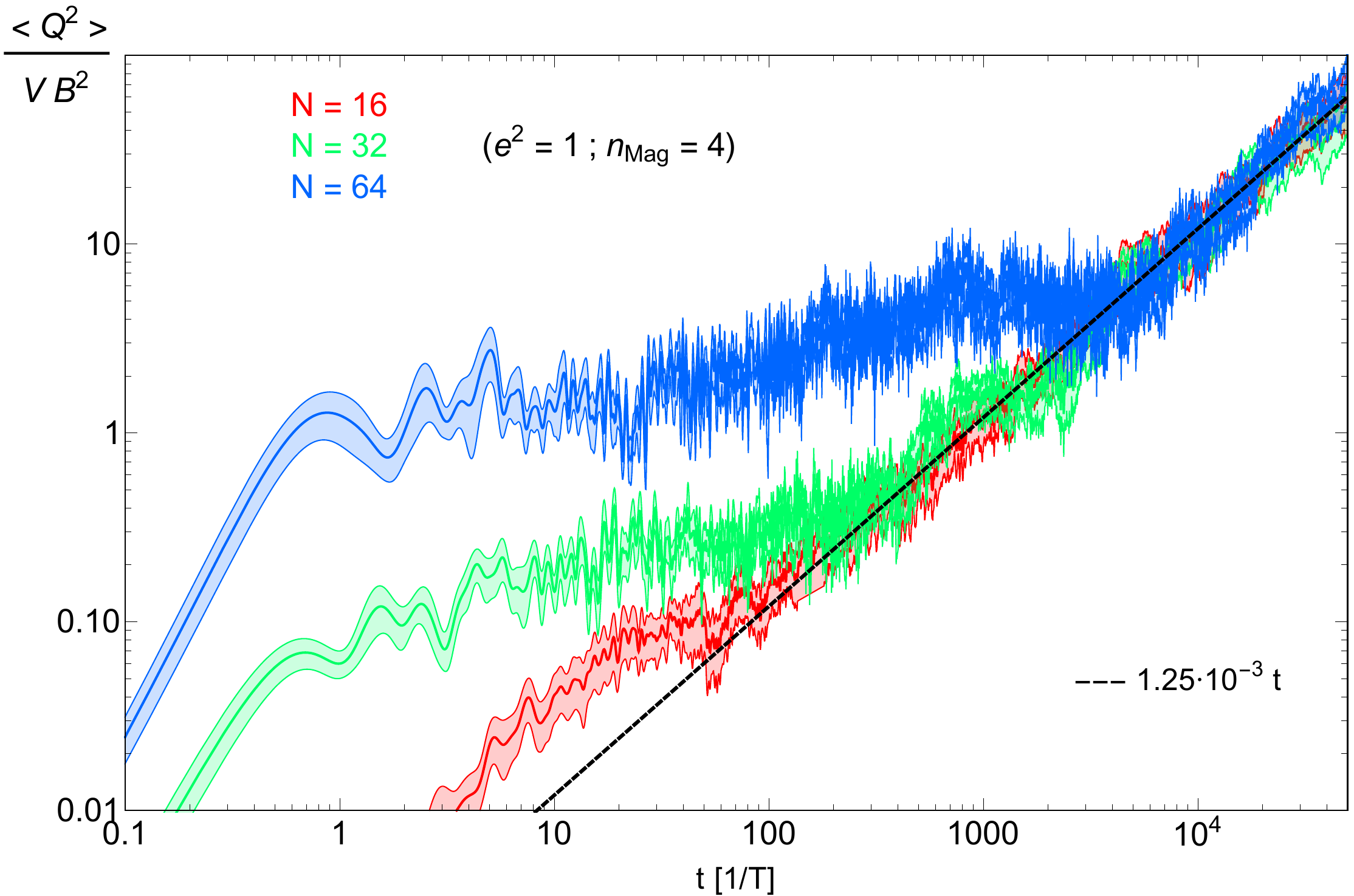}
\includegraphics[width=7.5cm]{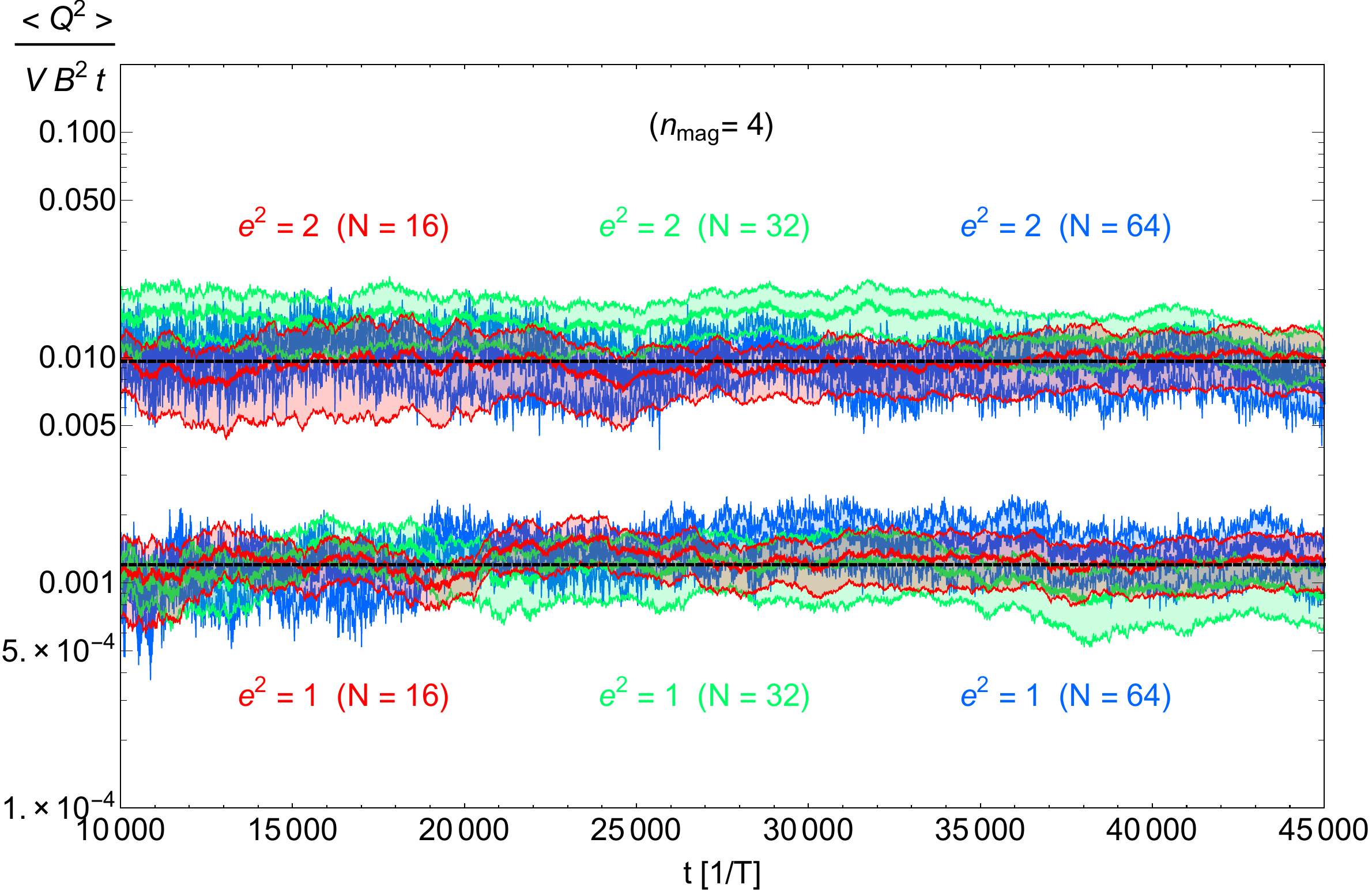}
\end{center}
\caption{Growth of $\langle Q^2\rangle$ in the presence of an external magnetic field with $n_{\rm mag} = 4$, for a variety of gauge couplings and volumes. {\it Left}: Volume and magnetic field normalized correlator, ${\langle Q^2\rangle\over V B^2}$, for a fix coupling $e^2 = 1$, sampling the volume $V = (N\cdot dx)^3$ for $N = 16, 32$ and $64$. The straight line corresponds to a fit to the data, denoting a linear growth well sustained once diffusion is onset. {\it Right}: Volume, magnetic field and time normalized correlator, ${\langle Q^2\rangle\over V B^2 t}$, shown at times $t \gg t_{\rm diff}$, once diffusion has been well established for all the cases. The horizontal lines correspond to fits to the data.}
\label{fig:nMag4Q2overB2t_e21e22}
\end{figure}

In Fig.~\ref{fig:nMag4Q2overB2t_e21e22} we plot again the correlator $\langle Q^2\rangle$ in the presence of a external magnetic field with $n_{\rm mag} = 4$, for a variety of gauge couplings and volumes. In particular, in the left panel we plot the volume and magnetic field normalized correlator amplitude, ${\langle Q^2\rangle\over V B^2}$, for a fix coupling $e^2 = 1$, varying the volume $V = (N\cdot dx)^3$ for $N = 16, 32$, $64$ and $128$ ($Tdx = 1$). Based on theoretical considerations, we expect the correlator to be directly proportional to external magnetic fields as $\langle Q^2\rangle \propto B^2$. We then expect the slopes of ${\langle Q^2\rangle\over V B^2}$ for different volumes $N$, to sit on top of each other (once in the diffusion regime). This is precisely what we observe in the left panel of Fig.~\ref{fig:nMag4Q2overB2t_e21e22}, within the statistical error, after a time $t \gtrsim \mathcal{O}(10^3) T^{-1}$. As $n_{\rm mag} = 4$ is the same for all cases depicted, but $N$ is different, the magnetic field $B \propto {n_{\rm mag}\over N^2}$ is weaker the larger the volume simulated. Hence, the diffusion process starts later the weaker it is the external magnetic field $B$ (i.e.~the larger the $N$). The left panel of Fig.~\ref{fig:nMag4Q2overB2t_e21e22} highlights therefore very clearly the technical problem we discussed before: for large volume simulations the required time needed to reach diffusion becomes unfeasible. Our simulations for the same parameters and $N = 128$, showed that the diffusion stage had not been reached yet at the end of the simulation time. That is why, for $n_{\rm mag} = 4$, we do not include the $N = 128$ case in the figure, nor in the parameter fit analysis we present later.

In the right panel of Fig.~\ref{fig:nMag4Q2overB2t_e21e22} we plot ${\langle Q^2\rangle\over V B^2 t}$ at large times, once the diffusion behavior is well established. If, as expected, the correlator grows linearly in time as $\langle Q^2\rangle \propto t$, then ${\langle Q^2\rangle\over V B^2 t}$ should turn into horizontal plots, lying on top of each other for a given gauge coupling strength $e^2$, independently of the physical magnetic field $B(n_{\rm mag},N)$ and of the volume $V(N,dx)$. This is precisely what we observe, within the statistical error, in the right panel of Fig.~\ref{fig:nMag4Q2overB2t_e21e22}. To quantify deviations from the linearly growth behavior, we could introduce a new parameter ${\langle Q^2\rangle} \propto t^{1+\alpha}$. However, our attempts do so, have shown clearly that within the statistical error, one can take safely take $\alpha = 0$ for every set of values of $(B,V,e^2)$ considered. 

Let us note that the numerical correlators $\langle Q^2(t) \rangle$ clearly grow with the gauge coupling. Based on the same scaling arguments as before, we expect ${\langle Q^2\rangle\over V B^2 t} \propto (e^2)^3$. In order to quantify possible deviations from this scaling, we have take into account that $\langle Q^2 \rangle$ grows in time. We have defined an amplitude for the correlator in diffusion as
\begin{eqnarray}\label{eq:Q2aymp1}
\Gamma_{\rm diff} \equiv {1\over (t-t_{\rm diff})}\int_{t_{\rm diff}}^t dt'{\langle Q^2\rangle(t')\over Vt'}\,,
\end{eqnarray}
and assigned two uncertainties to this quantity, a statistical error as
\begin{eqnarray}\label{eq:StatError}
\Delta\Gamma_{\rm stat} \equiv {1\over (t-t_{\rm diff})}\int_{t_{\rm diff}}^t dt'{\Delta_{Q^2}(t')\over Vt'}\,,
\end{eqnarray}
and an error due to the oscillatory behavior as
\begin{eqnarray}\label{eq:OscError}
\Delta\Gamma_{\rm osc} \equiv \sqrt{{1\over(t-t_{\rm diff})}\int_{t_{\rm diff}}^t dt'\left({\langle Q^2\rangle(t')\over Vt'} - \Gamma_{\rm diff}\right)^2}\,.
\end{eqnarray}
We will assign as the error of $\Gamma_{\rm diff}$ in the diffusion regime, the maximum of the previous two errors, 
\begin{equation}\label{eq:MaxErrorDef}
\Delta\Gamma_{\rm diff} \equiv {\rm max}\lbrace\Delta\Gamma_{\rm stat},\Delta\Gamma_{\rm osc}\rbrace \,.
\end{equation}

\begin{figure}[t]
\begin{center}
\includegraphics[width=7.5cm]{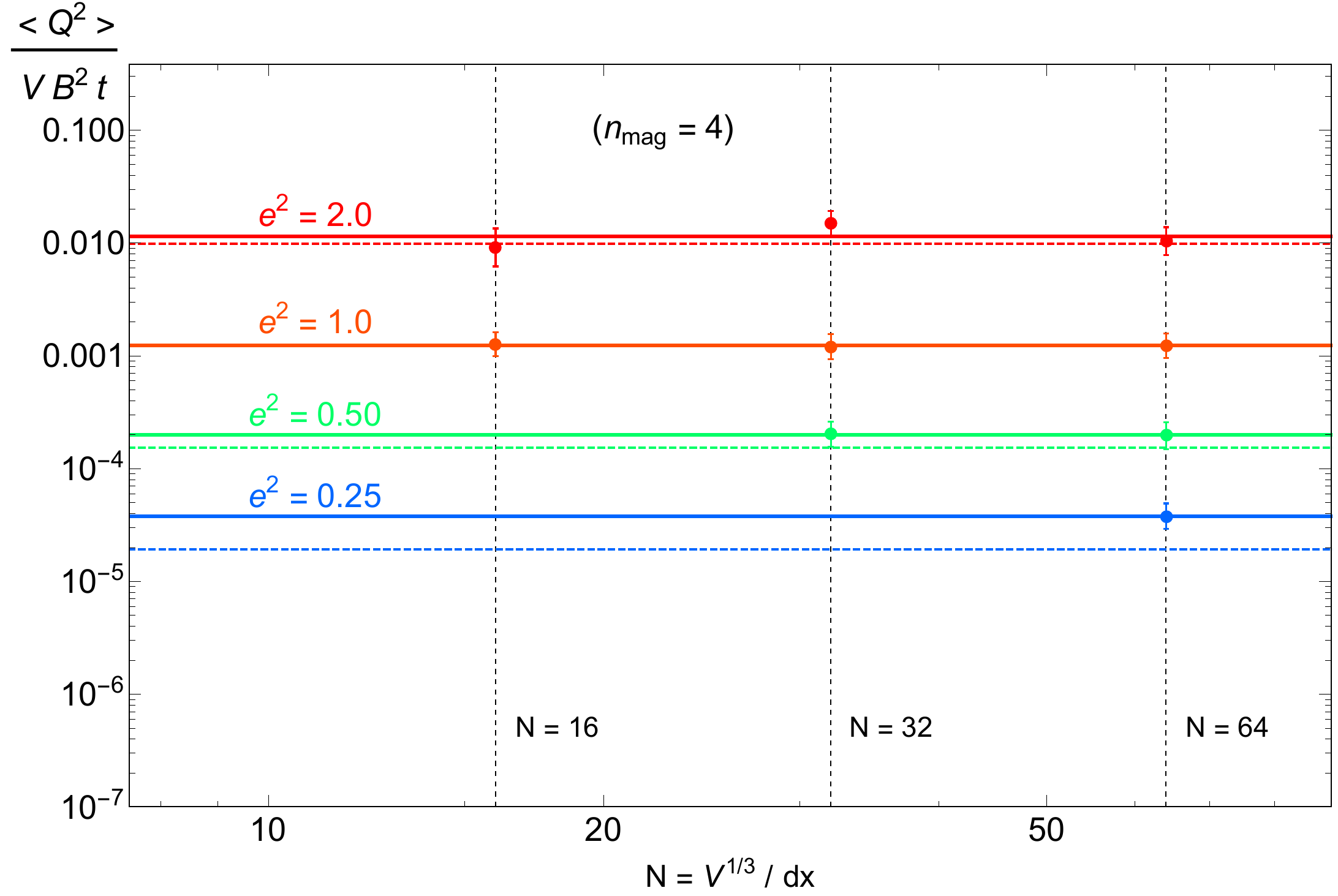}
\includegraphics[width=7.5cm]{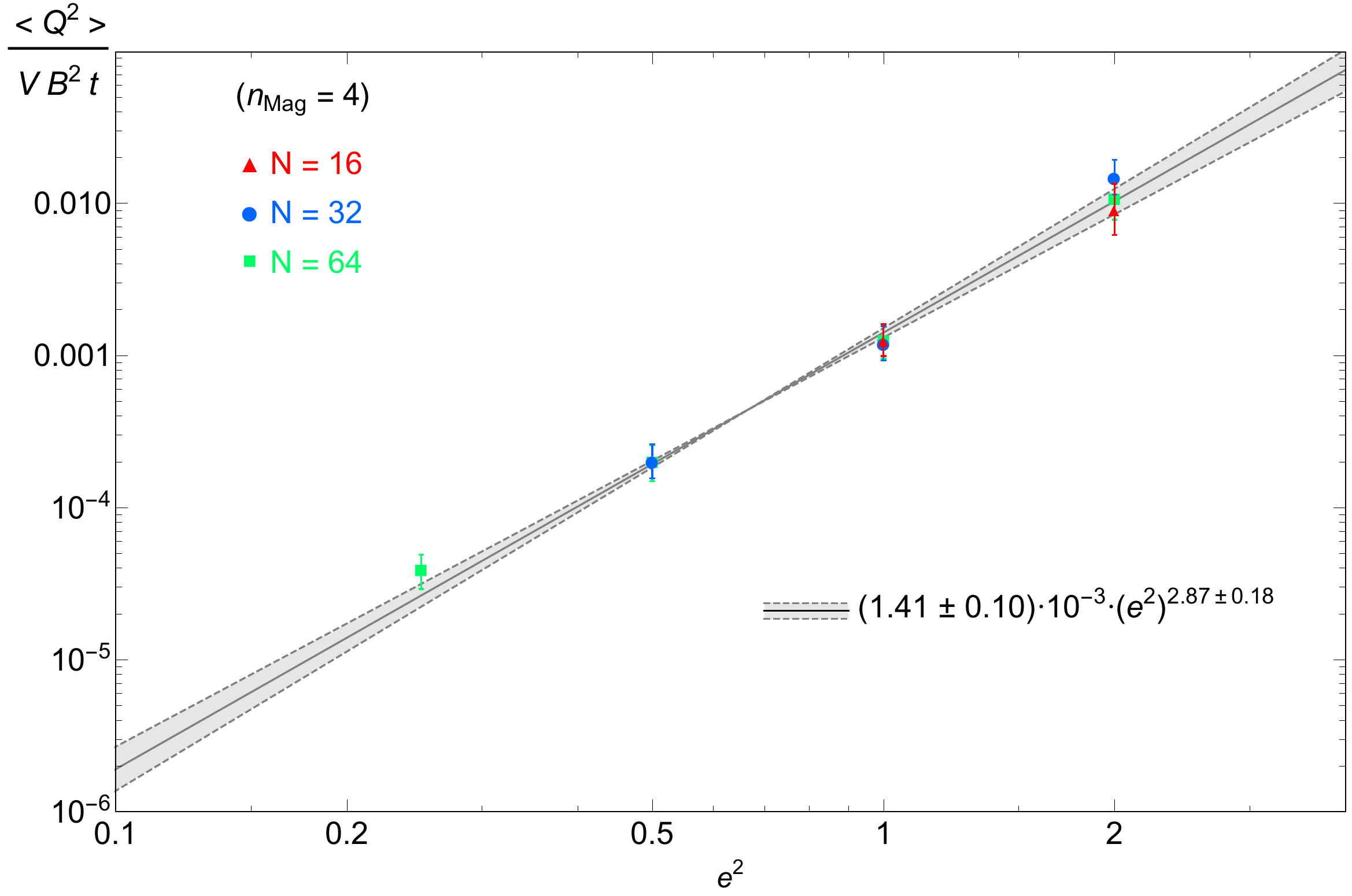}
\end{center}
\caption{Amplitude ${\langle Q^2\rangle_{\rm diff}\over {B^2Vt}} \equiv {\Gamma_{\rm diff}\over B^2}$ for $n_{\rm mag} = 4$ (where $B \equiv 2\pi n_{\rm mag}/N^2$), for all the values $(e^2,N)$ considered. {\it Left}: ${\Gamma_{\rm diff}\over B^2}$ versus $N$. {\it Right}: ${\Gamma_{\rm diff}\over B^2}$ versus $e^2$.}
\label{fig:Q2nMag4_vs_allN_allE2}
\end{figure}

In the left panel of Fig.~\ref{fig:Q2nMag4_vs_allN_allE2} we present $\Gamma_{\rm diff} \pm \Delta\Gamma_{\rm diff}$ for all our data sample for $n_{\rm mag}= 4$ (i.e.~considering all our volume and coupling constants pool). For each volume $N$, we have found a best fit to the data points $\lbrace e^2,\Gamma_{\rm diff} \rbrace$ of the form $\Gamma_{\rm diff}^{(N)} = A_N\cdot(e^2)^{m_N}B^2$. We obtain
\begin{eqnarray}\label{eq:GammasNmag4}
n_{\rm mag} = 4~~\rightarrow~~\left\lbrace\begin{array}{l}
\Gamma_{\rm diff}^{(16)} \simeq 1.26\cdot 10^{-3}\cdot(e^2)^{2.85}\cdot B^2\,,~~N = 16\\
\Gamma_{\rm diff}^{(32)} \simeq 1.53\cdot 10^{-3}\cdot(e^2)^{3.10}\cdot B^2\,,~~N = 32\\
\Gamma_{\rm diff}^{(64)} \simeq 1.41\cdot 10^{-3}\cdot(e^2)^{2.70}\cdot B^2\,,~~N = 64\\
\end{array}\right.~~~~\,.
\end{eqnarray}

We have then averaged over the different fits, weighting each case by the number of gauge coupling values for each $N$. We find an averaged fit of the form $\Gamma \equiv {\langle Q^2\rangle\over Vt} = (\bar A \pm \Delta A)\cdot (e^2)^{\bar m \pm \Delta m}$, 
\begin{eqnarray}\label{eq:GammaDiffnMag4}
\Gamma_{\rm diff} \simeq (1.41 \pm 0.10)\cdot 10^{-3}\cdot (e^2)^{2.87 \pm 0.18}\cdot B^2\,,~~~~~ (n_{\rm mag} = 4)\,,
\end{eqnarray}
where the errors in the amplitude and the exponent obtained by $\bar A \equiv \sum_N \#_N A_N$, $\bar m \equiv \sum_N \#_N m_N$, $(\Delta A)^2 \equiv \sum_N \#_N (A_N-\bar A)^2$, and $(\Delta m)^2 \equiv \sum_N \#_N (m_N-\bar m)^2$, with the weights $\#_{16} = 2/9$, $\#_{32} = 3/9$ and $\#_{64} = 4/9$, corresponding to the gauge coupling values $e^2 = 1, 2$ for $N = 16$, $e^2 = 0.5,1,2$ for $N = 32$ and $e^2 = 0.25,0.5,1,2$ for $N = 64$. In the left panel of Fig.~\ref{fig:Q2nMag4_vs_allN_allE2} we show ${\langle Q^2\rangle_{\rm diff}\over {B^2Vt}} \equiv {\Gamma_{\rm diff}\over B^2}$ versus $N$. For a given gauge coupling value we find the amplitudes of this quantity to be similar to each other within the statistical error, independently of the volume. Solid horizontal lines indicate the best fit to ${\Gamma_{\rm diff}\over B^2}$ for a fixed value of $e^2$, whereas dashed lines indicate the expected amplitude based on the theoretical scaling ${\Gamma_{\rm diff}\over B^2} \propto e^6$ (taking the amplitude for $e^2 = 1$ as a reference). In the right panel of Fig.~\ref{fig:Q2nMag4_vs_allN_allE2} we show ${\Gamma_{\rm diff}\over B^2}$ as a function of $e^2$. Consistently with the left panel, the amplitude is of the same order for a given gauge coupling $e^2$, independently of the volume. The solid line in the right panel represents the average fit Eq.~(\ref{eq:GammaDiffnMag4}) from the best fits Eqs.~(\ref{eq:GammasNmag4}) to each choice of $N$. The shade area between the dashed lines represents the dispersion among such best fits. We conclude that the theoretical expectation ${\Gamma_{\rm diff}\over B^2} \propto (e^2)^3$ is verified well within the associated errors. The best fit of the data actually prefers a scaling as ${\Gamma_{\rm diff}\over B^2} \propto (e^2)^{2.87}$, see Eq.~(\ref{eq:GammaDiffnMag4}), but the uncertainty in the exponent encompasses well the case ${\Gamma_{\rm diff}\over B^2} \propto (e^2)^{3}$.

\subsection{Magnetic field dependence} 
\label{sec:MagneticFieldScaling}

In this section we will first present an analysis similar to that in section~\ref{sec:GaugeCouplingScaling}, applied to the data obtained from our simulations with stronger magnetic fields, corresponding to the seeds $n_{\rm mag} = 16$ and $64$. This implies the use of magnetic fields $4$ and $16$ times larger, respectively, than in the previous section. This analysis introduces yet a $B^2$ factor in the fits, based on the theoretically expected scaling $\langle Q^2 \rangle \propto B^2$. In order words, these fits are obtained over the normalized correlators ${\langle Q^2\rangle_{\rm diff}\over {B^2Vt}} \equiv {\Gamma_{\rm diff}\over B^2}$ vs $e^2$. In order to quantify numerically the deviations of our data set from the theoretically expected behavior $\propto B^2$, we will then re-analyze the data, fixing the gauge coupling value and fitting the correlator amplitudes against the magnetic field seed $n_{\rm mag}$. Finally, we will perform another re-analysis of the data, based on a multi-parameter fit of the correlator amplitudes against all parameters involved, the volume, the magnetic field seed, and the gauge coupling.

\begin{figure}[t]
\begin{center}
\includegraphics[width=7.5cm]{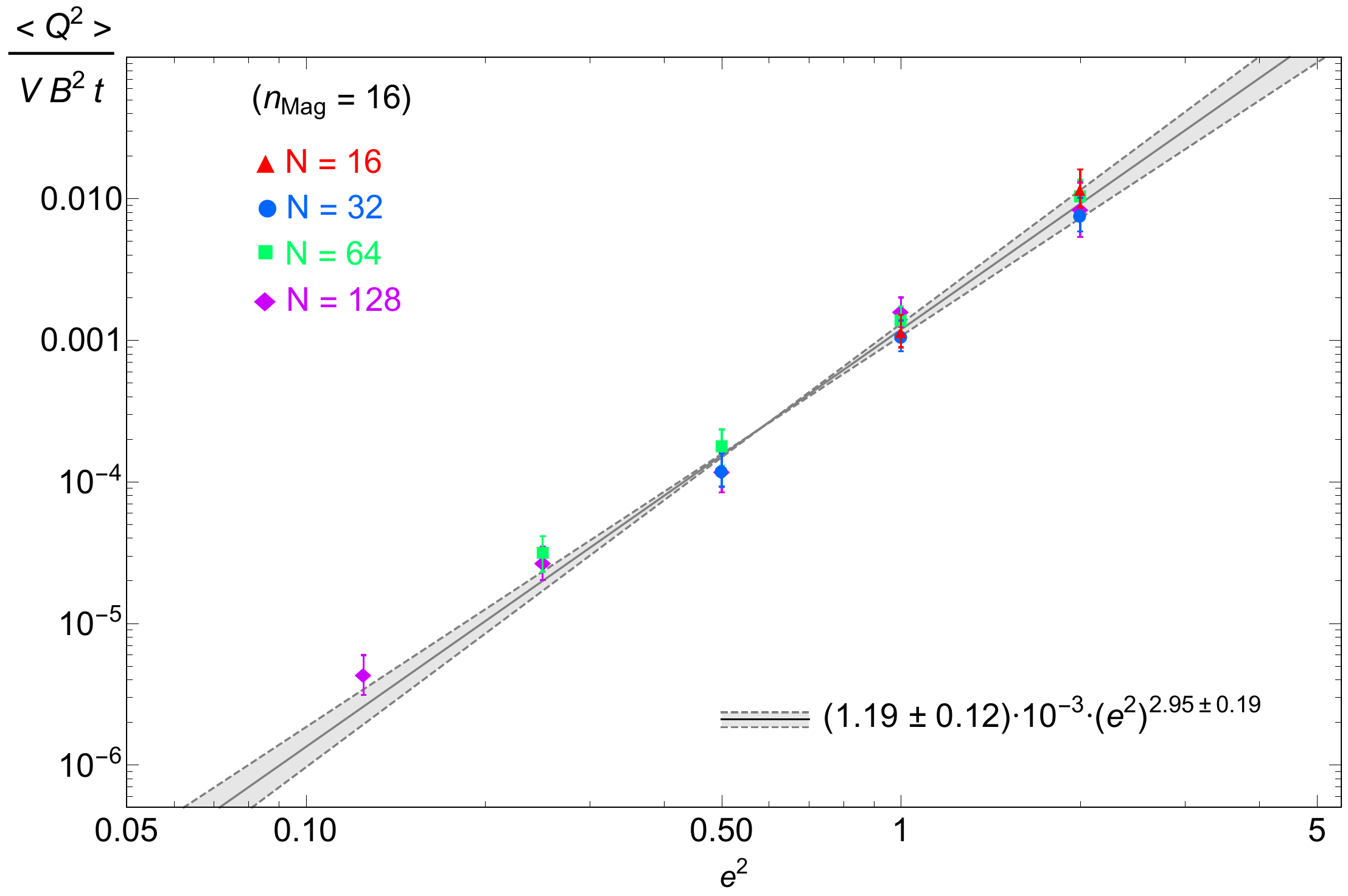}
\includegraphics[width=7.5cm]{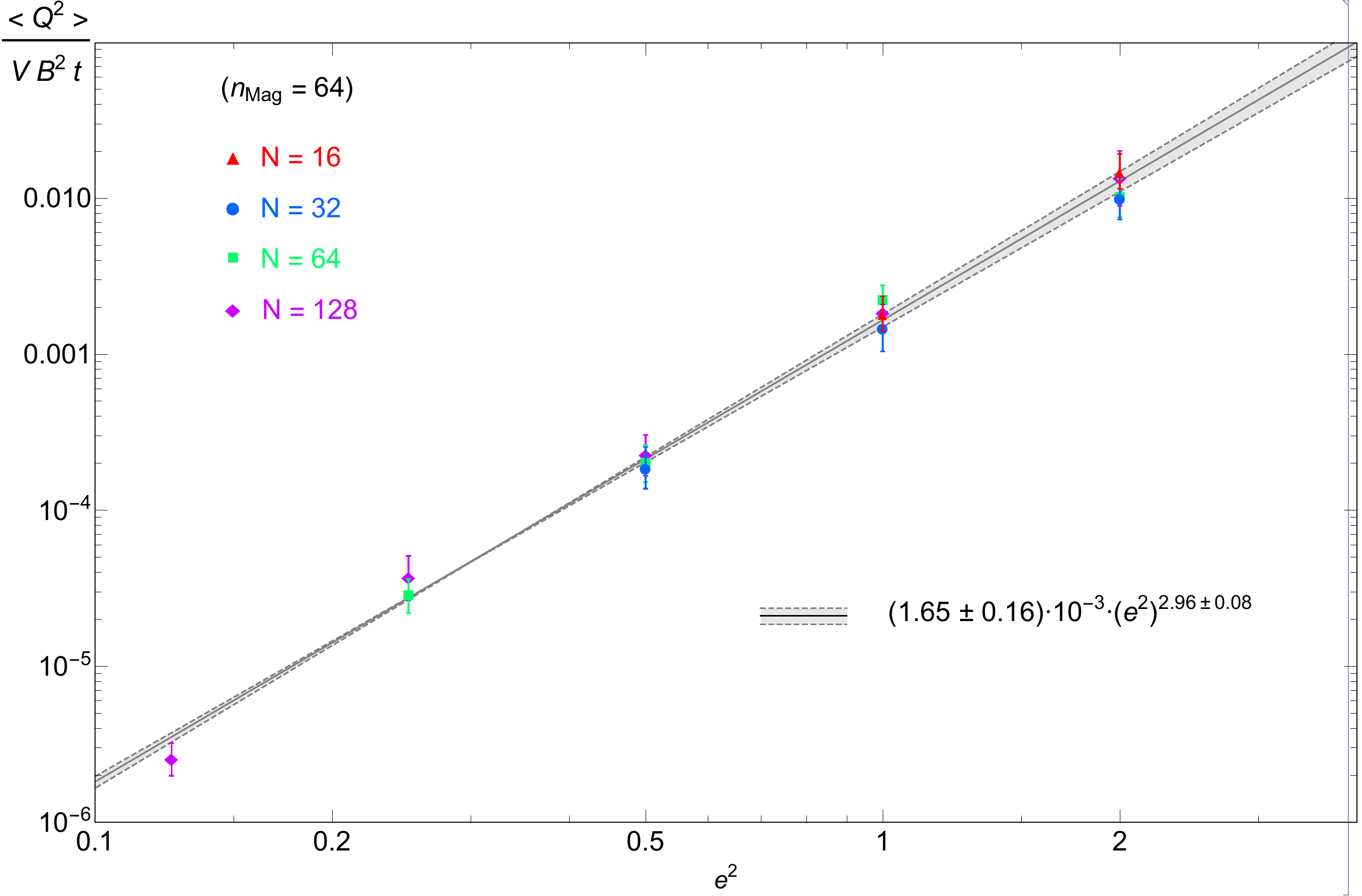}
\end{center}
\caption{${\Gamma_{\rm diff}\over B^2}$ versus $e^2$ for $n_{\rm mag} = 16$ (left panel) and $n_{\rm mag} = 64$ (right panel), for all our $(e^2,N)$ values. Solid straight lines represent an average [Eq.~(\ref{eq:GammaDiffnMag16}) for the left panel, and Eq.~(\ref{eq:GammaDiffnMag64}) for the right panel] over the best fits to the data for each volume $N$. Shade areas represent the dispersion among such best fits, see Eqs.~(\ref{eq:GammasNmag16}) and Eqs.~(\ref{eq:GammasNmag64}).}
\label{fig:Q2nMag16and64_allN_vs_e2}
\end{figure}

Let us then first look at the scaling of the correlators vs gauge coupling, for $n_{\rm mag} = 16$ and $64$, so that we can compare these results with Eqs.~(\ref{eq:GammasNmag4}), (\ref{eq:GammaDiffnMag4}) obtained for $n_{\rm mag} = 4$. We have analyzed our $n_{\rm mag} = 16$ and $n_{\rm mag} = 64$ data set in identical manner as we did for $n_{\rm mag} = 4$ in section~\ref{sec:GaugeCouplingScaling}: for each pair of values $(n_{\rm mag},N)$ we have determined $\Gamma_{\rm diff} \pm \Delta\Gamma_{\rm diff}$. Then, for each magnetic seed $n_{\rm mag}$, we have found a best fit to the data points $(e^2,\Gamma_{\rm diff})$ for each value $N$, of the form $\Gamma_{\rm diff}^{(N)} = A_N\cdot(e^2)^{m_N}B^2$. The one difference now is that for the larger magnetic seeds $n_{\rm mag} = 16, 64$, we have added data from our simulations with volumes $N^3 = 128^3$, as in this occasion (contrary to $n_{\rm mag} = 4$), the system was reaching the diffusion regime before the end of the simulation time. We obtain
\begin{eqnarray}\label{eq:GammasNmag16}
n_{\rm mag} = 16~~\rightarrow~~\left\lbrace\begin{array}{l}
\Gamma_{\rm diff}^{(16)} \simeq 1.16\cdot 10^{-3}\cdot(e^2)^{3.35}\cdot B^2\,,~~~N = 16\\
\Gamma_{\rm diff}^{(32)} \simeq 1.00\cdot 10^{-3}\cdot(e^2)^{3.00}\cdot B^2\,,~~~N = 32\\
\Gamma_{\rm diff}^{(64)} \simeq 1.34\cdot 10^{-3}\cdot(e^2)^{2.93}\cdot B^2\,,~~~N = 64\\
\Gamma_{\rm diff}^{(128)} \simeq 1.21\cdot 10^{-3}\cdot(e^2)^{2.77}\cdot B^2\,,~~N = 128\\
\end{array}\right.
\end{eqnarray}
\begin{eqnarray}\label{eq:GammasNmag64}
n_{\rm mag} = 64~~\rightarrow~~\left\lbrace\begin{array}{l}
\Gamma_{\rm diff}^{(16)} \simeq 1.82\cdot 10^{-3}\cdot(e^2)^{3.04}\cdot B^2\,,~~~N = 16\\
\Gamma_{\rm diff}^{(32)} \simeq 1.40\cdot 10^{-3}\cdot(e^2)^{2.87}\cdot B^2\,,~~~N = 32\\
\Gamma_{\rm diff}^{(64)} \simeq 1.61\cdot 10^{-3}\cdot(e^2)^{2.88}\cdot B^2\,,~~~N = 64\\
\Gamma_{\rm diff}^{(128)} \simeq 1.80\cdot 10^{-3}\cdot(e^2)^{3.04}\cdot B^2\,,~~N = 128\\
\end{array}\right.\,
\end{eqnarray}
For each $n_{\rm mag}$ value, we have averaged over the different fits, weighting each case by the number of possible gauge coupling values for each $N$ [recall Eqs.~(\ref{eq:Param1})-(\ref{eq:Param4}) and the related discussion]. This is the same procedure as we did for the $n_{\rm mag} = 4$ data, except that now we include all gauge coupling cases from the simulations for $N = 128$. We find an averaged fit of the form $\Gamma_{\rm diff} \equiv {\langle Q^2\rangle\over Vt} = (\bar A \pm \Delta A)\cdot (e^2)^{\bar m \pm \Delta m}$ as [for completeness, we reproduce again the fit to $n_{\rm mag} = 4$ Eq.~(\ref{eq:GammaDiffnMag4})] 
\begin{eqnarray}\label{eq:GammaDiffAllnMag}
\Gamma_{\rm diff} &\simeq& (1.41 \pm 0.10)\cdot 10^{-3}\cdot (e^2)^{2.87 \pm 0.18}\cdot B^2\,,~~~~~ (n_{\rm mag} = 4)\,,\\
\Gamma_{\rm diff} &\simeq& (1.19\pm 0.12)\cdot 10^{-3}\cdot (e^2)^{2.95 \pm 0.19}\cdot B^2\,,~~~~~ (n_{\rm mag} = 16)\,,
\label{eq:GammaDiffnMag16}\\
\Gamma_{\rm diff} &\simeq& (1.65\pm 0.16)\cdot 10^{-3}\cdot (e^2)^{2.96 \pm 0.08}\cdot B^2\,,~~~~~ (n_{\rm mag} = 64)\,,
\label{eq:GammaDiffnMag64}
\end{eqnarray}
where the errors in the amplitude and the exponent are obtained by $\bar A \equiv \sum_N \#_N A_N$, $\bar m \equiv \sum_N \#_N m_N$, $(\Delta A)^2 \equiv \sum_N \#_N (A_N-\bar A)^2$, and $(\Delta m)^2 \equiv \sum_N \#_N (m_N-\bar m)^2$, with weights $\#_{16} = 2/14$, $\#_{32} = 3/14$, $\#_{64} = 4/14$ and $\#_{128} = 5/14$, corresponding to the gauge coupling values $e^2 = 1, 2$ for $N = 16$, $e^2 = 0.5,1,2$ for $N = 32$, $e^2 = 0.25,0.5,1,2$ for $N = 64$, and finally $e^2 = 2, 1, 1/2, 1/4$ and $1/8$ for $N = 128$.

In the left panel of Fig.~\ref{fig:Q2nMag16and64_allN_vs_e2} we present $\Gamma_{\rm diff} \pm \Delta\Gamma_{\rm diff}$ as a function of $e^2$, using all our data sample for $n_{\rm mag}= 16$ (i.e.~considering all our volume and coupling constants pool). The solid line represents the average fit Eq.~(\ref{eq:GammaDiffnMag16}) from the best fits Eqs.~(\ref{eq:GammasNmag16}) to each choice of $N$. In the right panel of Fig.~\ref{fig:Q2nMag16and64_allN_vs_e2} we present the analogous plot for our data sample for $n_{\rm mag}= 64$. The solid line represents the average fit Eq.~(\ref{eq:GammaDiffnMag64}) from the best fits Eqs.~(\ref{eq:GammasNmag64}) to each choice of $N$. The shade area between the dashed lines in both left and right panels, represents the dispersion among the best fits for each $n_{\rm mag}$ case. We conclude from these fits that the theoretical expectation ${\Gamma_{\rm diff}\over B^2} \propto (e^2)^3$ is still well verified within the associated errors for both $n_{\rm mag} = 16$ and $n_{\rm mag} = 64$. The best fit of the data for $n_{\rm mag} = 16$ prefers a scaling as ${\Gamma_{\rm diff}\over B^2} \propto (e^2)^{2.96}$, whereas for $n_{\rm mag} = 64$ it prefers ${\Gamma_{\rm diff}\over B^2} \propto (e^2)^{2.95}$, see Eqs.~(\ref{eq:GammaDiffnMag16})-(\ref{eq:GammaDiffnMag64}). In both cases, the uncertainty in the exponents encompasses well, nonetheless, the theoretical scaling.

\begin{figure}[t]
\begin{center}
\includegraphics[width=12.0cm]{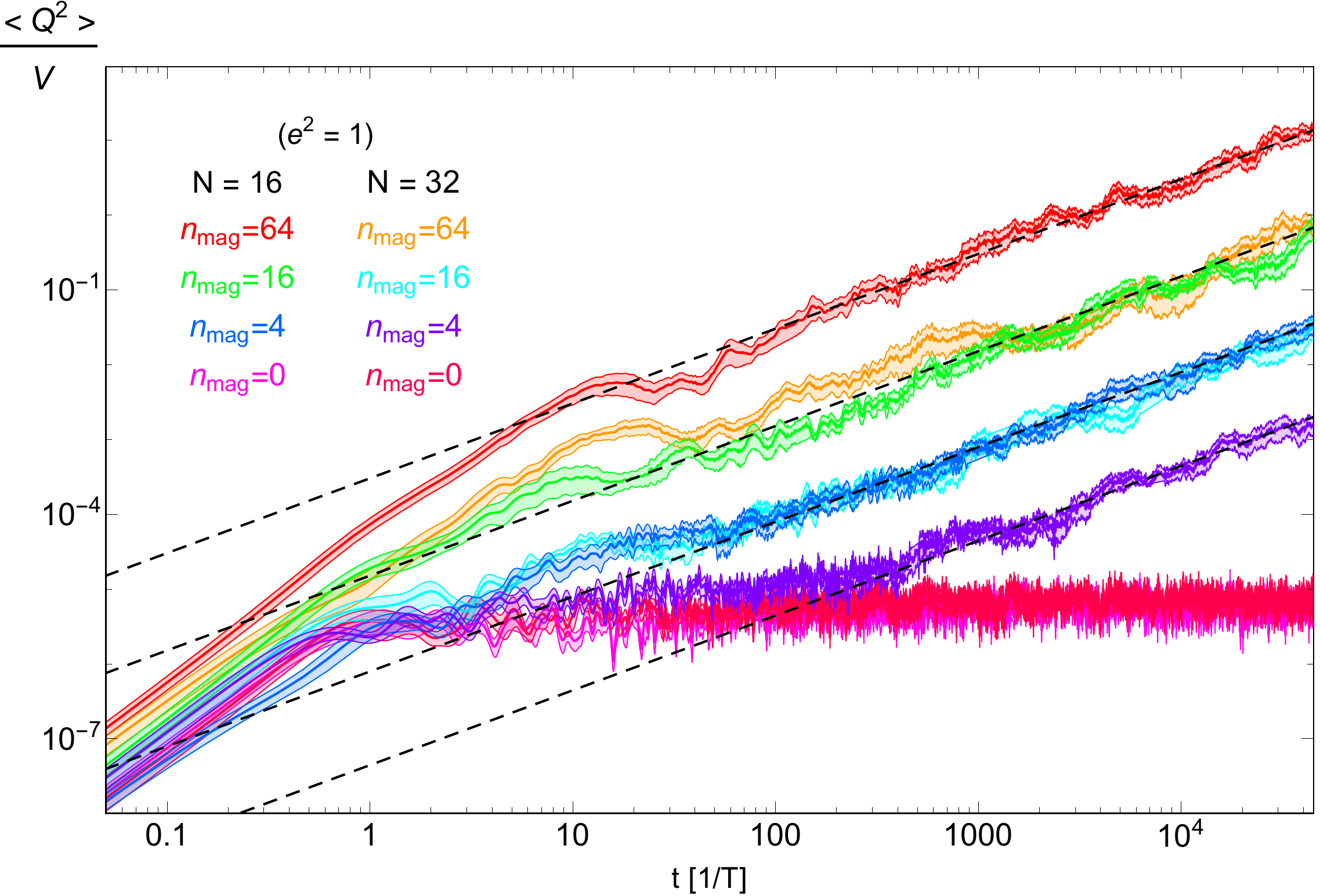}
\end{center}
\caption{Correlator $\langle Q^2 \rangle/V$ for a fixed gauge coupling $e^2 = 1$, sampling different volumes  with $N = 16$ and $32$, and magnetic fields with seed $n_{\rm mag} = 4, 16$ and $64$. For comparison we also show $\langle Q^2 \rangle/V$ for $e^2 = 1$ and $n_{\rm mag} = 0$, for two volumes $N = 16, 32$. The straight lines are just meant to guide the eye (they are not fits), representing a linear growth $\propto t$ with different slopes for each case.}
\label{fig:Allmag1panel}
\end{figure}

Let us turn our attention now to the dependence of the correlator $\langle Q^2 \rangle/V$ on the strength of the magnetic field. In Fig.~\ref{fig:Allmag1panel} we show $\langle Q^2 \rangle/V$ for a fixed gauge coupling $e^2 = 1$, different volumes ($N = 16, 32$) and magnetic seeds ($n_{\rm mag} = 4,16,64$). If the scaling behavior $\langle Q^2 \rangle \propto B^2$ is basically correct, we expect some of the signals to overlap, as it is clearly appreciated in the figure. This is because the magnetic field scales as $B \propto n_{\rm mag}/N^2$, and hence, theoretically, the correlator scales as $\langle Q^2 \rangle/V \propto B^2 \propto n_{\rm mag}^2/N^4$. Thus, the signal for a given pair of parameter values $(n_{\rm mag},N)$ = $(n_1,N_1)$, should be, in principle, of the same size as the signal for other parameter values $(n_2,N_2)$, for as long as $n_2/n_1 = (N_2/N_1)^2$. Our set of parameter values $\lbrace e^2,N \rbrace$ allow precisely for these type of combinations, for instance when comparing the signal obtained for $(n_{\rm mag},N)$ = $(16,16)$ [green signal in the figure] versus that for $(n_{\rm mag},N)$ = $(64,32)$ [yellow signal], or the signal for $(n_{\rm mag},N)$ = $(4,16)$ [dark blue signal] versus that for $(n_{\rm mag},N)$ = $(16,32)$ [light blue signal]. Within the statistical error these two pairs clearly overlap, once the diffusion regimen is well onset. As a guideline for the eye we also include straight lines depicting a linear growth, demonstrating that a linear growth in time remains a robust feature even in the presence of a strong magnetic field. For comparison, we also show the signal in the absence of magnetic field ($n_{\rm mag} = 0$), for $e^2 = 1$ and volumes $N = 16, 32$. In this case the growth of $\langle Q^2 \rangle / V$ ceases very soon, reaching a given amplitude fixed by the strength of $e^2 = 1$, independently of $N$. It is instructive to note the behavior of the correlator $\langle Q^2 \rangle / V$ for the parameters $(n_{\rm mag},N)$ = $(4,32)$, which corresponds to the weakest magnetic field case in Fig.~\ref{fig:Allmag1panel} [purple signal]. As the magnetic field is relatively weak in this case, the correlator saturates initially to a similar amplitude to the signal with no external magnetic field, before it finally starts the diffusive regime after a time $\Delta t \sim 300~T^{-1}$.

\begin{figure}[t]
\begin{center}
\includegraphics[width=7.5cm]{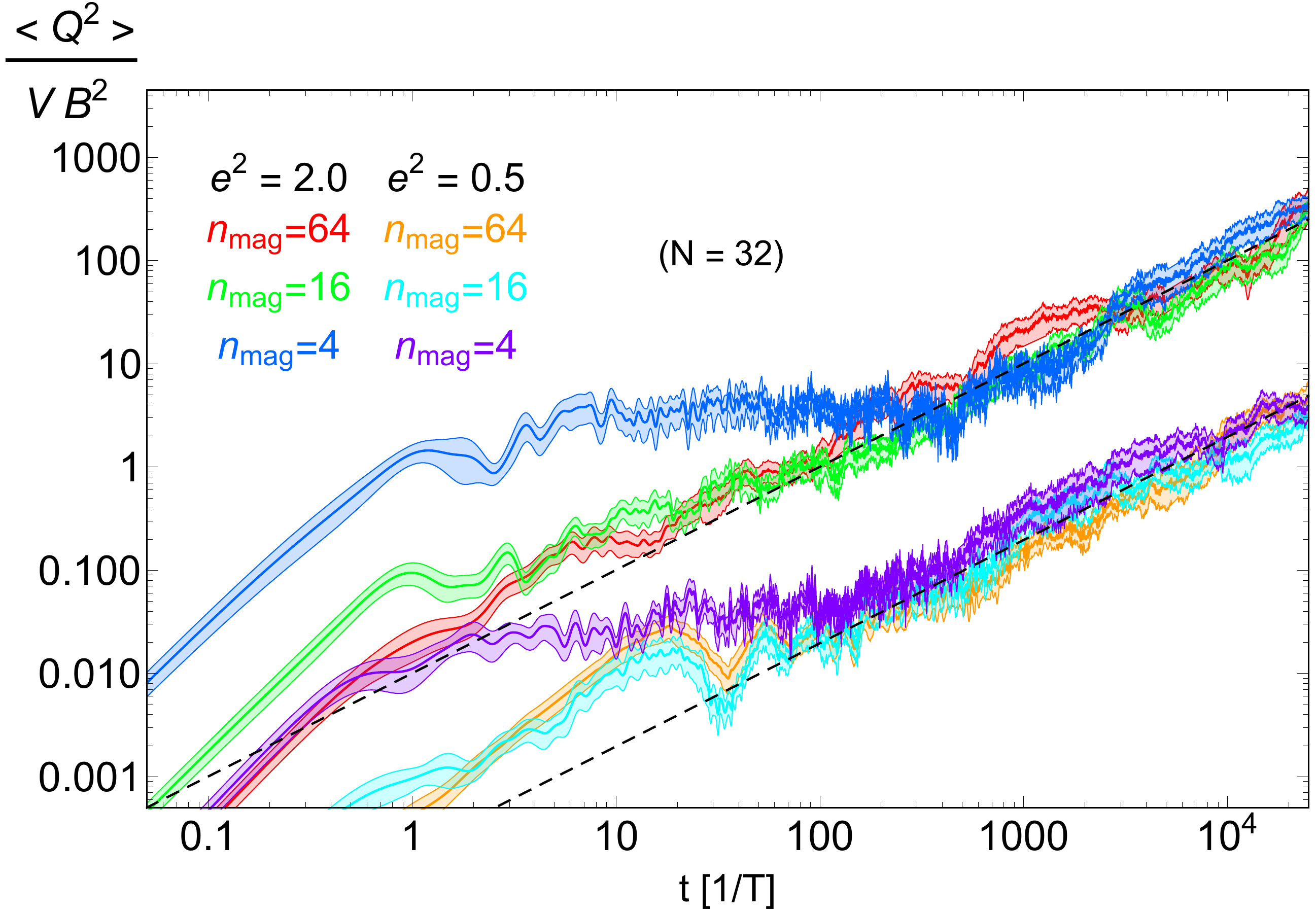}
\includegraphics[width=7.5cm]{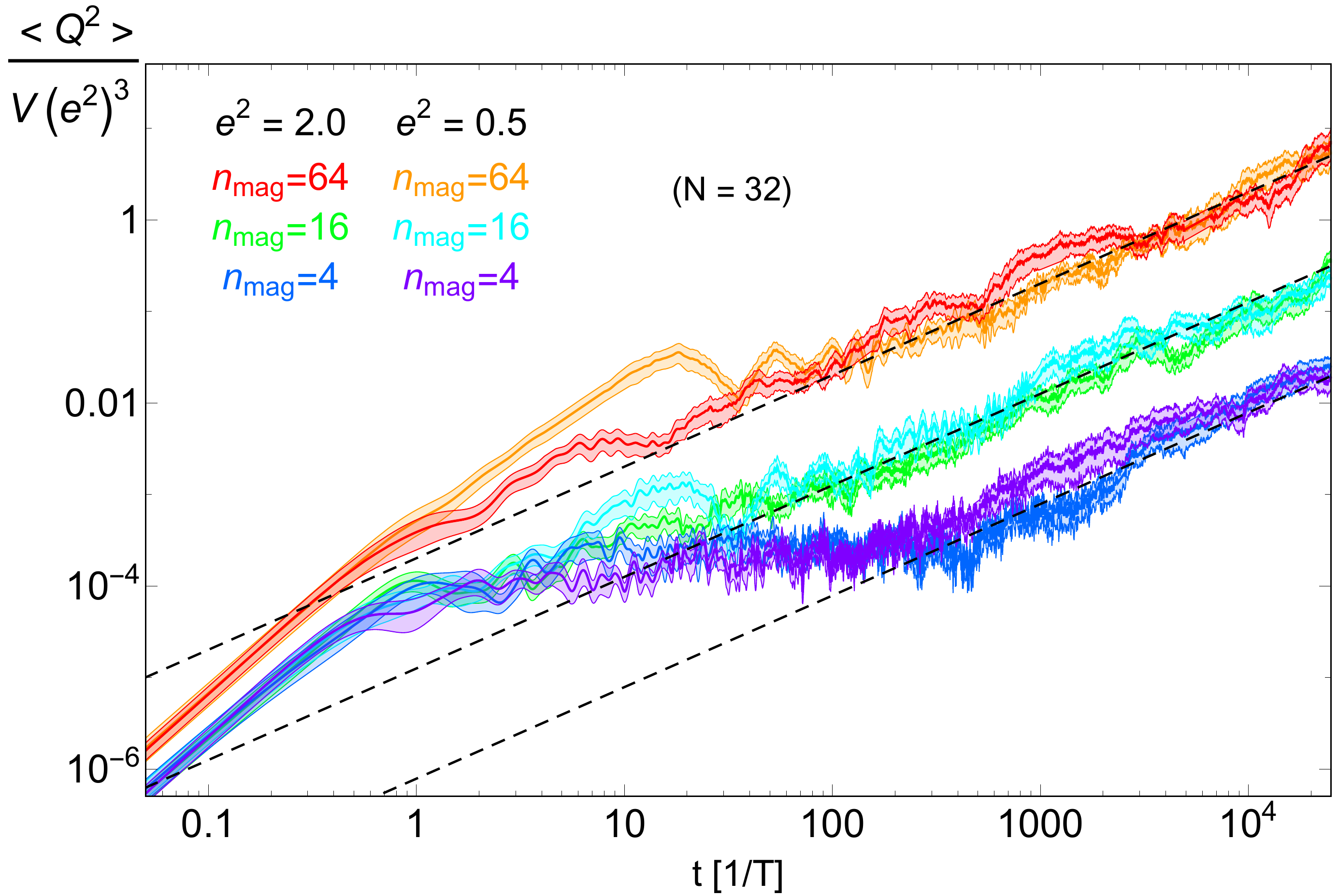}
\includegraphics[width=7.5cm]{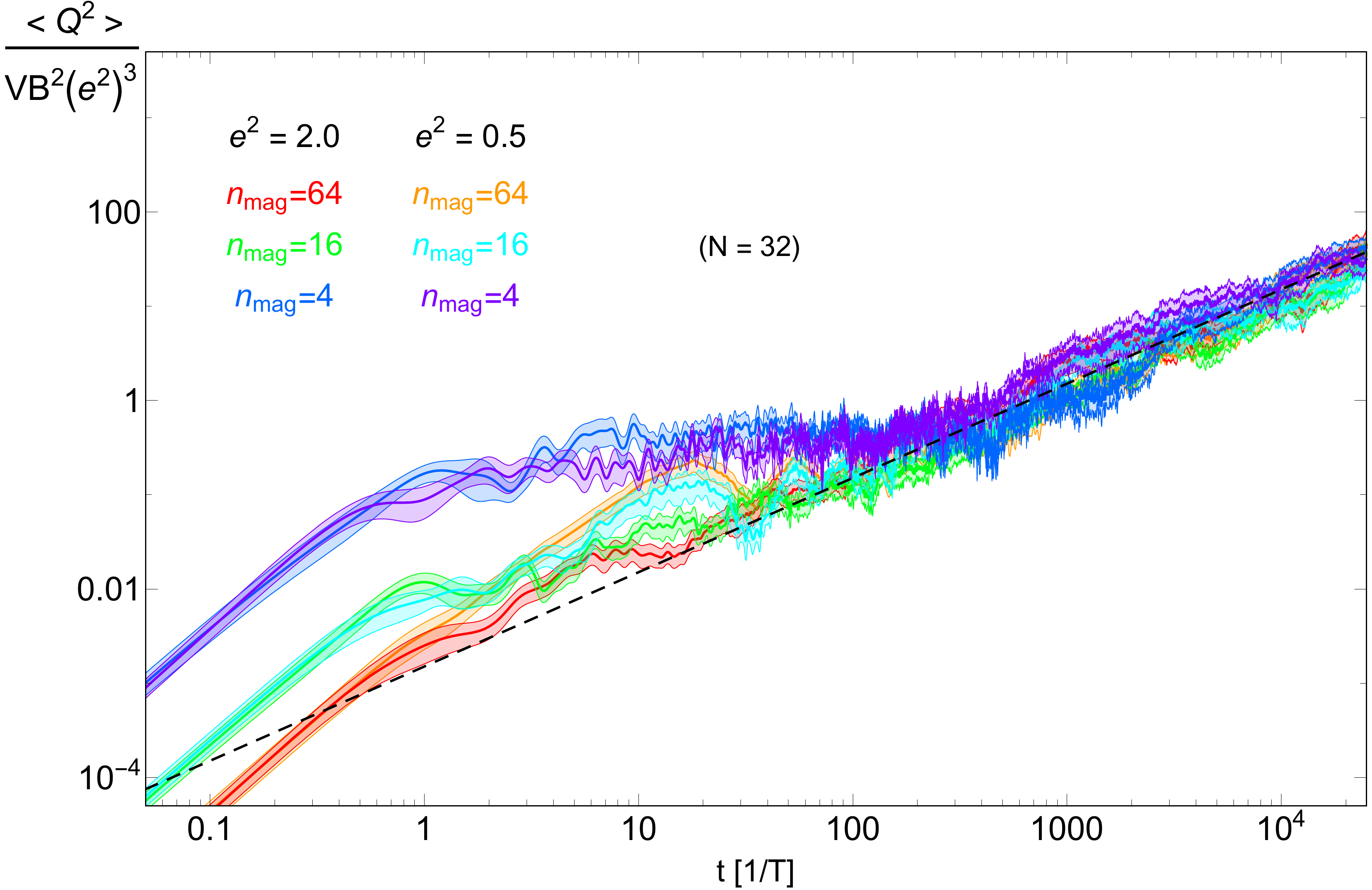}
\includegraphics[width=7.5cm]{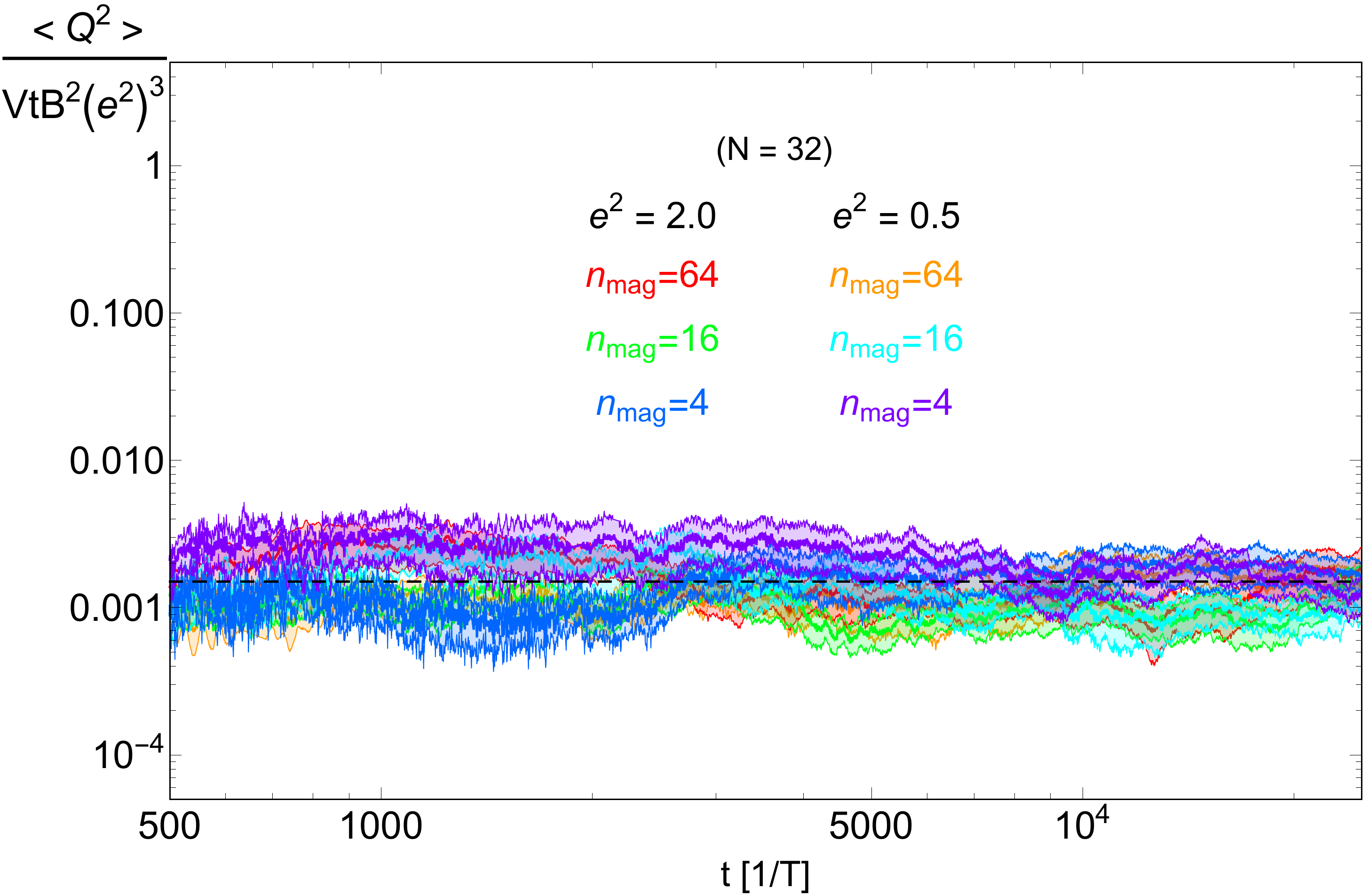}
\end{center}
\caption{Correlator $\langle Q^2 \rangle$ for various magnetic seeds $n_{\rm mag} = 4, 16$ and $64$ and gauge coupling values $e^2 = 0.5$ and $2$, for a fixed lattice size $N = 32$ ($Tdx = 1$). {\it Top Left}: ${\langle Q^2 \rangle\over VB^2}$. {\it Top Right}: ${\langle Q^2 \rangle\over V(e^2)^2}$. {\it Bottom Left}: ${\langle Q^2 \rangle\over V(Be^2)^2}$. {\it Bottom Right}: ${\langle Q^2 \rangle\over Vt(Be^2)^2}$.}
\label{fig:Allmag4panels}
\end{figure}

In Fig.~\ref{fig:Allmag4panels} we present a series of panels where we plot the correlator $\langle Q^2 \rangle$ for various magnetic seeds ($n_{\rm mag} = 4,16,64$) and gauge coupling values ($e^2 = 0.5, 2$), for a fixed lattice size $N = 32$ ($Tdx = 1$). The color coding is actually shared by the four panels of the figure, as the plots of each panel are based in the same data. This way we can compare easily the effect of the different normalization of the correlators by simply comparing the signals of the same color from panel to panel. In the top left panel we show ${\langle Q^2 \rangle\over VB^2}$. If the scaling $\langle Q^2 \rangle \propto B^2$ is roughly correct, the six signals considered should be split in two groups, a higher amplitude corresponding to the highest gauge coupling $e^2 = 2$, and a lower amplitude corresponding to $e^2 = 0.5$. The top left panel clearly shows such two-group splitting pattern, indicating that a scaling $\propto B^2$ is roughly correct [we will quantify this shortly]. In the top right panel of we plot ${\langle Q^2 \rangle\over V(e^2)^3}$. Based on the scaling ${\langle Q^2 \rangle}/B^2 \propto (e^2)^2$, we expect the signals to split in three groups, depending on the strength of $n_{\rm mag}$ (given that $N$ is common to all of them). For each magnetic seed $n_{\rm mag} = 4, 16$ and $64$ we expect to see a pair of overlapping correlators. Such three-group pattern is clearly observed in the top right panel. In the bottom left panel we plot ${\langle Q^2 \rangle\over VB^2(e^2)^3}$. Based on the scaling ${\langle Q^2 \rangle} \propto B^2(e^2)^3$, we expect the whole set of signals to overlap into a single growing pattern. This is precisely what we observe in this panel, where all signals overlap within a scatter of an order of magnitude, compatible with the statistical errors. For further clarity, we plot again these set of overlapping signals in the bottom right panel, but in this occasion normalizing by a linear time growth as well, ${\langle Q^2 \rangle\over VtB^2(e^2)^3}$. If the scaling ${\langle Q^2 \rangle} \propto B^2(e^2)^3t$ is correct, then all the signals should not only overlap but also remain constant in time. Within the statistical error, this is precisely the pattern we observe in the bottom right panel, where all signals oscillate and scatter around a common value $\sim 10^{-3}$ (it is actually appreciated in the figure that the correlator amplitudes for $n_{\rm mag} = 4$ only turn into a flat $pleteau$ only at late times, of the order of $Tt \sim few \times 10^3$).

Finally, let us recall again that all fits as a function of $e^2$ obtained so far, have been derived assuming always an exact magnetic field scaling as $\langle Q^2 \rangle \propto B^2$. Even though our plots in Fig.~\ref{fig:Allmag4panels} clearly show that, if not exact, such scaling is roughly correct, we will test now this simple scaling behavior against our data. In order to do so, we divide our data sample in subsets so that for a given gauge coupling $e^2$, we collect the amplitudes $\Gamma_{\rm diff}$ as a function of volume $N$ and magnetic seed $n_{\rm mag}$. Given a pair of values $(e^2,N)$, we can then fit $\Gamma_{\rm diff}$ as a function of $n_{\rm mag}$, theoretically expecting that $\langle Q^2 \rangle \propto n_{\rm mag}^2$. For each coupling $e^2$ we have found a best fit for each volume $N$ considered, like
\begin{eqnarray}\label{eq:GammasE22}
e^2 = 2~~\rightarrow~~\left\lbrace\begin{array}{l}
\Gamma_{\rm diff}^{(16)} \simeq 1.14\cdot 10^{-2}\cdot B^{2.18}\,,~~N = 16\\
\Gamma_{\rm diff}^{(32)} \simeq 7.50\cdot 10^{-3}\cdot B^{1.86}\,,~~N = 32\\
\Gamma_{\rm diff}^{(64)} \simeq 9.67\cdot 10^{-3}\cdot B^{1.98}\,,~~N = 64\\
\Gamma_{\rm diff}^{(128)} \simeq 1.17\cdot 10^{-2}\cdot B^{2.34}\,,~~N = 128\\
\end{array}\right.
\end{eqnarray}
\begin{eqnarray}\label{eq:GammasE21}
e^2 = 1~~\rightarrow~~\left\lbrace\begin{array}{l}
\Gamma_{\rm diff}^{(16)} \simeq 1.56\cdot 10^{-3}\cdot B^{2.13}\,,~~N = 16\\
\Gamma_{\rm diff}^{(32)} \simeq 1.46\cdot 10^{-3}\cdot B^{2.07}\,,~~N = 32\\
\Gamma_{\rm diff}^{(64)} \simeq 3.31\cdot 10^{-3}\cdot B^{2.21}\,,~~N = 64\\
\Gamma_{\rm diff}^{(128)} \simeq 2.68\cdot 10^{-3}\cdot B^{2.10}\,,~~N = 128\\
\end{array}\right.
\end{eqnarray}
\begin{eqnarray}\label{eq:GammasE205}
e^2 = 0.5~~\rightarrow~~\left\lbrace\begin{array}{l}
\Gamma_{\rm diff}^{(32)} \simeq 1.55\cdot 10^{-4}\cdot B^{1.97}\,,~~N = 32\\
\Gamma_{\rm diff}^{(64)} \simeq 1.93\cdot 10^{-4}\cdot B^{2.01}\,,~~N = 64\\
\Gamma_{\rm diff}^{(128)} \simeq 1.28\cdot 10^{-3}\cdot B^{2.47}\,,~~N = 128\\
\end{array}\right.
\end{eqnarray} 
\begin{eqnarray}\label{eq:GammasE2025}
e^2 = 0.25~~\rightarrow~~\left\lbrace\begin{array}{l}
\Gamma_{\rm diff}^{(64)} \simeq 2.16\cdot 10^{-5}\cdot B^{1.89}\,,~~N = 64\\
\Gamma_{\rm diff}^{(128)} \simeq 8.70\cdot 10^{-5}\cdot B^{2.23}\,,~~N = 128\\
\end{array}\right.
\end{eqnarray} 
Averaging over the best fits for each gauge coupling, we obtain
\begin{eqnarray}\label{eq:GammasBestFitE22}
\Gamma_{\rm diff} &\simeq& (1.15 \pm 0.35)\cdot 10^{-2}\cdot B^{2.07 \pm 0.18}\,,~~e^2 = 2.0\\
\label{eq:GammasBestFitE21}
\Gamma_{\rm diff} &\simeq& (2.22 \pm 0.80)\cdot 10^{-3}\cdot B^{2.13 \pm 0.05}\,,~~e^2 = 1.0\\
\label{eq:GammasBestFitE205}
\Gamma_{\rm diff} &\simeq& (1.74 \pm 0.19)\cdot 10^{-4}\cdot B^{1.99 \pm 0.02}\,,~~e^2 = 0.5\\
\label{eq:GammasBestFitE2025}
\Gamma_{\rm diff} &\simeq& (4.78 \pm 3.20)\cdot 10^{-5}\cdot B^{2.03 \pm 0.17}\,,~~e^2 = 0.25\,,
\end{eqnarray}
where the prefactor for $e^2 = 0.25$ exhibits a very large dispersion, as only two points from our data could be used, corresponding to correlators amplitudes obtained for $(e^2 = 0.25, N = 64)$ and $(e^2 = 0.25, N = 128)$. In general there is some dispersion with respect an exact $\propto B^2$ scaling, but in the cases $e^2 = 2.0$, $e^2 = 0.5$, $e^2 = 0.25$, the deviation is either not significant or a scaling $\propto B^2$ is well encompassed within the associated error. For $e^2 = 1.0$ there is however a systematic deviation of $few~\%$ with respect an exact quadratic scaling.


Let us note that whereas the fits Eqs.~(\ref{eq:GammaDiffAllnMag})-(\ref{eq:GammaDiffnMag64}) of $\Gamma_{\rm diff}$ vs $e^2$ are obtained for a fixed magnetic seed $n_{\rm mag}$, the new fits Eqs.~(\ref{eq:GammasBestFitE22})-(\ref{eq:GammasBestFitE2025}) of $\Gamma_{\rm diff}$ vs $B$ are obtained for a fixed gauge coupling $e^2$. A new approach we may consider is then a multi-dimensional fit to $\Gamma_{\rm diff}$ as function of both $e^2$ and $B$. Theoretically $\Gamma_{\rm diff} \propto B^2(e^2)^3$, so we can consider our data sample of $\Gamma_{\rm diff}$ values obtained for each set of parameter values $(e^2,N,n_{\rm mag})$, and attempt a multi-dimensional fit of the whole array of data to a functional form $\Gamma_{\rm diff} = A\cdot(e^2)^{m_e}(2\pi n_{\rm mag})^{m_B}N^{-2m_B+3m_N}$, where $m_e$ will characterize deviations from $\propto (e^2)^3$, $m_B$ from $\propto B^2$, and $m_N$ from $\langle Q^2 \rangle \propto V = N^3$. Using our full set of data obtained for $N = 16, 32, 64, 128$, $e^2 = 0.125, 0.25, 0.50, 1.0, 2.0$ and $n_{\rm mag} = 4, 16, 64$, and expressing the fit in terms of the physical magnetic field, we obtain\footnote{We have used {\tt Mathetica} 10.3 {\it LinearModelFit} routine over the $log$ of our data, i.e.~a fit of $\log_{10}(\langle Q^2\rangle/V)$ against a linear combination of $\log_{10}e^2, \log_{10}n_{\rm mag}, log_{10}N$.}
\begin{eqnarray}
\label{eq:GammaBestFitMultiple}
\Gamma_{\rm diff} \simeq 10^{-3.09 \pm 0.48}\cdot (e^2)^{2.86 \pm 0.12}\cdot B^{2.06 \pm 0.08}\cdot V^{0.04 \pm 0.13}\,,
\end{eqnarray} 
with the errors representing $95\%$ confidence level intervals. Our multi-dimensional fit indicates that there is no residual volume dependence, so that $\langle Q^2\rangle \propto V$ is pretty much exact. The scaling $\langle Q^2\rangle \propto B_{\rm cont}^2$ is totally compatible with our data set, within the statistical error, whereas a scaling $\langle Q^2\rangle \propto \alpha^3$ is only marginally compatible, as the multi-dimensional fit prefers an exponent slightly smaller.

To finalize, let us confront our numerical results from this section, with the analytical results presented before in Sect.~\ref{sec:diffCS}. The theoretical prediction for the diffusion rate is given by Eq.~(\ref{ralp3}), which can be written like
\be\label{eq:THgamma}
\Gamma_{\rm eff}^{\rm (th)} = {\alpha^2 B^2\over {\pi^2 (\sigma/T)}} \simeq 2.5 \cdot 10^{-5} \ (e^2)^3 B^2\,. 
\ee
As the theoretical estimation scales exactly as $\propto (e^2)^3$ and $\propto B^2$, in order to make a fair comparison with our numerical outcome, we present below the numerical diffusion rates obtained when assuming an exact scaling of the data as $\propto (e^2)^3B^2$,
\begin{eqnarray}\label{eq:GammaDiffAllnMagII}
\Gamma_{\rm diff} &\simeq& (1.51 \pm 0.39)\cdot 10^{-3}\cdot (e^2)^3\cdot B^2\,,~~~~~ (n_{\rm mag} = 4)\,,\\
\Gamma_{\rm diff} &\simeq& (1.36\pm 0.38)\cdot 10^{-3}\cdot (e^2)^3\cdot B^2\,,~~~~~ (n_{\rm mag} = 16)\,,
\label{eq:GammaDiffnMag16II}\\
\Gamma_{\rm diff} &\simeq& (1.69 \pm 0.32)\cdot 10^{-3}\cdot (e^2)^3\cdot B^2\,,~~~~~ (n_{\rm mag} = 64)\,.
\label{eq:GammaDiffnMag64II}
\end{eqnarray}
Here the mean values and errors have been obtained by identical procedure as that used in the fits Eqs.~(\ref{eq:GammaDiffAllnMag})-(\ref{eq:GammaDiffnMag64}), and Eqs.~(\ref{eq:GammasBestFitE22})-(\ref{eq:GammasBestFitE2025}). Whereas in Eqs.~(\ref{eq:GammaDiffAllnMag})-(\ref{eq:GammaDiffnMag64}) an exact scaling $\propto B^2$ was assumed (hence fitting the behavior vs $e^2$), and in Eqs.~(\ref{eq:GammasBestFitE22})-(\ref{eq:GammasBestFitE2025}) an exact scaling $\propto (e^2)^3$ was assumed instead (thus fitting the behavior vs $B$), now in Eqs.~(\ref{eq:GammaDiffAllnMagII})-(\ref{eq:GammaDiffnMag64II}) we have assumed both the scalings $\propto B^2$ and $\propto (e^2)^3$ simultaneously  (hence fitting only the dimensionless pre-factor). In order to make the comparison with Eq.~(\ref{eq:THgamma}), we then average over Eqs.~(\ref{eq:GammaDiffAllnMagII})-(\ref{eq:GammaDiffnMag64II}), and obtain
\begin{eqnarray}\label{eq:NUMgamma}
\Gamma_{\rm diff}^{\rm (num)} &\simeq& (1.54 \pm 0.21)\cdot 10^{-3}\cdot (e^2)^3\cdot B^2\,.
\end{eqnarray}
Comparing the prefactor from Eq.~(\ref{eq:NUMgamma}) with that in Eq.~(\ref{eq:THgamma}), we see that we obtain numerically a rate $\Gamma_{\rm diff}^{\rm (num)}/\Gamma_{\rm diff}^{\rm (th)} \simeq (58 \pm 9)$ times larger than the theoretical expectation\footnote{Of course, if we consider the multidimensional-fit Eq.~(\ref{eq:GammaBestFitMultiple}), the numerical rate can be considered to be closer to the theoretical, given the great dispersion obtained in the multidimensional fitted amplitude. This large dispersion is however only a consequence of having fitted the data against multiple parameters, and thus we consider more fair to make a comparison between theory and numerics by using Eq.~(\ref{eq:NUMgamma}), extracted from the data once we assume (for the shake of the comparison) a scaling $\Gamma_{\rm diff} \propto \alpha^3B_{\rm cont}^2$ as exact.}. This is of course one of the most relevant results of this paper.

\subsection{Conductivity}
\label{sec:conductivity}

The substantial numerical difference between the MHD result (\ref{eq:THgamma}) and (\ref{eq:NUMgamma}) calls for an additional analysis of this discrepancy, at least at the qualitative level. The only essential dynamical parameter that enters the Eq.~(\ref{eq:THgamma}) is the electric conductivity of the plasma, which was found for long ranged electric fields, with the scale exceeding the mean free path of particles in the plasma  $l \gg 1/(\alpha^2 T)$, see Eq.~(\ref{cond}). It is known \cite{Arnold:1997yb,Arnold:1998cy} that a similar quantity, the colour conductivity, plays an important role in non-Abelian sphaleron rate. So, it is natural to ask what is the reaction of our system to short ranged electric fields, with  $l \ll 1/(\alpha^2 T)$, potentially relevant for CS number diffusion.

Probably, the simplest way to address this question is to use the Kubo formula~\cite{Kubo:1957mj} for conductivity  \footnote{We thank Guy Moore for suggesting us to use it in our numerical analysis.}, which can be written, for homogeneous and isotropic plasma, as 
\be
\sigma = \int_0^\infty dt \int d^3 x\int_0^\beta d \lambda Tr\left[\rho j_i(x,t+ i \lambda) j_i(0,0)\right]
\label{sigmaq}
\ee
irrespectively of whether we take $i = 1, 2$ or $3$. Here $\rho$ is the equilibrium density matrix of the system, $j_i$ are the spatial components of the electric current (no summation over $i$ is assumed), $\beta=1/T$, and integration over $\lambda$ appears due to non-commutativity of different operators in the quantum case. For the classical field theory on the lattice with finite volume $V$, the formula can be written as 
\be
\sigma = \frac{1}{V T}  \left\langle\int_0^\infty dt \int d^3 x j_i(x,t)  \int d^3 y j_i(y,0)\right\rangle~,
\label{sigmacl}
\ee
where $ \langle ...\rangle$ is the ensemble average. It can be used as a basis for lattice numerical simulations. 

To get an idea of expected behavior, we have first computed the correlator (\ref{sigmacl}) in classical approximation for scalar free field theory. For this end the scalar field can be expressed as usual via creation ($a(k)$) and annihilation ($a^\dagger(k)$) operators, with the ensemble averages 
\be
 \langle a^\dagger(k)  a(k') \rangle = (2\pi)^3 2 k^0 \delta(k-k') n_B(k)~,
\label{aver}
\ee
where the Bose distribution function is taken to be $n_B(k)= T/k^0$, specific for classical statistics, and $k^0=\sqrt{k^2+m^2}$. Straightforward calculation of the integrand in (\ref{sigmacl}) gives
\be
\Sigma(t) = \frac{1}{V T}  \int d^3 x \int d^3 y \left\langle  j_i(x,t) j_i(y,0)\right\rangle = \frac{2 e^2 T}{3\pi^2}\int \frac{k^4 dk}{(k^2+m^2)^2} \cos^2(k^0 t)~.
\label{integrand}
\ee
This expression is ultra-violet sensitive, linearly divergent with momentum, so the lattice regularization provides a cutoff $k_{\max} \sim 1/dx$, where $dx$ is the lattice spacing. 

In the free field theory there is no damping, and the conductivity must be infinite, what is indeed the case since the time integral of (\ref{integrand}) is divergent. If interactions are included, we expect that the correlator in (\ref{integrand}) will be decaying exponentially with time, $\Sigma(t) \propto \exp{(-\gamma t)}$, and oscillating with the plasma frequency. If we take $\gamma \propto e^4 T$ (see discussion in in Sect.~\ref{sec:diffCS}), then parametrically  $\sigma \propto 1/e^2$, whereas the plasma frequency for long-ranged fluctuations for quantum scalar electrodynamics is $\omega = e T/3$ (see, e.g.~\cite{Kraemmer:1994az}).

\begin{figure}[t]
\begin{center}
\includegraphics[width=7.5cm]{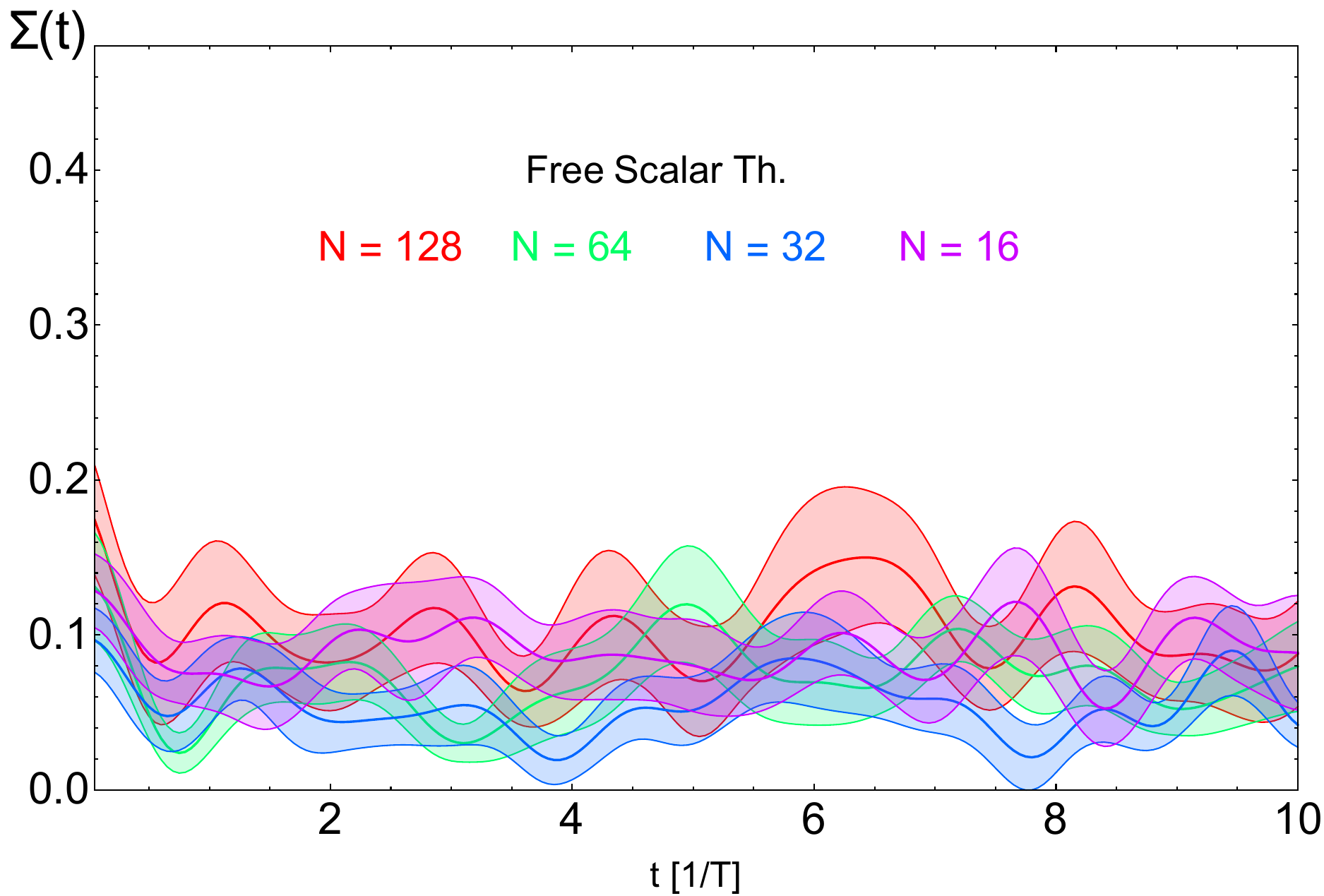}
\includegraphics[width=7.5cm]{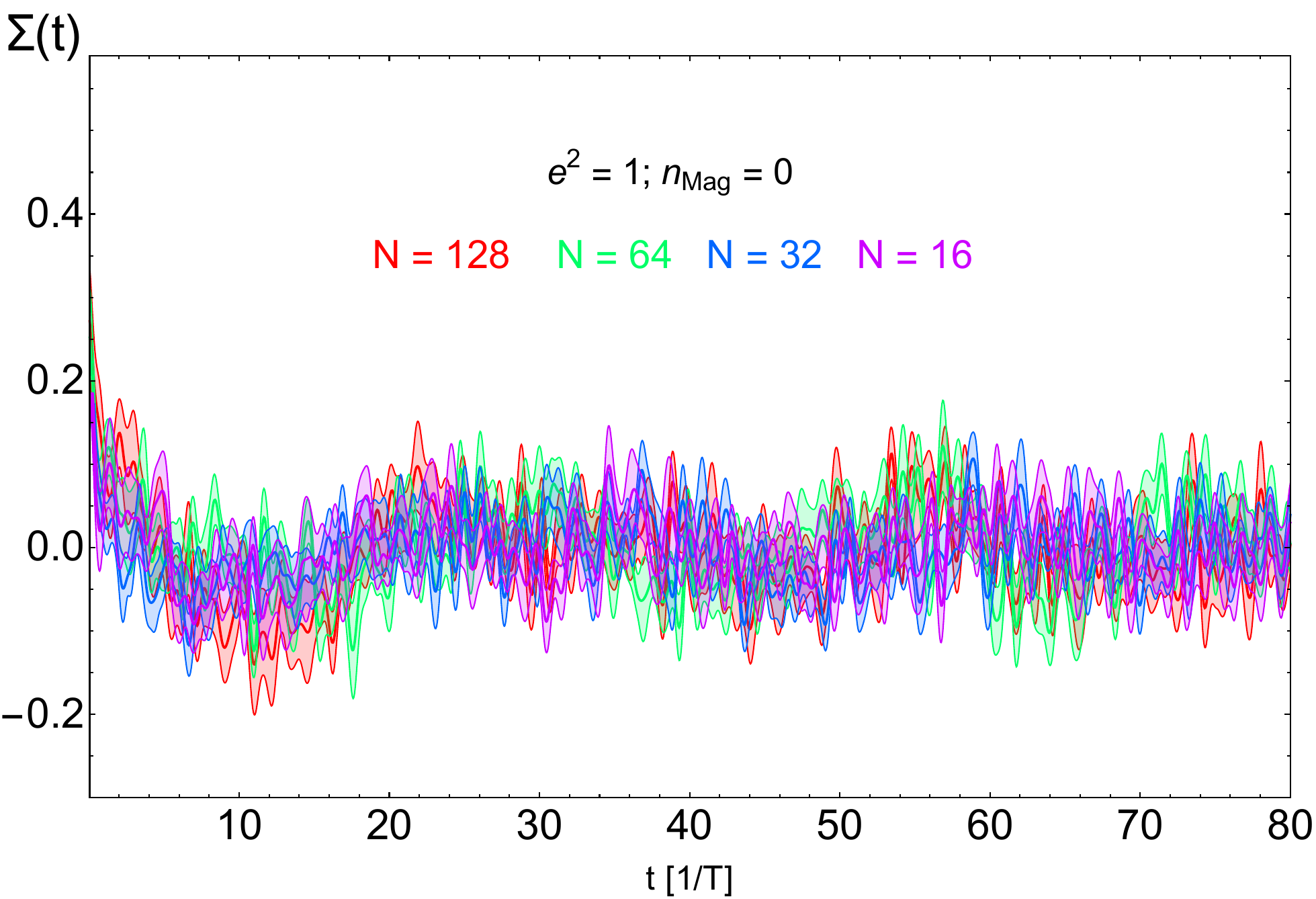}
\end{center}
\caption{We plot the integrand $\Sigma(t)$ of the conductivity $\sigma = \int dt \Sigma(t)$, both for a free scalar theory (left panel) and for an interacting scalar-gauge theory (in the absence of an external magnetic field, so $n_{\rm mag} = 0$) for $e^2 = 1$ and various volumes $N = 16, 32, 64$ and $128$. The bands around each solid line represent the statistical error in the estimation of the correlator, based on $N_R = 28$ realizations in each case.}
\label{fig:Conductivity}
\end{figure}

To test our numerical procedure we have made first simulations of a free scalar field theory and verified that the behavior of the correlator verifies indeed the structure of Eq.~(\ref{integrand}). See left panel of Fig.~\ref{fig:Conductivity}, where we plot $\Sigma(t)$ as obtained for different volumes $N = 16, 32, 64, 128$. As expected the initial amplitude falls to a constant positive value (the same, independently of $N$), and this is modulated by short scale oscillations.

We carried out also the simulations in interacting theory as well, the typical pictures of the integrand are presented in the right panel of Fig.~\ref{fig:Conductivity}. The short time oscillations associated with the UV cutoff $1/a$ are clearly seen. Also, contrary to the free field theory, it looks like there is some oscillating pattern with frequency considerably smaller than $1/T$, which can be attributed to the plasma frequency, and that after some time the oscillations occur around zero, indicating the dumping. We made no attempt to perform a complete quantitative study of the Kubo integral Eq.~(\ref{sigmacl}), which would include the parameter dependence and averaging over sufficiently large number of time histories, leaving it to future work. Still, an estimation ``by eye'' of the time of short time damping for $e^2=1$, gives $1/\gamma\sim 2$, leading to $\sigma\sim 0.5$, which smaller than that given by Eq.~(\ref{cond}) by a factor around $\sim 50$. If this value of conductivity is inserted into Eq.~(\ref{ralp3}), the result roughly agrees with the diffusion rate of the CS number we observe in our simulations, making the picture self-consistent. Thus, the high values of the diffusion rate of the CS number in the presence of magnetic fields can be explained by the small value of the conductivity at small time scales in the non-linear bosonic theory under investigation.

We do not know whether the same conclusion is true in real QED with fermions, because rather than physical quantum electrons, we have instead lattice classical scalar fields as the charged particles. We expect, though, that our results are applicable at their face value to the hypercharge U(1) field of the electroweak theory above the point of the electroweak crossover, as the theory does contain the relatively light scalar fields -- Higgses, which are charged under this U(1). We also also expect the enhancement of the CS diffusion rate in the electromagnetic sector right below the electroweak cross-over, due to the presence of electrically charged W-bosons, introducing similar non-linearities as the scalar field.

\section{Conclusions}
\label{sec:conclus}

In this work we have studied the non-conservation of fermion/chiral number in Abelian gauge theories at finite temperatures. We have considered different mechanisms of fermionic charge disappearance in the plasma, and analyzed these with both analytical and numerical techniques. We use a bosonic effective theory where a non-zero fermionic charge density is characterized by a chemical potential $\mu$, so that we add in the Hamiltonian a term $H_{\rm CS} \propto \mu Q$, with $Q \propto \int d^4x F_{\mu\nu}{\tilde F}^{\mu\nu}$ the Chern-Simons (CS) number. 

In sections~\ref{sec:Intro} and \ref{sec:analytics} we have provided a number of analytical results about the dynamical behavior of this system, both when $\mu = 0$ and $\mu \neq 0$, either in the presence or absence of an external magnetic field. In Section~\ref{sec:model} we have discussed the details of our modeling, including its lattice formulation. We then dedicate Section~\ref{sec:diff} to our numerical analysis, based on the outcome from the real time lattice simulations for $\mu = 0$. We plan to study numerically the case with $\mu \neq 0$ in a forthcoming publication.

In section~\ref{sec:diff} we have considered, in particular, the diffusion of the CS number in a system with an external magnetic field present, but in the absence of fermions, hence setting $\mu = 0$ as a fixed value at all times. In the case with no external magnetic field  $B$ the correlator $\left\langle Q^2 \right\rangle$ is a  constant, as expected. If $B\neq 0$ this correlator grows continuously and linearly in time. Based on the numerical outcome from the lattice simulations, we have obtained fits to the appropriate coefficients characterizing the behavior of $\left\langle Q^2 \right\rangle$ for different parameters of the theory and lattice volumes. We then compared these fits to the analytical estimates presented in sections~\ref{sec:Intro}, \ref{sec:analytics}. 

The diffusive behavior of $\left\langle Q^2 \right\rangle$ in the presence of an external magnetic field $B$, indicates that the mechanism for fermionic number non-conservation for $B \neq 0$ is due to fluctuations of the gauge fields, similarly as in the case of non-Abelian gauge theories. Our numerical determination of such diffusion rate when $B \neq 0$, $\Gamma \simeq \kappa \alpha^3 B^2$, is perhaps the most important result of the paper. We provide multiple fits to $\left\langle Q^2 \right\rangle$ in Sect.~\ref{sec:diff}, based on different assumptions about the scaling of the correlator with $\alpha$ and $B$. Within the statistical errors, we obtain that the rate scales, parametrically, with the fine-structure constant as $\propto \alpha^3$, and with magnetic field $\propto B^2$. We thus find that it scales with parameters as it was found long time ago, from equations of MHD modified to account for anomaly effects. However, the numerical amplitude of the rate we find is much larger than previously estimated, as large as $\Gamma_{\rm diff}^{\rm (num)}/\Gamma_{\rm diff}^{\rm (th)} \simeq 60$, when compared to the rate derived from MHD. We argue that this is an indication that the small scale fluctuations of the electromagnetic fields with the size $\sim 1/(\alpha T)$, are important and must be accounted for. The higher rate of CS diffusion implies a higher rate of chirality (or fermion number) non-conservation in the presence of a magnetic field. Quite a number of works, e.g.~\cite{Giovannini:1997eg,Kamada:2016eeb,Kamada:2016cnb,Joyce:1997uy,Boyarsky:2011uy} and those who used MHD for the chirality non-conservation rate [some of them written by the one of the authors (MS)], used the MHD approach for the study of combined evolution of the chirality/fermion number and magnetic fields. We believe that the effects found in them may have to be reconsidered in the view of our present findings. 

In this paper, we have also provided a number of arguments about the fact that, even in the absence of a magnetic field, the fluctuations of bosonic fields may provide a mechanism for chirality non-conservation, unrelated to instabilities. This mechanism would lead to the damping of the fermion number with a rate independent on the chemical potential. Unfortunately we were not able to test this hypothesis in our lattice simulations due to the lack of computer resources, but we plan to study this circumstance once we parallelize and vectorize our lattice code.

There is quite a number of points where our analysis can be refined and improved. We  studied only the classical theory without incorporating hard thermal loops and B\"odeker's random force~\cite{Bodeker:1999gx}, the state of art that has been achieved for investigations of the non-Abelian sphaleron rate~\cite{DOnofrio:2014rug}. The analysis of the classical electric conductivity of the plasma carried out in subsection~\ref{sec:conductivity} can be made more quantitative. Yet another important question is related to applicability of our results for CS diffusion rate to high temperature electrodynamics without any scalar fields. While the Abelian Higgs model  serves as a good approximation to the U(1) sector of the SM around the electroweak cross-over, this is not the case for smaller temperatures when all charged bosonic degrees of freedom are decoupled. We leave these aspects as points for future work.

\section*{Acknowledgements}

The authors thank  A. Boyarsky, G. Moore, O. Ruchayskiy and E. Sabancilar for helpful discussions. This work was supported by the ERC-AdG-2015 grant 694896.  The work of  M.S. was supported partially by the Swiss National Science Foundation. 

\bibliographystyle{utphys}
\bibliography{U1pheno}

\providecommand{\href}[2]{#2}\begingroup\raggedright\begin{thebibliography}{10}

\bibitem{Adler:1969gk}
S.~L. Adler, ``{Axial vector vertex in spinor electrodynamics},''
\href{http://dx.doi.org/10.1103/PhysRev.177.2426}{{\em Phys. Rev.} {\bfseries
  177} (1969) 2426--2438}.

\bibitem{Bell:1969ts}
J.~S. Bell and R.~Jackiw, ``{A PCAC puzzle: $\pi^0\to\gamma\gamma$ in the sigma
  model},''
\href{http://dx.doi.org/10.1007/BF02823296}{{\em Nuovo Cim.} {\bfseries A60}
  (1969) 47--61}.

\bibitem{Veneziano:1979ec}
G.~Veneziano, ``{U(1) Without Instantons},''
\href{http://dx.doi.org/10.1016/0550-3213(79)90332-8}{{\em Nucl. Phys.}
  {\bfseries B159} (1979) 213--224}.

\bibitem{Witten:1979vv}
E.~Witten, ``{Current Algebra Theorems for the U(1) Goldstone Boson},''
\href{http://dx.doi.org/10.1016/0550-3213(79)90031-2}{{\em Nucl. Phys.}
  {\bfseries B156} (1979) 269--283}.

\bibitem{tHooft:1976rip}
G.~'t~Hooft, ``{Symmetry Breaking Through Bell-Jackiw Anomalies},''
\href{http://dx.doi.org/10.1103/PhysRevLett.37.8}{{\em Phys. Rev. Lett.}
  {\bfseries 37} (1976) 8--11}.

\bibitem{tHooft:1976snw}
G.~'t~Hooft, ``{Computation of the Quantum Effects Due to a Four-Dimensional
  Pseudoparticle},'' \href{http://dx.doi.org/10.1103/PhysRevD.18.2199.3,
  10.1103/PhysRevD.14.3432}{{\em Phys. Rev.} {\bfseries D14} (1976)
  3432--3450}.
[Erratum: Phys. Rev.D18,2199(1978)].

\bibitem{Klinkhamer:1984di}
F.~R. Klinkhamer and N.~S. Manton, ``{A Saddle Point Solution in the
  Weinberg-Salam Theory},''
\href{http://dx.doi.org/10.1103/PhysRevD.30.2212}{{\em Phys. Rev.} {\bfseries
  D30} (1984) 2212}.

\bibitem{Kuzmin:1985mm}
V.~A. Kuzmin, V.~A. Rubakov, and M.~E. Shaposhnikov, ``{On the Anomalous
  Electroweak Baryon Number Nonconservation in the Early Universe},''
\href{http://dx.doi.org/10.1016/0370-2693(85)91028-7}{{\em Phys. Lett.}
  {\bfseries 155B} (1985) 36}.

\bibitem{McLerran:1990de}
L.~D. McLerran, E.~Mottola, and M.~E. Shaposhnikov, ``{Sphalerons and Axion
  Dynamics in High Temperature {QCD}},''
\href{http://dx.doi.org/10.1103/PhysRevD.43.2027}{{\em Phys. Rev.} {\bfseries
  D43} (1991) 2027--2035}.

\bibitem{Callan:1976je}
C.~G. Callan, Jr., R.~F. Dashen, and D.~J. Gross, ``{The Structure of the Gauge
  Theory Vacuum},''
\href{http://dx.doi.org/10.1016/0370-2693(76)90277-X}{{\em Phys. Lett.}
  {\bfseries 63B} (1976) 334--340}.

\bibitem{Jackiw:1976pf}
R.~Jackiw and C.~Rebbi, ``{Vacuum Periodicity in a Yang-Mills Quantum
  Theory},''
\href{http://dx.doi.org/10.1103/PhysRevLett.37.172}{{\em Phys. Rev. Lett.}
  {\bfseries 37} (1976) 172--175}.

\bibitem{Giovannini:1997eg}
M.~Giovannini and M.~E. Shaposhnikov, ``{Primordial hypermagnetic fields and
  triangle anomaly},'' \href{http://dx.doi.org/10.1103/PhysRevD.57.2186}{{\em
  Phys. Rev.} {\bfseries D57} (1998) 2186--2206},
\href{http://arxiv.org/abs/hep-ph/9710234}{{\ttfamily arXiv:hep-ph/9710234
  [hep-ph]}}.

\bibitem{Kamada:2016eeb}
K.~Kamada and A.~J. Long, ``{Baryogenesis from decaying magnetic helicity},''
  \href{http://dx.doi.org/10.1103/PhysRevD.94.063501}{{\em Phys. Rev.}
  {\bfseries D94} no.~6, (2016) 063501},
\href{http://arxiv.org/abs/1606.08891}{{\ttfamily arXiv:1606.08891
  [astro-ph.CO]}}.

\bibitem{Kamada:2016cnb}
K.~Kamada and A.~J. Long, ``{Evolution of the Baryon Asymmetry through the
  Electroweak Crossover in the Presence of a Helical Magnetic Field},''
  \href{http://dx.doi.org/10.1103/PhysRevD.94.123509}{{\em Phys. Rev.}
  {\bfseries D94} no.~12, (2016) 123509},
\href{http://arxiv.org/abs/1610.03074}{{\ttfamily arXiv:1610.03074 [hep-ph]}}.

\bibitem{Joyce:1997uy}
M.~Joyce and M.~E. Shaposhnikov, ``{Primordial magnetic fields, right-handed
  electrons, and the Abelian anomaly},''
  \href{http://dx.doi.org/10.1103/PhysRevLett.79.1193}{{\em Phys. Rev. Lett.}
  {\bfseries 79} (1997) 1193--1196},
\href{http://arxiv.org/abs/astro-ph/9703005}{{\ttfamily arXiv:astro-ph/9703005
  [astro-ph]}}.

\bibitem{Boyarsky:2011uy}
A.~Boyarsky, J.~Frohlich, and O.~Ruchayskiy, ``{Self-consistent evolution of
  magnetic fields and chiral asymmetry in the early Universe},''
  \href{http://dx.doi.org/10.1103/PhysRevLett.108.031301}{{\em Phys. Rev.
  Lett.} {\bfseries 108} (2012) 031301},
\href{http://arxiv.org/abs/1109.3350}{{\ttfamily arXiv:1109.3350
  [astro-ph.CO]}}.

\bibitem{Fukushima:2008xe}
K.~Fukushima, D.~E. Kharzeev, and H.~J. Warringa, ``{The Chiral Magnetic
  Effect},'' \href{http://dx.doi.org/10.1103/PhysRevD.78.074033}{{\em Phys.
  Rev.} {\bfseries D78} (2008) 074033},
\href{http://arxiv.org/abs/0808.3382}{{\ttfamily arXiv:0808.3382 [hep-ph]}}.

\bibitem{Rubakov:1985nk}
V.~A. Rubakov, ``{On the Electroweak Theory at High Fermion Density},''
\href{http://dx.doi.org/10.1143/PTP.75.366}{{\em Prog. Theor. Phys.} {\bfseries
  75} (1986) 366}.

\bibitem{Redlich:1984md}
A.~N. Redlich and L.~C.~R. Wijewardhana, ``{Induced Chern-simons Terms at High
  Temperatures and Finite Densities},''
\href{http://dx.doi.org/10.1103/PhysRevLett.54.970}{{\em Phys. Rev. Lett.}
  {\bfseries 54} (1985) 970}.

\bibitem{Niemi:1985yp}
A.~J. Niemi and G.~W. Semenoff, ``{Quantum Holonomy and the Chiral Gauge
  Anomaly},'' \href{http://dx.doi.org/10.1103/PhysRevLett.55.927}{{\em Phys.
  Rev. Lett.} {\bfseries 55} (1985) 927}.
[Erratum: Phys. Rev. Lett.55,2627(1985)].

\bibitem{fradkin}
E.~Fradkin, ``{Green’s Function Method in Quantum Field Theory and Quantum
  Statistics},'' {\em Proc. P.N. Lebedev Inst} {\bfseries 29} (1965) 7.

\bibitem{Kalashnikov:1979kq}
O.~K. Kalashnikov and V.~V. Klimov, ``{Infrared Behavior of the Polarization
  Operator in Scalar Electrodynamics at Finite Temperature},''
\href{http://dx.doi.org/10.1016/0370-2693(80)90182-3}{{\em Phys. Lett.}
  {\bfseries 95B} (1980) 423--425}.

\bibitem{Kajantie:1996qd}
K.~Kajantie, M.~Laine, K.~Rummukainen, and M.~E. Shaposhnikov, ``{A
  Nonperturbative analysis of the finite T phase transition in SU(2) x U(1)
  electroweak theory},''
  \href{http://dx.doi.org/10.1016/S0550-3213(97)00164-8}{{\em Nucl. Phys.}
  {\bfseries B493} (1997) 413--438},
\href{http://arxiv.org/abs/hep-lat/9612006}{{\ttfamily arXiv:hep-lat/9612006
  [hep-lat]}}.

\bibitem{Baym:1997gq}
G.~Baym and H.~Heiselberg, ``{The Electrical conductivity in the early
  universe},'' \href{http://dx.doi.org/10.1103/PhysRevD.56.5254}{{\em Phys.
  Rev.} {\bfseries D56} (1997) 5254--5259},
\href{http://arxiv.org/abs/astro-ph/9704214}{{\ttfamily arXiv:astro-ph/9704214
  [astro-ph]}}.

\bibitem{Akamatsu:2013pjd}
Y.~Akamatsu and N.~Yamamoto, ``{Chiral Plasma Instabilities},''
  \href{http://dx.doi.org/10.1103/PhysRevLett.111.052002}{{\em Phys. Rev.
  Lett.} {\bfseries 111} (2013) 052002},
\href{http://arxiv.org/abs/1302.2125}{{\ttfamily arXiv:1302.2125 [nucl-th]}}.

\bibitem{Linde:1980ts}
A.~D. Linde, ``{Infrared Problem in Thermodynamics of the Yang-Mills Gas},''
\href{http://dx.doi.org/10.1016/0370-2693(80)90769-8}{{\em Phys. Lett.}
  {\bfseries 96B} (1980) 289--292}.

\bibitem{Kajantie:1997tt}
K.~Kajantie, M.~Laine, K.~Rummukainen, and M.~E. Shaposhnikov, ``{3-D SU(N) +
  adjoint Higgs theory and finite temperature QCD},''
  \href{http://dx.doi.org/10.1016/S0550-3213(97)00425-2}{{\em Nucl. Phys.}
  {\bfseries B503} (1997) 357--384},
\href{http://arxiv.org/abs/hep-ph/9704416}{{\ttfamily arXiv:hep-ph/9704416
  [hep-ph]}}.

\bibitem{Khlebnikov:1988sr}
S.~{\relax Yu}. Khlebnikov and M.~E. Shaposhnikov, ``{The Statistical Theory of
  Anomalous Fermion Number Nonconservation},''
\href{http://dx.doi.org/10.1016/0550-3213(88)90133-2}{{\em Nucl. Phys.}
  {\bfseries B308} (1988) 885--912}.

\bibitem{Arnold:1987mh}
P.~B. Arnold and L.~D. McLerran, ``{Sphalerons, Small Fluctuations and Baryon
  Number Violation in Electroweak Theory},''
\href{http://dx.doi.org/10.1103/PhysRevD.36.581}{{\em Phys. Rev.} {\bfseries
  D36} (1987) 581}.

\bibitem{Carson:1990jm}
L.~Carson, X.~Li, L.~D. McLerran, and R.-T. Wang, ``{Exact Computation of the
  Small Fluctuation Determinant Around a Sphaleron},''
\href{http://dx.doi.org/10.1103/PhysRevD.42.2127}{{\em Phys. Rev.} {\bfseries
  D42} (1990) 2127--2143}.

\bibitem{Arnold:1996dy}
P.~B. Arnold, D.~Son, and L.~G. Yaffe, ``{The Hot baryon violation rate is O
  (alpha-w**5 T**4)},'' \href{http://dx.doi.org/10.1103/PhysRevD.55.6264}{{\em
  Phys. Rev.} {\bfseries D55} (1997) 6264--6273},
\href{http://arxiv.org/abs/hep-ph/9609481}{{\ttfamily arXiv:hep-ph/9609481
  [hep-ph]}}.

\bibitem{Bodeker:1998hm}
D.~Bodeker, ``{On the effective dynamics of soft nonAbelian gauge fields at
  finite temperature},''
  \href{http://dx.doi.org/10.1016/S0370-2693(98)00279-2}{{\em Phys. Lett.}
  {\bfseries B426} (1998) 351--360},
\href{http://arxiv.org/abs/hep-ph/9801430}{{\ttfamily arXiv:hep-ph/9801430
  [hep-ph]}}.

\bibitem{Arnold:1998cy}
P.~B. Arnold, D.~T. Son, and L.~G. Yaffe, ``{Effective dynamics of hot, soft
  nonAbelian gauge fields. Color conductivity and log(1/alpha) effects},''
  \href{http://dx.doi.org/10.1103/PhysRevD.59.105020}{{\em Phys. Rev.}
  {\bfseries D59} (1999) 105020},
\href{http://arxiv.org/abs/hep-ph/9810216}{{\ttfamily arXiv:hep-ph/9810216
  [hep-ph]}}.

\bibitem{Moore:1998zk}
G.~D. Moore, ``{The Sphaleron rate: Bodeker's leading log},''
  \href{http://dx.doi.org/10.1016/S0550-3213(99)00746-4}{{\em Nucl. Phys.}
  {\bfseries B568} (2000) 367--404},
\href{http://arxiv.org/abs/hep-ph/9810313}{{\ttfamily arXiv:hep-ph/9810313
  [hep-ph]}}.

\bibitem{Grigoriev:1988bd}
D.~{\relax Yu}. Grigoriev and V.~A. Rubakov, ``{Soliton Pair Creation at Finite
  Temperatures. Numerical Study in (1+1)-dimensions},''
\href{http://dx.doi.org/10.1016/0550-3213(88)90466-X}{{\em Nucl. Phys.}
  {\bfseries B299} (1988) 67--78}.

\bibitem{Grigoriev:1989ub}
D.~{\relax Yu}. Grigoriev, V.~A. Rubakov, and M.~E. Shaposhnikov,
  ``{Topological transitions at finite temperatures: a real time numerical
  approach},''
\href{http://dx.doi.org/10.1016/0550-3213(89)90553-1}{{\em Nucl. Phys.}
  {\bfseries B326} (1989) 737--757}.

\bibitem{Ambjorn:1990pu}
J.~Ambjorn, T.~Askgaard, H.~Porter, and M.~E. Shaposhnikov, ``{Sphaleron
  transitions and baryon asymmetry: A Numerical real time analysis},''
\href{http://dx.doi.org/10.1016/0550-3213(91)90341-T}{{\em Nucl. Phys.}
  {\bfseries B353} (1991) 346--378}.

\bibitem{Bodeker:1999gx}
D.~Bodeker, G.~D. Moore, and K.~Rummukainen, ``{Chern-Simons number diffusion
  and hard thermal loops on the lattice},''
  \href{http://dx.doi.org/10.1103/PhysRevD.61.056003}{{\em Phys. Rev.}
  {\bfseries D61} (2000) 056003},
\href{http://arxiv.org/abs/hep-ph/9907545}{{\ttfamily arXiv:hep-ph/9907545
  [hep-ph]}}.

\bibitem{Moore:1999fs}
G.~D. Moore and K.~Rummukainen, ``{Classical sphaleron rate on fine
  lattices},'' \href{http://dx.doi.org/10.1103/PhysRevD.61.105008}{{\em Phys.
  Rev.} {\bfseries D61} (2000) 105008},
\href{http://arxiv.org/abs/hep-ph/9906259}{{\ttfamily arXiv:hep-ph/9906259
  [hep-ph]}}.

\bibitem{Moore:1998swa}
G.~D. Moore, ``{Measuring the broken phase sphaleron rate nonperturbatively},''
  \href{http://dx.doi.org/10.1103/PhysRevD.59.014503}{{\em Phys. Rev.}
  {\bfseries D59} (1999) 014503},
\href{http://arxiv.org/abs/hep-ph/9805264}{{\ttfamily arXiv:hep-ph/9805264
  [hep-ph]}}.

\bibitem{DOnofrio:2014rug}
M.~D'Onofrio, K.~Rummukainen, and A.~Tranberg, ``{Sphaleron Rate in the Minimal
  Standard Model},''
  \href{http://dx.doi.org/10.1103/PhysRevLett.113.141602}{{\em Phys. Rev.
  Lett.} {\bfseries 113} no.~14, (2014) 141602},
\href{http://arxiv.org/abs/1404.3565}{{\ttfamily arXiv:1404.3565 [hep-ph]}}.

\bibitem{Moore:1996qs}
G.~D. Moore, ``{Motion of Chern-Simons number at high temperatures under a
  chemical potential},''
  \href{http://dx.doi.org/10.1016/S0550-3213(96)00445-2}{{\em Nucl. Phys.}
  {\bfseries B480} (1996) 657--688},
\href{http://arxiv.org/abs/hep-ph/9603384}{{\ttfamily arXiv:hep-ph/9603384
  [hep-ph]}}.

\bibitem{Long:2013tha}
A.~J. Long, E.~Sabancilar, and T.~Vachaspati, ``{Leptogenesis and Primordial
  Magnetic Fields},''
  \href{http://dx.doi.org/10.1088/1475-7516/2014/02/036}{{\em JCAP} {\bfseries
  1402} (2014) 036},
\href{http://arxiv.org/abs/1309.2315}{{\ttfamily arXiv:1309.2315
  [astro-ph.CO]}}.

\bibitem{Long:2016uez}
A.~J. Long and E.~Sabancilar, ``{Chiral Charge Erasure via Thermal Fluctuations
  of Magnetic Helicity},''
  \href{http://dx.doi.org/10.1088/1475-7516/2016/05/029}{{\em JCAP} {\bfseries
  1605} no.~05, (2016) 029},
\href{http://arxiv.org/abs/1601.03777}{{\ttfamily arXiv:1601.03777 [hep-th]}}.

\bibitem{Figueroa:2017qmv}
D.~G. Figueroa and M.~Shaposhnikov, ``{Lattice implementation of Abelian gauge
  theories with Chern–Simons number and an axion field},''
  \href{http://dx.doi.org/10.1016/j.nuclphysb.2017.12.001}{{\em Nucl. Phys.}
  {\bfseries B926} (2018) 544--569},
\href{http://arxiv.org/abs/1705.09629}{{\ttfamily arXiv:1705.09629 [hep-lat]}}.

\bibitem{Moore:1996wn}
G.~D. Moore, ``{Improved Hamiltonian for Minkowski Yang-Mills theory},''
  \href{http://dx.doi.org/10.1016/S0550-3213(96)00497-X}{{\em Nucl. Phys.}
  {\bfseries B480} (1996) 689--728},
\href{http://arxiv.org/abs/hep-lat/9605001}{{\ttfamily arXiv:hep-lat/9605001
  [hep-lat]}}.

\bibitem{Kajantie:1998rz}
K.~Kajantie, M.~Laine, J.~Peisa, K.~Rummukainen, and M.~E. Shaposhnikov, ``{The
  Electroweak phase transition in a magnetic field},''
  \href{http://dx.doi.org/10.1016/S0550-3213(98)00854-2}{{\em Nucl. Phys.}
  {\bfseries B544} (1999) 357--373},
\href{http://arxiv.org/abs/hep-lat/9809004}{{\ttfamily arXiv:hep-lat/9809004
  [hep-lat]}}.

\bibitem{Laine:1997dy}
M.~Laine and A.~Rajantie, ``{Lattice continuum relations for 3-D SU(N) + Higgs
  theories},'' \href{http://dx.doi.org/10.1016/S0550-3213(97)00709-8}{{\em
  Nucl. Phys.} {\bfseries B513} (1998) 471--489},
\href{http://arxiv.org/abs/hep-lat/9705003}{{\ttfamily arXiv:hep-lat/9705003
  [hep-lat]}}.

\bibitem{Dimopoulos:1997cz}
P.~Dimopoulos, K.~Farakos, and G.~Koutsoumbas, ``{Three-dimensional lattice
  U(1) gauge Higgs model at low m(H)},''
  \href{http://dx.doi.org/10.1007/s100520050116}{{\em Eur. Phys. J.} {\bfseries
  C1} (1998) 711--719},
\href{http://arxiv.org/abs/hep-lat/9703004}{{\ttfamily arXiv:hep-lat/9703004
  [hep-lat]}}.

\bibitem{Kajantie:1997vc}
K.~Kajantie, M.~Karjalainen, M.~Laine, and J.~Peisa, ``{Masses and phase
  structure in the Ginzburg-Landau model},''
  \href{http://dx.doi.org/10.1103/PhysRevB.57.3011}{{\em Phys. Rev.} {\bfseries
  B57} (1998) 3011--3016},
\href{http://arxiv.org/abs/cond-mat/9704056}{{\ttfamily arXiv:cond-mat/9704056
  [cond-mat]}}.

\bibitem{Kajantie:1997hn}
K.~Kajantie, M.~Karjalainen, M.~Laine, and J.~Peisa, ``{Three-dimensional U(1)
  gauge + Higgs theory as an effective theory for finite temperature phase
  transitions},'' \href{http://dx.doi.org/10.1016/S0550-3213(98)00064-9}{{\em
  Nucl. Phys.} {\bfseries B520} (1998) 345--381},
\href{http://arxiv.org/abs/hep-lat/9711048}{{\ttfamily arXiv:hep-lat/9711048
  [hep-lat]}}.

\bibitem{Bevis:2006mj}
N.~Bevis, M.~Hindmarsh, M.~Kunz, and J.~Urrestilla, ``{CMB power spectrum
  contribution from cosmic strings using field-evolution simulations of the
  Abelian Higgs model},''
  \href{http://dx.doi.org/10.1103/PhysRevD.75.065015}{{\em Phys. Rev.}
  {\bfseries D75} (2007) 065015},
\href{http://arxiv.org/abs/astro-ph/0605018}{{\ttfamily arXiv:astro-ph/0605018
  [astro-ph]}}.

\bibitem{Figueroa:2012kw}
D.~G. Figueroa, M.~Hindmarsh, and J.~Urrestilla, ``{Exact Scale-Invariant
  Background of Gravitational Waves from Cosmic Defects},''
  \href{http://dx.doi.org/10.1103/PhysRevLett.110.101302}{{\em Phys. Rev.
  Lett.} {\bfseries 110} no.~10, (2013) 101302},
\href{http://arxiv.org/abs/1212.5458}{{\ttfamily arXiv:1212.5458
  [astro-ph.CO]}}.

\bibitem{Daverio:2015nva}
D.~Daverio, M.~Hindmarsh, M.~Kunz, J.~Lizarraga, and J.~Urrestilla,
  ``{Energy-momentum correlations for Abelian Higgs cosmic strings},''
  \href{http://dx.doi.org/10.1103/PhysRevD.95.049903,
  10.1103/PhysRevD.93.085014}{{\em Phys. Rev.} {\bfseries D93} no.~8, (2016)
  085014}, \href{http://arxiv.org/abs/1510.05006}{{\ttfamily arXiv:1510.05006
  [astro-ph.CO]}}.
[Erratum: Phys. Rev.D95,no.4,049903(2017)].

\bibitem{Hindmarsh:2017qff}
M.~Hindmarsh, J.~Lizarraga, J.~Urrestilla, D.~Daverio, and M.~Kunz, ``{Scaling
  from gauge and scalar radiation in Abelian Higgs string networks},''
  \href{http://dx.doi.org/10.1103/PhysRevD.96.023525}{{\em Phys. Rev.}
  {\bfseries D96} no.~2, (2017) 023525},
\href{http://arxiv.org/abs/1703.06696}{{\ttfamily arXiv:1703.06696
  [astro-ph.CO]}}.

\bibitem{Figueroa:2015rqa}
D.~G. Figueroa, J.~Garcia-Bellido, and F.~Torrenti, ``{Decay of the standard
  model Higgs field after inflation},''
  \href{http://dx.doi.org/10.1103/PhysRevD.92.083511}{{\em Phys. Rev.}
  {\bfseries D92} no.~8, (2015) 083511},
\href{http://arxiv.org/abs/1504.04600}{{\ttfamily arXiv:1504.04600
  [astro-ph.CO]}}.

\bibitem{Enqvist:2015sua}
K.~Enqvist, S.~Nurmi, S.~Rusak, and D.~Weir, ``{Lattice Calculation of the
  Decay of Primordial Higgs Condensate},''
  \href{http://dx.doi.org/10.1088/1475-7516/2016/02/057}{{\em JCAP} {\bfseries
  1602} no.~02, (2016) 057},
\href{http://arxiv.org/abs/1506.06895}{{\ttfamily arXiv:1506.06895
  [astro-ph.CO]}}.

\bibitem{Figueroa:2016wxr}
D.~G. Figueroa and F.~Torrenti, ``{Parametric resonance in the early
  Universe—a fitting analysis},''
  \href{http://dx.doi.org/10.1088/1475-7516/2017/02/001}{{\em JCAP} {\bfseries
  1702} no.~02, (2017) 001},
\href{http://arxiv.org/abs/1609.05197}{{\ttfamily arXiv:1609.05197
  [astro-ph.CO]}}.

\bibitem{Easther:2006gt}
R.~Easther and E.~A. Lim, ``{Stochastic gravitational wave production after
  inflation},'' \href{http://dx.doi.org/10.1088/1475-7516/2006/04/010}{{\em
  JCAP} {\bfseries 0604} (2006) 010},
\href{http://arxiv.org/abs/astro-ph/0601617}{{\ttfamily arXiv:astro-ph/0601617
  [astro-ph]}}.

\bibitem{GarciaBellido:2007dg}
J.~Garcia-Bellido and D.~G. Figueroa, ``{A stochastic background of
  gravitational waves from hybrid preheating},''
  \href{http://dx.doi.org/10.1103/PhysRevLett.98.061302}{{\em Phys. Rev. Lett.}
  {\bfseries 98} (2007) 061302},
\href{http://arxiv.org/abs/astro-ph/0701014}{{\ttfamily arXiv:astro-ph/0701014
  [astro-ph]}}.

\bibitem{GarciaBellido:2007af}
J.~Garcia-Bellido, D.~G. Figueroa, and A.~Sastre, ``{A Gravitational Wave
  Background from Reheating after Hybrid Inflation},''
  \href{http://dx.doi.org/10.1103/PhysRevD.77.043517}{{\em Phys. Rev.}
  {\bfseries D77} (2008) 043517},
\href{http://arxiv.org/abs/0707.0839}{{\ttfamily arXiv:0707.0839 [hep-ph]}}.

\bibitem{Dufaux:2007pt}
J.~F. Dufaux, A.~Bergman, G.~N. Felder, L.~Kofman, and J.-P. Uzan, ``{Theory
  and Numerics of Gravitational Waves from Preheating after Inflation},''
  \href{http://dx.doi.org/10.1103/PhysRevD.76.123517}{{\em Phys. Rev.}
  {\bfseries D76} (2007) 123517},
\href{http://arxiv.org/abs/0707.0875}{{\ttfamily arXiv:0707.0875 [astro-ph]}}.

\bibitem{Dufaux:2008dn}
J.-F. Dufaux, G.~Felder, L.~Kofman, and O.~Navros, ``{Gravity Waves from
  Tachyonic Preheating after Hybrid Inflation},''
  \href{http://dx.doi.org/10.1088/1475-7516/2009/03/001}{{\em JCAP} {\bfseries
  0903} (2009) 001},
\href{http://arxiv.org/abs/0812.2917}{{\ttfamily arXiv:0812.2917 [astro-ph]}}.

\bibitem{Figueroa:2011ye}
D.~G. Figueroa, J.~Garcia-Bellido, and A.~Rajantie, ``{On the
  Transverse-Traceless Projection in Lattice Simulations of Gravitational Wave
  Production},'' \href{http://dx.doi.org/10.1088/1475-7516/2011/11/015}{{\em
  JCAP} {\bfseries 1111} (2011) 015},
\href{http://arxiv.org/abs/1110.0337}{{\ttfamily arXiv:1110.0337
  [astro-ph.CO]}}.

\bibitem{Figueroa:2016ojl}
D.~G. Figueroa, J.~García-Bellido, and F.~Torrentí, ``{Gravitational wave
  production from the decay of the standard model Higgs field after
  inflation},'' \href{http://dx.doi.org/10.1103/PhysRevD.93.103521}{{\em Phys.
  Rev.} {\bfseries D93} no.~10, (2016) 103521},
\href{http://arxiv.org/abs/1602.03085}{{\ttfamily arXiv:1602.03085
  [astro-ph.CO]}}.

\bibitem{Figueroa:2017vfa}
D.~G. Figueroa and F.~Torrenti, ``{Gravitational wave production from
  preheating: parameter dependence},''
  \href{http://dx.doi.org/10.1088/1475-7516/2017/10/057}{{\em JCAP} {\bfseries
  1710} no.~10, (2017) 057},
\href{http://arxiv.org/abs/1707.04533}{{\ttfamily arXiv:1707.04533
  [astro-ph.CO]}}.

\bibitem{Hindmarsh:2013xza}
M.~Hindmarsh, S.~J. Huber, K.~Rummukainen, and D.~J. Weir, ``{Gravitational
  waves from the sound of a first order phase transition},''
  \href{http://dx.doi.org/10.1103/PhysRevLett.112.041301}{{\em Phys. Rev.
  Lett.} {\bfseries 112} (2014) 041301},
\href{http://arxiv.org/abs/1304.2433}{{\ttfamily arXiv:1304.2433 [hep-ph]}}.

\bibitem{Hindmarsh:2015qta}
M.~Hindmarsh, S.~J. Huber, K.~Rummukainen, and D.~J. Weir, ``{Numerical
  simulations of acoustically generated gravitational waves at a first order
  phase transition},'' \href{http://dx.doi.org/10.1103/PhysRevD.92.123009}{{\em
  Phys. Rev.} {\bfseries D92} no.~12, (2015) 123009},
\href{http://arxiv.org/abs/1504.03291}{{\ttfamily arXiv:1504.03291
  [astro-ph.CO]}}.

\bibitem{Cutting:2018tjt}
D.~Cutting, M.~Hindmarsh, and D.~J. Weir, ``{Gravitational waves from vacuum
  first-order phase transitions: from the envelope to the lattice},''
\href{http://arxiv.org/abs/1802.05712}{{\ttfamily arXiv:1802.05712
  [astro-ph.CO]}}.

\bibitem{Ambjorn:1995xm}
J.~Ambjorn and A.~Krasnitz, ``{The Classical sphaleron transition rate exists
  and is equal to 1.1 (alpha(w) T)**4},''
  \href{http://dx.doi.org/10.1016/0370-2693(95)01157-L}{{\em Phys. Lett.}
  {\bfseries B362} (1995) 97--104},
\href{http://arxiv.org/abs/hep-ph/9508202}{{\ttfamily arXiv:hep-ph/9508202
  [hep-ph]}}.

\bibitem{Arnold:1997yb}
P.~B. Arnold, ``{Hot B violation, the lattice, and hard thermal loops},''
  \href{http://dx.doi.org/10.1103/PhysRevD.55.7781}{{\em Phys. Rev.} {\bfseries
  D55} (1997) 7781--7796},
\href{http://arxiv.org/abs/hep-ph/9701393}{{\ttfamily arXiv:hep-ph/9701393
  [hep-ph]}}.

\bibitem{Kubo:1957mj}
R.~Kubo, ``{Statistical mechanical theory of irreversible processes. 1. General
  theory and simple applications in magnetic and conduction problems},''
\href{http://dx.doi.org/10.1143/JPSJ.12.570}{{\em J. Phys. Soc. Jap.}
  {\bfseries 12} (1957) 570--586}.

\bibitem{Kraemmer:1994az}
U.~Kraemmer, A.~K. Rebhan, and H.~Schulz, ``{Resummations in hot scalar
  electrodynamics},'' \href{http://dx.doi.org/10.1006/aphy.1995.1023}{{\em
  Annals Phys.} {\bfseries 238} (1995) 286--331},
\href{http://arxiv.org/abs/hep-ph/9403301}{{\ttfamily arXiv:hep-ph/9403301
  [hep-ph]}}.

\bibitem{Figueroa:2019jsi}
D.~G. Figueroa, A.~Florio, and M.~Shaposhnikov, ``{Chiral charge dynamics in
  Abelian gauge theories at finite temperature},''
  \href{http://dx.doi.org/10.1007/JHEP10(2019)142}{{\em JHEP} {\bfseries 10}
  (2019) 142}, \href{http://arxiv.org/abs/1904.11892}{{\ttfamily
  arXiv:1904.11892 [hep-th]}}.

\bibitem{Arnold:2000dr}
P.~B. Arnold, G.~D. Moore, and L.~G. Yaffe, ``{Transport coefficients in high
  temperature gauge theories. 1. Leading log results},''
  \href{http://dx.doi.org/10.1088/1126-6708/2000/11/001}{{\em JHEP} {\bfseries
  11} (2000) 001},
\href{http://arxiv.org/abs/hep-ph/0010177}{{\ttfamily arXiv:hep-ph/0010177
  [hep-ph]}}.

\bibitem{Arnold:2003zc}
P.~B. Arnold, G.~D. Moore, and L.~G. Yaffe, ``{Transport coefficients in high
  temperature gauge theories. 2. Beyond leading log},''
  \href{http://dx.doi.org/10.1088/1126-6708/2003/05/051}{{\em JHEP} {\bfseries
  05} (2003) 051},
\href{http://arxiv.org/abs/hep-ph/0302165}{{\ttfamily arXiv:hep-ph/0302165
  [hep-ph]}}.

\end{thebibliography}\endgroup

\newpage
\section{\it Erratum}

To confront the numerical results of $\Gamma_{\rm diff}$ with the analytical results from Sect.~2.3, we originally considered the theoretical prediction for the diffusion rate given by Eq.~(2.17), which we re-wrote in Eq.~(4.25). However, we have more recently found in Ref.~\cite{Figueroa:2019jsi} (Ref.~I below) that Eq.~(2.17), based on Eq.~(2.16), has an extra factor $2$. The correct expression for Eq.~(2.16) should rather read $\Gamma_5 = 6 \frac{\Gamma}{T^3}$ (instead of $\Gamma_5 = 12 \frac{\Gamma}{T^3}$), see Appendix B in Ref.~\cite{Figueroa:2019jsi} for details. Furthermore, to make the comparison between the theoretical rate and our lattice prediction, we also used Eq.~(1.11) for the MHD conductivity. The conductivity prediction has been however refined in Refs.~\cite{Arnold:2000dr,Arnold:2003zc} (Refs.~II, III below). Putting all together, we conclude in Ref.~\cite{Figueroa:2019jsi} that the effective diffusion rate expected in MHD, can be written as
\begin{eqnarray}\label{eq:GammaDiffMHD}
\Gamma^{\rm (th)}_{\rm diff} \simeq 4.1\cdot 10^{-5}\log({17.6/e^2})\,e^6B^2\,.
\end{eqnarray}
Comparing the theoretical prediction Eq.~(\ref{eq:GammaDiffMHD}) [say for $e^2 = 1$] with a re-analysis of the numerical diffusion rate $\Gamma_{\rm diff}$ (by weighting the mean values of our data with the error $\Delta\Gamma_{\rm diff}$, c.f.~Eq.~(4.8), and without assuming an enforcement of a fixed exponent in the scaling of $\Gamma$ with $e^2$), we obtain now in Ref.~\cite{Figueroa:2019jsi}
\begin{eqnarray}
{\Gamma^{\rm (num)}_{\rm diff}\over\Gamma_{\rm diff}^{\rm (th)}}\Big|_{e^2 = 1} = 11.2\substack{+6.9 \\ -4.3}\,.
\label{eq:ratios}
\end{eqnarray}
This computation reduces by a factor $\sim 5-6$ our original claim in the discrepancy between theory and numerics: we still obtain that the numerically extracted rates are larger than the MHD counterpart by a factor $\mathcal{O}(10)$, but this factor is rather $\sim 11$ instead of $\sim 58$, as originally claimed. The reduction from a factor $\sim 58$ down to $\sim 11$ is a combined effect of correcting a factor 2 in Eq.~(2.16) [this leads to a ratio $\sim 29$] and a factor $\sim 2.6$  when comparing the numerical result against the theoretical prediction Eq.~(\ref{eq:GammaDiffMHD}), instead of Eq.~(1.11) [this leads to the final ratio $\sim 11$]. The errors in the new ratio are also larger as the numerical fit we use now exhibits larger errors, given that we do not fix the scaling power of $\Gamma_{\rm diff}$ with $e^2$. If we enforce a scaling as $\Gamma_{\rm diff} \propto e^6$, as we did originally in the main tex of the article, we still obtain a similar ratio $11.4\substack{+3.0 \\ -2.4}$, albeit with smaller errors.

\vspace*{1cm}

\noindent------------------\vspace*{0.5cm}

\noindent [I] Daniel G. Figueroa, Adrien Florio, Mikhail Shaposhnikov, {\it Chiral charge dynamics in Abelian gauge theories at finite temperature}, JHEP 10 (2019) 142 (1904.11892 [hep-th])\vspace*{0.2cm}

\noindent [II] P. B. Arnold, G. D. Moore and L. G. Yaffe, {\it Transport coefficients in high temperature gauge theories. 1. Leading log results}, JHEP 11 (2000) 001 [hep-ph/0010177].\vspace*{0.2cm}

\noindent [III] P. B. Arnold, G. D. Moore and L. G. Yaffe, {\it Transport coefficients in high temperature gauge theories. 2. Beyond leading log}, JHEP 05 (2003) 051 [hep-ph/0302165]

\end{document}